\documentclass[twocolumn]{aastex63}
\usepackage{rotating}

\newcommand\tstrut{\rule{0pt}{3.4ex}}
\newcommand\bstrut{\rule[-2.4ex]{0pt}{0pt}}
\newcommand{\msun}{$M_{\odot}$}

\shorttitle{Wide WD+WD Ages}
\shortauthors{Heintz et al.}

\graphicspath{{./}}
\begin{document}

\title{Testing White Dwarf Age Estimates using Wide Double White Dwarf Binaries from Gaia EDR3}

\correspondingauthor{Tyler M. Heintz}
\email{tmheintz@bu.edu}

\author[0000-0003-3868-1123]{Tyler M. Heintz}
\affiliation{Department of Astronomy \& Institute for Astrophyiscal Research, Boston University, 725 Commonwealth Ave, Boston, MA, 02215, USA}

\author[0000-0001-5941-2286]{J. J. Hermes}
\affiliation{Department of Astronomy \& Institute for Astrophyiscal Research, Boston University, 725 Commonwealth Ave, Boston, MA, 02215, USA}

\author[0000-0002-6871-1752]{Kareem El-Badry}
\affiliation{Center for Astrophysics $|$ Harvard \& Smithsonian, 60 Garden Street, Cambridge, MA 02138, USA}
\affiliation{Harvard Society of Fellows, 78 Mount Auburn Street, Cambridge, MA 02138}
%\affiliation{Max-Planck Institute for Astronomy, K\"onigstuhl 17, D-69117 Heidelberg, Germany}

\author{Charlie Walsh}
\affiliation{Department of Astronomy \& Institute for Astrophyiscal Research, Boston University, 725 Commonwealth Ave, Boston, MA, 02215, USA}

\author[0000-0002-4284-8638]{Jennifer L. van Saders}
\affiliation{Institute for Astronomy, University of Hawai'i, 2680 Woodlawn Drive, Honolulu, HI 96822, USA}

\author[0000-0002-8925-057X]{C.~E.~Fields}
\altaffiliation{Feynman Fellow}
\affiliation{Center for Theoretical Astrophysics, Los Alamos National Laboratory, Los Alamos, NM 87545, USA}
\affiliation{Computer, Computational, and Statistical Sciences Division, Los Alamos National Laboratory, Los Alamos, NM 87545, USA}
\affiliation{X Computational Physics Division, Los Alamos National Laboratory, Los Alamos, NM 87545, USA}

\author[0000-0002-4284-8638]{Detlev Koester}
\affiliation{Institut f{\"u}r Theoretische Physik und Astrophysik, University of Kiel, 24098 Kiel, Germany}

\begin{abstract}
White dwarf (WD) stars evolve simply and predictably, making them reliable age indicators. However, self-consistent validation of the methods for determining WD total ages has yet to be widely performed. This work uses 1565 wide ($>100$\,au) WD+WD binaries and 24 new triples containing at least two WDs to test the accuracy and validity of WD total age determinations. For these 1589 wide double-WD binaries and triples, we derive total ages of each WD using photometric data from all-sky surveys, in conjunction with Gaia parallaxes and current hydrogen-atmosphere WD models. Ignoring initial-to-final-mass relations and considering only WD cooling ages, we find that roughly $21-36\%$ of the more massive WDs in a system have a shorter cooling age. Since more massive WDs should be born as more massive main-sequence stars, we attribute this unphysical disagreement as evidence of prior mergers or the presence of an unresolved companion, suggesting that roughly $21-36\%$ of wide WD+WD binaries were once triples. Among the 423 wide WD+WD pairs that pass high-fidelity cuts, we find that 25\% total age uncertainties are generally appropriate for WDs with masses $>0.63$\,$M_{\sun}$ and temperatures $<12{,}000\,K$, and provide suggested inflation factors for age uncertainties for higher-mass WDs. Overall, WDs return reliable stellar ages, but we detail cases where total ages are least reliable, especially for WDs $<0.63$\,$M_{\sun}$.
\end{abstract}


\section{Introduction} \label{sec:intro}

White dwarf (WD) stars are the end stages of all stars with initial masses less than roughly $8-10$\,$M_{\sun}$ \citep{2015MNRAS.446.2599D} and have been used for decades as stellar chronometers (see review by \citealt{2001PASP..113..409F}). Since WDs no longer undergo core fusion, their evolution is generally a cooling problem. The development of robust cooling models (e.g., \citealt{1995PASP..107.1047B}) enables WDs to act as precise and accurate age indicators. However, to determine the total age of a WD, the time from the zero-age main sequence (ZAMS) to the current state of the WD is required. The progenitor lifetimes from ZAMS to the WD phase are determined through the use of a semi-empirical initial-final mass relation (IFMR, \citealt{2008MNRAS.387.1693C}, \citealt{2008ApJ...676..594K}, \citealt{2009ApJ...692.1013S}, \citealt{2018ApJ...866...21C}) in conjunction with stellar evolution model grids, such as MESA \citep{2016ApJS..222....8D} and PARSEC \citep{2012MNRAS.427..127B}. 
 
The total ages of WDs can be used to date a variety of astronomical objects, including wide binary companions, such as M dwarfs (e.g., \citealt{2019ApJ...870....9F,2021ApJS..253...58Q}, \citealt{2021AJ....161..277K}), brown dwarfs \citep{2020ApJ...891..171Z}, and a growing number of planet-host stars \citep{2021MNRAS.507.4132M}. WDs can also age-date larger populations, such as stars in the disk (e.g., \citealt{1987ApJ...315L..77W}), or even the halo of our Galaxy \citep{2019MNRAS.482..965K,2021MNRAS.502.1753T}. WD ages can also be used to study the star formation history of the Milky Way (e.g., \citealt{2014ApJ...791...92T}). An understanding of the limitations and reliability of WD total ages is crucial to the inferences derived from these studies.
 
The total age derived for a WD is strongly dependent on the IFMR used, especially for the lowest-mass WDs ($M<0.7$\,$M_{\sun}$). However, the low-mass region of the IFMR is poorly sampled because nearby clusters containing at least one WD are not old enough. Attempts have been made to alleviate this problem by using wide binary companions to WDs \citep{2012ApJ...746..144Z, 2021ApJ...923..181B} and the distribution of WDs in the color-magnitude diagram \citep{2018ApJ...860L..17E}, but this still results in a large range of predicted progenitor masses across different IFMRs. Further, the IFMR may show evidence of a kink where the mass-loss relationship is no longer monotonic \citep{2020NatAs...4.1102M}. The IFMR is likely also sensitive to the metallicity and initial rotation of the progenitor star \citep{2019ApJ...871L..18C}. These parameters are not easy to determine, and thus assumptions for their values must be made.
 
Widely separated double WD binaries (WD+WD binaries) are ideal systems for checking the viability of our existing methods for determining WD total ages. It is safe to assume the two WDs bound to each binary system formed from the same collapsing molecular cloud at nearly the same time with the same chemical composition \citep{2001ApJ...556..265W, 2009ApJ...704..531K}. Wide WD+WD binaries that are sufficiently separated ($>100$\,au) have not experienced any significant mass transfer events or other interactions between the two stars during their lifetimes. At intermediate separations (1-100\,au) where tidal forces, mass transfer, and accretion are not a significant source of interaction, it has been shown that the interaction between a companion and the circumstellar disk can affect the angular momentum evolution of the star \citep{2012AJ....144...93M}. In the absence of any interactions between the two stars, any methods for determining the total age of a WD should independently return the same total age for both components of the binary.
 
Since the launch of the Gaia spacecraft (\citealt{2016A&A...595A...1G}, \citealt{2018A&A...616A...1G}, \citealt{2021A&A...649A...1G}), the sample of wide WD+WD binary systems has increased by an order of magnitude (\citealt{2015ApJ...815...63A}, \citealt{2018MNRAS.480.4884E}, \citealt{Tian}, \citealt{2021MNRAS.506.2269E}). This increase in systems presents the opportunity for a large statistical study into the accuracy of the models and methods used to determine the total ages of WDs, which is the focus of this work. In Section 2, we present the sample selection of resolved WD+WD binary systems, including 23 new WD+WD+MS systems and one bound WD+WD+WD system. Section 3 discusses the techniques used to determine their ages. In Section 4, we discuss comparisons of the total ages in each binary. We end with a summary of the findings and results in Section 5.

\begin{figure*}[t]
    \centering
    \includegraphics[width=1.95\columnwidth]{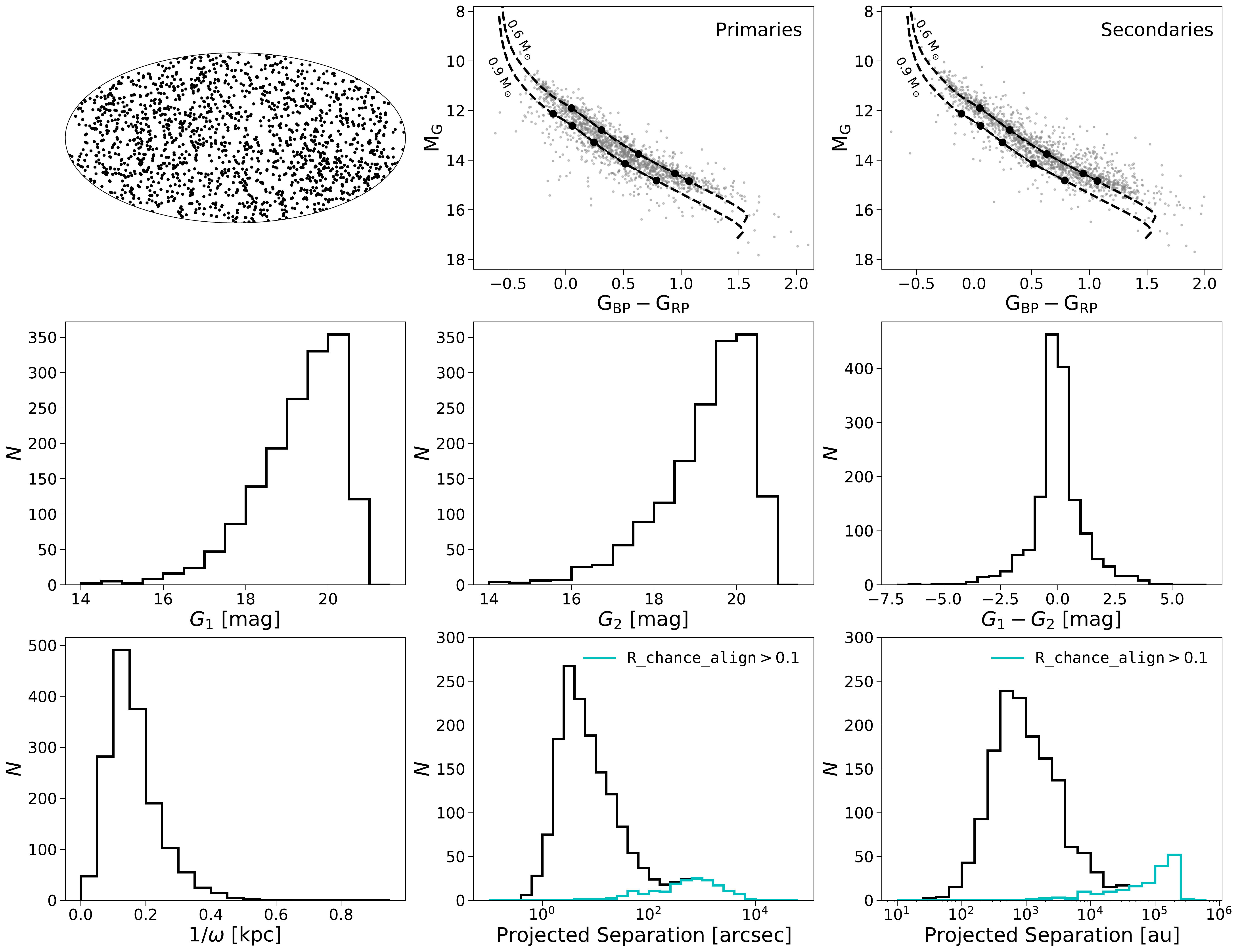}
    \caption{Basic properties of the full sample of 3179 WDs in wide WD+WD binaries from \citet{2021MNRAS.506.2269E} used in this work, including 23 of our newly discovered wide WD+WD+MS triples and one WD+WD+WD triple: distribution in galactic coordinates, color-magnitude diagrams for the primaries and secondaries, apparent magnitude distributions, magnitude difference, distances, and projected angular and physical separations. The primaries are defined to be the more massive WD component. A large range of the WD cooling tracks is covered by the sample, which allows us to test WD total ages across a representative sample of WD masses and effective temperatures. Cooling tracks for a 0.6\,$M_{\sun}$ and 0.9\,$M_{\sun}$ WD model from \citet{2020ApJ...901...93B} are shown for reference on the color-magnitude diagrams, as well as dots representing WD cooling ages of 0.5, 1, 2, 4, and 6 Gyr for each. The separation distribution of pairs with $\texttt{R\_chance\_align}>0.1$ is shown in cyan for reference in the bottom right panels. The upturn in the distribution at high separations is caused by these possible chance alignments.}
    \label{fig:sample}
\end{figure*}

\section{Sample Selection}\label{sec:sample}
Obtaining a large sample of wide double WD pairs is crucial for this work. The improved parallaxes and proper motions for more than 1 billion sources all-sky with the release of Gaia EDR3 (\citealt{2021A&A...649A...1G}) significantly increased the potential yield of searches for wide common-proper-motion pairs. With this larger sample we can delve into the systematic errors present in WD total age determinations.

\subsection{Binary Sample}
For this work, the base of our sample is composed of the 1565 wide WD+WD binaries from \citet{2021MNRAS.506.2269E}. Candidate pairs in this work were selected to have projected separations less than a parsec, parallaxes that are consistent within $3\sigma$, and proper motions consistent within $2\sigma$ of a bound orbit. This sample is restricted to objects with fractional parallax uncertainties less than 20\% for both components, with most systems within 300\,pc. The sample covers a wide range of magnitudes, spanning $13<G<21$\,mag, with roughly 68\% of the 3130 individual WDs in the sample fainter than $G>19$\,mag.
 
\citet{2021MNRAS.506.2269E} calculate a chance alignment factor (\texttt{R\_chance\_align}) for each candidate binary pair through the use of a shifted Gaia catalog. Conducting a search for binaries in the shifted catalog only produces chance alignments, and thus can be used as a way to determine the likelihood that a binary pair is a chance alignment (see their Sec 3). This parameter can be used to filter out low-fidelity wide binaries.

\subsection{Triple Sample}

To further increase the sample of wide double WDs, we conduct a search for wide, resolved triple star systems containing at least two WDs. In \citet{2021MNRAS.506.2269E}, all triple systems are intentionally removed from the sample by the neighboring pairs cut (see their Sec 2.1). In this work, we take the sample of binary pairs from \citet{2021MNRAS.506.2269E} before the cleaning of the sample and remove the neighboring pairs cut, but retain the cut on neighboring sources of $N < 30$, where a neighboring source is defined, following \citet{2021MNRAS.506.2269E}, as a source with (i) projected on-sky separation less than 5 parsecs; (ii) proper motion difference less than 5 $\mathrm{km\ s^{-1}}$ with a $2\sigma$ tolerance; (iii) parallaxes consistent within $2\sigma$.

Next, we make a list of systems of various sizes by combining all binaries with at least one common source between them. We iterate this process until each system is independent of each other.
 
To clean this new sample of possible wide systems, we use the classifier for spurious astrometric solutions provided in \citet{2021arXiv210111641R}. We require each component of the triple to have the \texttt{fidelity\_v1}  parameter greater than 0.5. This assures that each source has a reliable astrometric solution. We are only concerned with finding systems with three components in this extended list and throw out any systems with a size not equal to three. (Repeating this exercise for $N=4,5,\ldots$, we do not find any systems with a size greater than $3$ in this extended list of wide systems with at least two WDs.)
 
With these cuts, we identify 23 new WD+WD+MS triples and one WD+WD+WD triple. This list includes the first resolved triple WD system found by \citet{2019MNRAS.483..901P}. All WD+WD+MS and WD+WD+WD triple systems are detailed in Table~\ref{tab:triples}. All but three of the triples in our sample are hierarchical, where we have defined a hierarchical triple as one in which the largest separation is greater than five times the smallest separation. An excess of hierarchical triples is expected since the hierarchy improves dynamical stability in the system \citep{1983ApJ...268..319H}.

The total ages on the non-hierarchical systems (WDWDMS EDR3 3681220906103896192, WDWDMS EDR3 6756233901661753472, WDWDMS EDR3 929001808377561216) can provide interesting constraints on the timescales of dissipation of such systems. In both WDWDMS EDR3 3681220906103896192 and WDWDMS EDR3 6756233901661753472, one of the WDs does not have a total age estimate since it has a mass of $0.2$\msun and $0.35$\msun, respectively, but the total ages of the other WDs in these triples are $92^{+3}_{-5}$ Myr and $2.9^{+0.4}_{-0.2}$ Gyr, respectively. The weighted mean total age of the WDs in WDWDMS EDR3 929001808377561216 is $3.3^{+3.1}_{-1.3}$ Gyr. Further information is found in Table~\ref{tab:triples}.

Although two of these non-heirarchical wide triples seem too old ($>2.5$\,Gyr) to still be bound, it has recently been shown that the unstable phase of these triples on average lasts $10^3-10^4$ crossing times, and can even last as long as $10^{6}-10^{7}$ \citep{2021arXiv210804272T}. Assuming the crossing time is of similar magnitude to the outermost orbital period, the two older systems would have a crossing time of $\sim{10^4}$ years, indicating the older systems have experienced $\sim{10^5}$ crossing times. Of the 21 systems that are hierarchical, 19 systems have an inner pair of two WDs, although this is likely an observational bias caused by the high luminosity ratio of a MS star to a WD.

Our search is sensitive to wide triple WD systems, but we do not find any new wide WD+WD+WD systems. We do find a possible candidate for a new WD+WD+WD system (Gaia EDR3 source IDs: 261249666477351552, 261249670772776832, 261249636413169920) but one component has a \texttt{fidelity\_v1} parameter from \citet{2021arXiv210111641R} of 0.01. Still, all three WDs have significant parallaxes and proper motions that are in better than $2\sigma$ agreement. With all components $G>20$\,mag, this potential new WD+WD+WD is significantly fainter than the first published triple WD system. This limits the parallaxes of each WD to 10\%-15\% uncertainties. The system lies somewhere between $100-130$\,pc, with the weighted mean distance being $118\pm8$\,pc. This possible WD triple has a similar hierarchy as the first such identified system in \cite{2019MNRAS.483..901P}; the system has an inner pair  separated by $\approx$$170$\,au and an outer component separated by $\approx$$6500$\,au. (The first triple WD system has an inner separation of $\approx$$300$\,au and outer separation of $\approx$$6400$\,au.) Follow-up analysis is needed to confirm the true nature of this possible WD+WD+WD system.

\subsection{Full Sample}\label{sec:sample_full}

Various parameter distributions of the full sample, including a color-magnitude diagram (CMD) of all wide double WDs, is shown in Figure~\ref{fig:sample}. Our full sample of wide WD+WD pairs spans a large range of the WD cooling sequence, covering many masses and effective temperatures. The sample also spans many different evolutionary stages. In fact, we expect roughly 430 WDs (13\%) to be mostly crystallized, based on the 80\% crystallization boundary from \citet{2019Natur.565..202T}. However, many of our targets are quite faint; the majority of sources are fainter than $G>19.5$\,mag (see Figure~\ref{fig:sample}). The separation distribution shows an excess of high separation pairs. These pairs are likely chance alignments and can be removed with a cut on \texttt{R\_chance\_align} from \citet{2021MNRAS.506.2269E}, as discussed in Section~\ref{sec:results}.
 
In preparation for the eventual total age determination, we gather additional photometric data on each WD in order to sample a larger range of wavelengths than just the three broad photometric bands in Gaia EDR3. We use the crossmatches provided by the Gaia team for EDR3 (\citealt{2021gdr3.reptE...9M}) to get photometric data from the Sloan Digital Sky Survey (SDSS), the Panoramic Survey Telescope and Rapid Response System (PanSTARRS), the SkyMapper Southern Survey, and the Two Micron All Sky Survey (2MASS). We find that 32\%, 70\%, 23\%, and 4\% of the WDs in our sample have photometry in SDSS, PanSTARRS, Skymapper, and 2MASS, respectively.

\section{Total Age Determination}
WD total ages have two necessary components that need to be determined: the time the star has spent as a WD (its cooling age) and its progenitor main-sequence and giant branch lifetimes. To calculate the cooling age of a WD, atmospheric parameters of the WD must be determined, which can be derived using either spectroscopic or photometric data in conjunction with model atmospheres. Fundamentally, the WD surface gravity and temperature must be determined.

Spectroscopic observations are available in the literature for roughly 300 of the WDs in our sample, so we lack spectroscopy for the vast majority of our 3179 objects. Therefore, we resort to conducting spectral energy distribution (SED) fitting of the average fluxes in various photometric bands from all sky surveys. In practice, SED fitting is sensitive to the radius and temperature of a WD. This radius can be converted into a surface gravity through the use of the mass-radius relationship for WDs \citep{2020ApJ...901...93B}.

Once a surface gravity and temperature are determined, an estimate of the mass of the WD can be calculated using the same mass-radius relationship. This estimate of the mass, with the use of an IFMR and stellar evolutionary grids, can be used to generate an estimate of the lifetime of the ZAMS progenitor. We discuss this process in more detail below.

\subsection{WD Atmospheric Parameters}
\label{sec:atm}

To determine masses and effective temperatures of the WDs in our sample, we use available photometry from Gaia EDR3 and the all-sky surveys discussed in Section \ref{sec:sample_full}, in conjunction with a fitting technique similar to the photometric technique described in \cite{2019ApJ...876...67B}.
 
Reddening effects from interstellar gas and dust can be important for the WDs in our sample (the majority are at a distance beyond 150\,pc). Thus, the observed magnitudes in each band are dereddened following the process outlined in \citealt{2019MNRAS.482.4570G}, which we summarize here. We use reddening maps from \cite{2011ApJ...737..103S} to determine the extinction coefficient in the Johnson $V$ band, $A_{{\rm V}}$. Then, the $A_{{\rm V}}$ values are converted to extinction coefficients in the other bands using the appropriate $R$ values. We assume the material is concentrated along the galactic plane and has a scale height of 200\,pc. When parameterised this way, the dereddened magnitude in a given band $x$ is determined by,
\begin{equation}
    M_{{\rm x}} = M_{{\rm x,obs}} - A_{{\rm x}}\left(1-\exp \left(-\frac{\mathrm{sin\left(|b|\right)}}{200\omega}\right)\right) 
\end{equation}
where $M_{{\rm x,obs}}$ is the observed magnitude, $A_{{\rm x}}$ is the extinction coefficient, $b$ is the galactic latitude, and $\omega$ is the parallax in arcseconds. A more sophisticated reddening determination can be used similar to what is done in \cite{2021MNRAS.508.3877G}, but in comparisons with atmospheric parameters derived in their work, we find no systematic shifts and the parameters are in good agreement (as demonstrated below).
 
The observed, dereddened magnitudes are converted to average fluxes using the appropriate zero points for each bandpass. We apply the $G$-band magnitude correction for all sources with a six-parameter astrometric solution provided in \cite{2021A&A...649A...1G}. To account for the offset of SDSS magnitudes from the AB magnitude standard, we apply an offset of $-$0.04 and 0.02 mags in the $u$ band and $z$ band, respectively \citep{2006AJ....132..676E}. For each survey, we remove photometry that has been flagged for various issues. For SDSS, we remove photometry that has been flagged with \texttt{EDGE}, \texttt{PEAKCENTER}, \texttt{SATUR}, and \texttt{NOTCHECKED}. We remove photometry from PanSTARRS that uses rank detections less than one. For SkyMapper, we only use photometry that has no raised flags. We also enforce a minimum uncertainty in each bandpass of 0.03\,mag to account for systematic errors in the conversion of magnitudes to average fluxes. This also gives equal weight to each bandpass and one point with a small error will not dominate the behavior of the fits. 

We generate synthetic averages fluxes in a given bandpass using synthetic hydrogen atmosphere WD (spectral type DA) spectra from \cite{2010MmSAI..81..921K}, convolved with the appropriate filter transmission profiles. The observed and synthetic fluxes are adjusted to a distance of 10\,pc using the weighted mean of the parallaxes of the two components of the binary, or the weighted mean of all three components for our triples. To adjust the synthetic fluxes to 10\,pc, the radius of the WD must be calculated from a mass-radius relation. We use the WD evolutionary models from \cite{2020ApJ...901...93B} to convert the temperature and surface gravity of a given model to a radius. The photometric uncertainties are then added in quadrature with the uncertainty on this weighted mean parallax. The use of the weighted parallax in these systems allows for more precise determinations of the atmospheric parameters due to its higher precision. For the sample of triples, the precision of the parallax is six times better on average, because the MS companions are much brighter. This leads to the WDs in these triples having smaller mass uncertainties than what is expected given their brightness (an average improvement of 40\% for the brighter WD component and 15\% for the fainter WD component). 
%(for the sample of triples, the precision of the parallax is six times better on average).
 
Due to the lack of spectral information for the majority of our objects, we assume all of our WDs are DAs. Although fitting a non-DA WD with a DA model can introduce systematic mass errors on the order of $10-15$\% \citep{2012ApJS..199...29G}, we are forced to choose a model. A DA model is the simplest assumption, considering $>$65\% of WDs in magnitude-limited samples within Gaia are DAs \citep{2013ApJS..204....5K}. 
 
\begin{figure}[t!]
     \centering
     \gridline{\fig{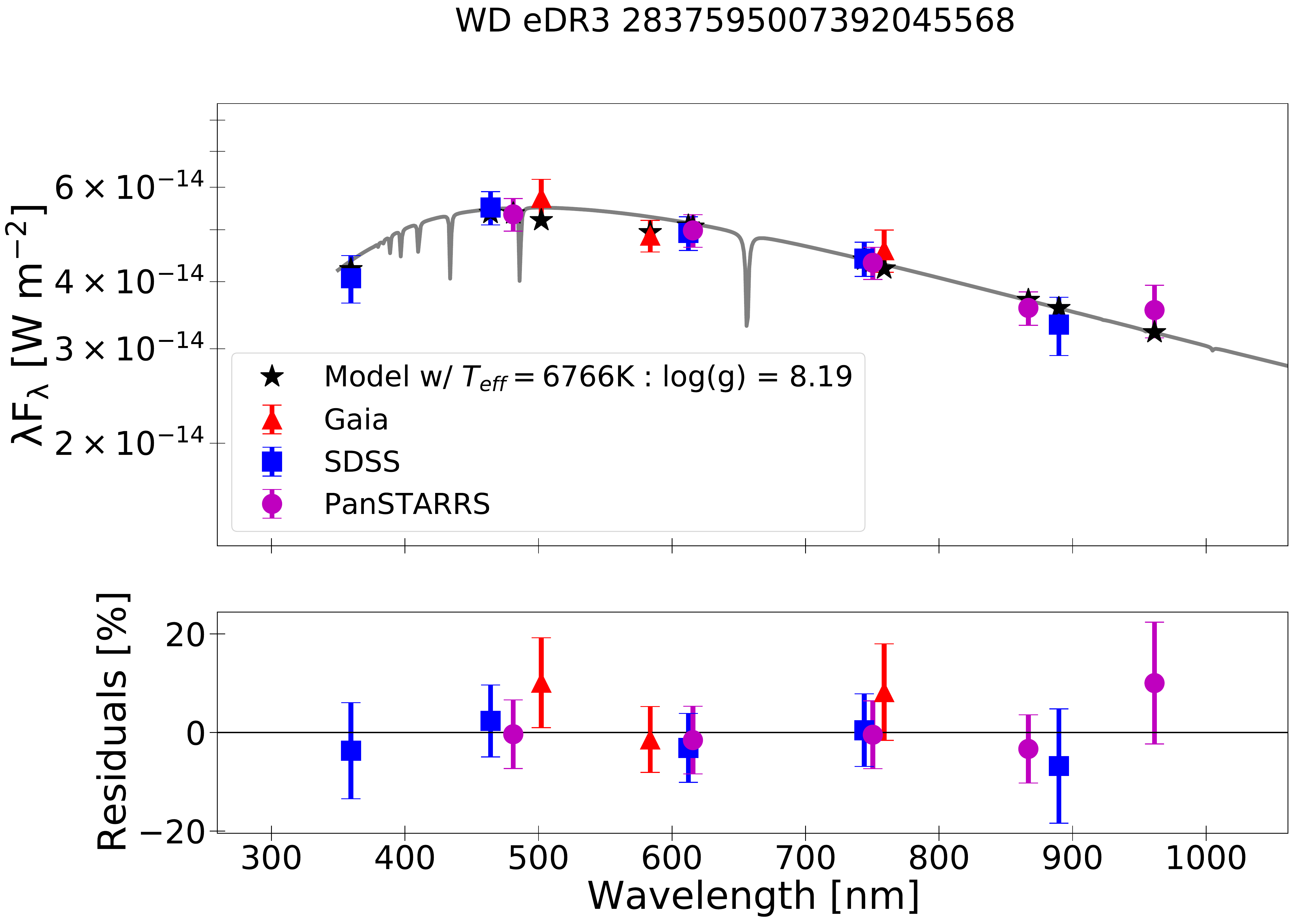}{0.99\columnwidth}{}} \gridline{\fig{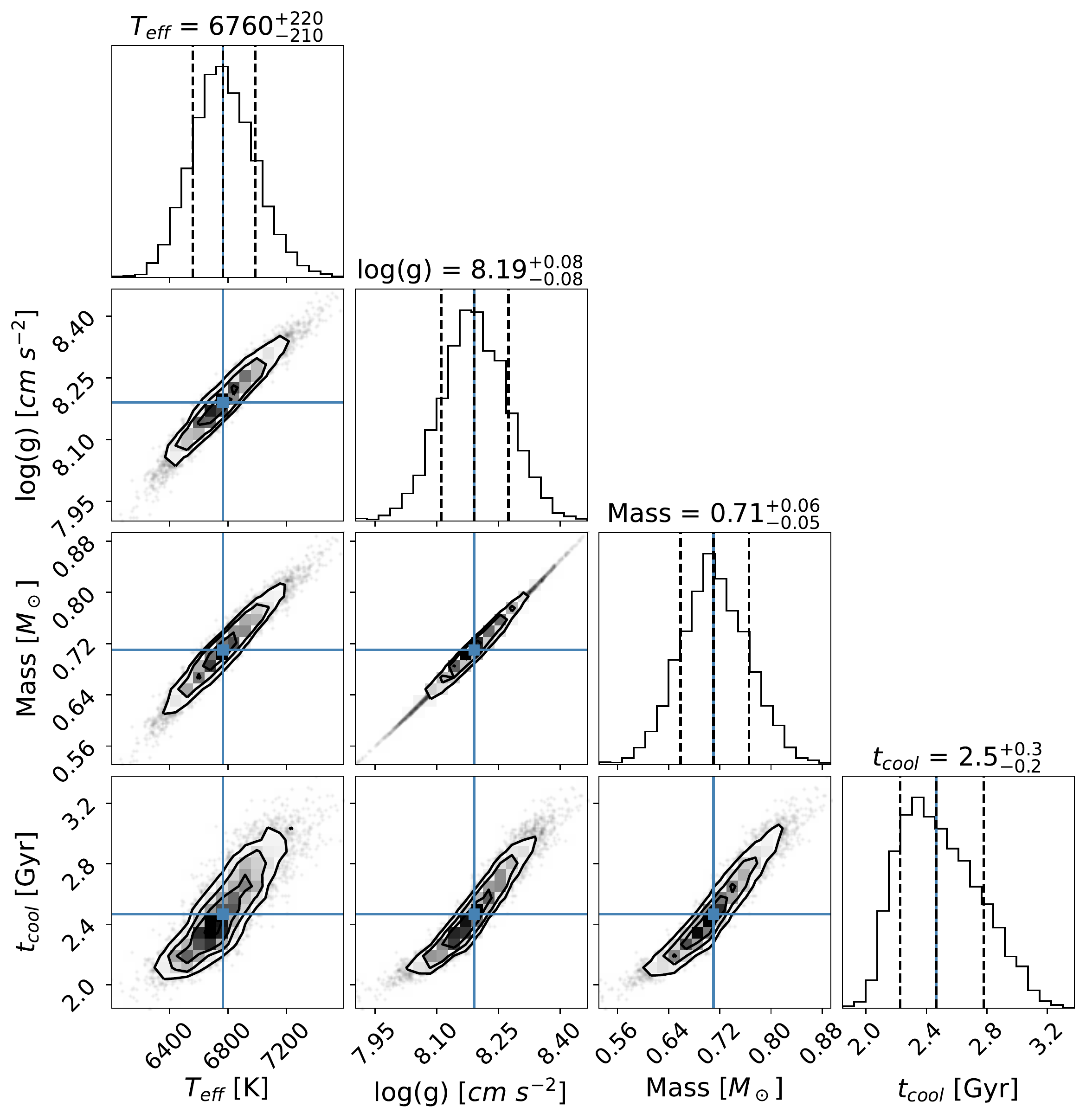}{0.99\columnwidth}{}}
     \caption{An example fit to get atmospheric parameters of our WDs. The top two panels shows the SED, including all observed photometry and a representative atmospheric model, as well as the residual from the fit. The posterior distributions from the MCMC fitting code used in this work are shown in the bottom part of the figure.}
\label{fig:example_fit}
\end{figure}

We use the python package emcee \citep{2013PASP..125..306F}, which uses an affine invariant Markov Chain Monte Carlo \citep{2010CAMCS...5...65G}, to get the posterior range on the temperature and surface gravity for each WD in the sample. We use the $50^{\mathrm{th}}$ percentile of these posteriors as an estimate of the best fit. We use flat priors on both effective temperature and surface gravities with bounds defined from the limits of the models ($3000-40{,}000$\,K and $\log{g}$ of $6.0-9.5$ dex) and assume Gaussian errors. Using the WD evolutionary models from \citet{2020ApJ...901...93B}, which assume thick H layers, we convert the surface gravities and temperatures to masses and cooling ages. The lower and upper errors on these parameters are taken as the 16th and 84th percentiles, respectively. To account for any unknown systematics, we set a lower limit on the uncertainty of the atmospheric parameters derived from the fits of 0.038 dex for surface gravity and 1.2\% in temperature \citep{2005ApJS..156...47L}. An example of a fit can be seen in Figure~\ref{fig:example_fit}.

As a check on our methods, we compare the atmospheric parameters we derive to the parameters from the 100-pc WD sample of \citet{2020ApJ...898...84K}, and to the full Gaia EDR3 WD catalog from \citet{2021MNRAS.508.3877G}, which were also determined from the photometric technique we employ. The comparison to the 100-pc WD sample from \citet{2020ApJ...898...84K} reveals no significant systematic differences between the two works. The atmospheric parameters on average agree within $0.9\sigma$ (100\,K) and $0.6\sigma$ (0.03\,dex) for effective temperature and surface gravity, respectively. We observe some small dispersion at higher temperatures, which we expect since our work incorporates interstellar reddening whereas \citet{2020ApJ...898...84K} does not. Comparing to the Gaia EDR3 WD catalog, we find that the atmospheric parameters on average agree within $0.4\sigma$ (300\,K) and $0.35\sigma$ (0.1\, dex) for effective temperature and surface gravity, respectively. The quoted uncertainties we find are significantly smaller than the ones in \citet{2021MNRAS.508.3877G} due to the use of more photometric points in the SED and the weighted parallax. Although we have smaller uncertainties and use different methods to determine the amount of interstellar reddening, we do not find any systematic shifts in derived WD atmospheric parameters.
  
We also search for spectral types of our objects that have been published in the literature, finding 305 objects with previously reported spectra. The sources for these spectral types can be found in the catalog of WD pairs described in Table~\ref{tab:description}. 

We supplement these previously published spectral types with 57 spectra taken on the DeVeny Spectrograph mounted on the 4.3-m Lowell Discovery Telescope (LDT, \citealt{2014SPIE.9147E..2NB}) in Happy Jack, Arizona. Using a 300 line $\mathrm{mm}^{-1}$ grating we obtain a roughly 4.5 \AA\ resolution, and rotate the position angle of the slit to capture both WDs simultaneously in our observations. We extracted the spectra using a quick-look pipeline written using the ccdproc and specutils packages from the AstroPy package \citep{2018AJ....156..123A}, and wavelength calibrated the spectra with a HgArNe lamp.  We do not attempt to fit these spectra and only conduct a visual inspection for the presence of hydrogen absorption features, and include their spectral type in the sp\_type column described in  Table~\ref{tab:description}. Future work will further analyze atmospheric parameters derived from this spectroscopy. 

Comparing the derived atmospheric parameters of the previously identified DAs with the parameters derived in this work, we find that the effective temperatures agree well but that the determination of the surface gravities from spectroscopic observations vary from our own values derived from photometric information. We find that the atmospheric parameters of our work and the spectroscopic parameters on average agree within $0.9\sigma$ (300\,K) and $1.3\sigma$ (0.1\,dex) for effective temperature and surface gravity, respectively. We also find that the spectroscopic fits find higher surface gravities for the WDs on average and thus hotter temperatures as well. Although the absolute differences are similar to the other comparisons, the quoted uncertainties from the spectroscopic parameters are smaller and result in a stronger disagreement in surface gravity ($1.3\sigma$ compared to $0.9\sigma$ and $0.35\sigma$).

\subsection{Initial Masses and Progenitor Lifetimes}
To get the total ages of the WDs in our sample, we need an estimate of the lifetime of the ZAMS progenitor. The first step towards this is to determine an initial main-sequence mass for each of the WDs in our sample. To get these masses, an initial-final mass relation is needed. In this work, we use the theoretical IFMR from \citet{2016ApJ...823...46F}, which was generated by running a suite of solar metallicity stellar evolution models from the Modules for Experiments in Stellar Astrophysics (MESA, \citealt{2011ApJS..192....3P}, \citealt{2013ApJS..208....4P}, \citealt{2015ApJS..220...15P}) starting at the pre-MS phase through the thermally pulsing asymptotic giant branch to the final WD phase. This allows us to remain self-consistent within both our IFMR and our determination of ZAMS progenitor lifetimes using the same open-source stellar evolution code and allows us to use a more theoretically motivated shape for the IFMR.

\begin{figure}
     \centering
     \includegraphics[width=0.975\columnwidth]{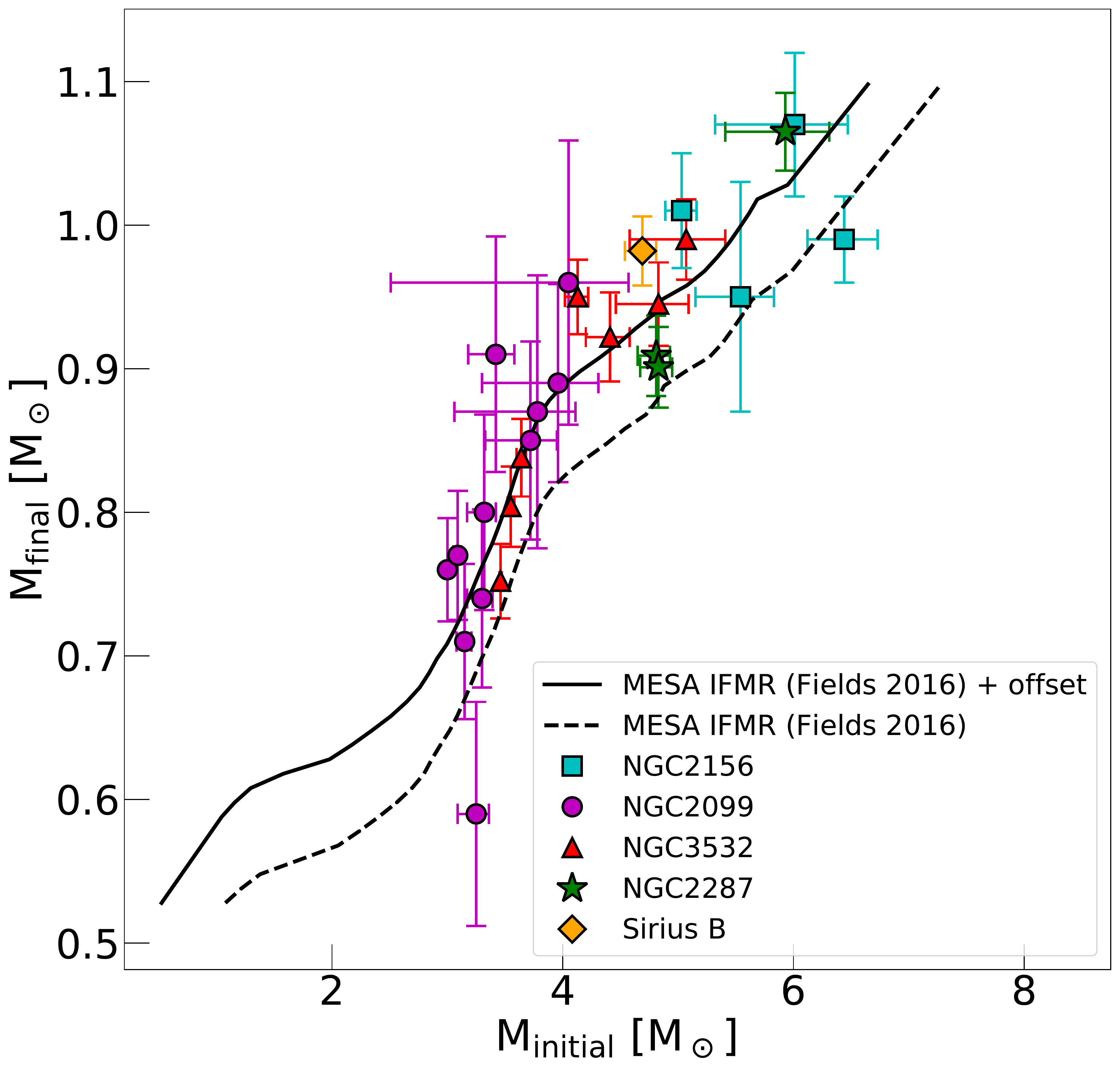}
     \caption{The initial-to-final mass relation (IFMR) used in this work to connect a WD mass to a single ZAMS progenitor mass. The dashed line shows the theoretical IFMR from MESA calculated by \citet{2016ApJ...823...46F}. The observational data points are WDs in solar metallicity clusters along with Sirius B from \citet{2008MNRAS.387.1693C}, \citet{2015ApJ...807...90C} and \citet{2016ApJ...818...84C}. The solid line shows our adopted IFMR, which include a constant y-axis offset of $+0.0636$\,$M_{\sun}$ found as a best-fit to the cluster observations (the values for which are in Table~\ref{tab:IFMR}).}
     \label{fig:IFMR_fit}
 \end{figure}

The theoretical IFMR (shown as the dashed line in Figure~\ref{fig:IFMR_fit}) lies below what is observed for WDs in near solar metallicity clusters, an effect that is still not fully understood, though might involve uncertainties in the amount of convective overshoot that is used in the models \citep{2016ApJ...823...46F} or uncertainties in nuclear reaction rates that can affect the core mass (e.g., \citealt{2017RvMP...89c5007D}). To account for this systematic effect, we fit a positive offset to the IFMR using the same WDs in solar metallicity clusters. We find the best-fitting offset to apply to the final WD mass is $+0.0636$\,$M_{\sun}$. The points of this shifted IFMR can be found in Table~\ref{tab:IFMR}.

Using this modified IFMR (the solid line in Figure~\ref{fig:IFMR_fit}), we generate progenitor MS masses for all the WDs in our sample. We convert these MS masses into progenitor lifetimes by interpolating the solar metallicity, zero rotation MESA evolutionary tracks of \citet{2016ApJS..222....8D} and \citet{2016ApJ...823..102C} and summing the main-sequence and giant branch lifetimes.
 
The uncertainties on the progenitor main-sequence masses and progenitor lifetimes are determined by taking the $1\sigma$ uncertainties on the WD mass, determined from the model-atmosphere fitting described in Section~\ref{sec:atm}, which determine an upper- and lower-limit on the MS mass. Then the uncertainty on the progenitor lifetime is quoted as the difference between these $1\sigma$ bounds and the determined value. Total ages are then determined through the sum of the cooling age and the progenitor lifetime, and the uncertainties on total age are calculated by adding the uncertainties on cooling age and progenitor lifetime in quadrature. A table of all double WD pairs with atmospheric parameters and total ages can be accessed at \href{https://sites.bu.edu/buwd/files/2022/05/Table_A1.csv}{https://sites.bu.edu/buwd/files/2022/05/Table\_A1.csv}; a description of the table columns and data can be seen in Table~\ref{tab:description}.

As a check, we compare the progenitor lifetimes from our MESA IFMR to ones derived from the three-piece, cluster-calibrated IFMR from \cite{2018ApJ...866...21C}. We find that the progenitor lifetimes from both IFMRs are within 25\% of each other on average. The difference between the total ages is on average 10\% when comparing the two IFMRs. Overall, we find that the use of the IFMR from \cite{2018ApJ...866...21C} does not significantly change the conclusions discussed in subsequent sections.

\subsection{Precision of Total Ages}\label{sec:age_precision}
% Discuss the reliability of WD Masses and uncertainties on those masses
The WD mass is the dominant factor for determining the total age of the WDs in our sample, and subsequently, the uncertainties in the WD mass strongly influence the degree of precision on the total ages. 

 \begin{figure*}
     \centering
     \includegraphics[width=1.99\columnwidth]{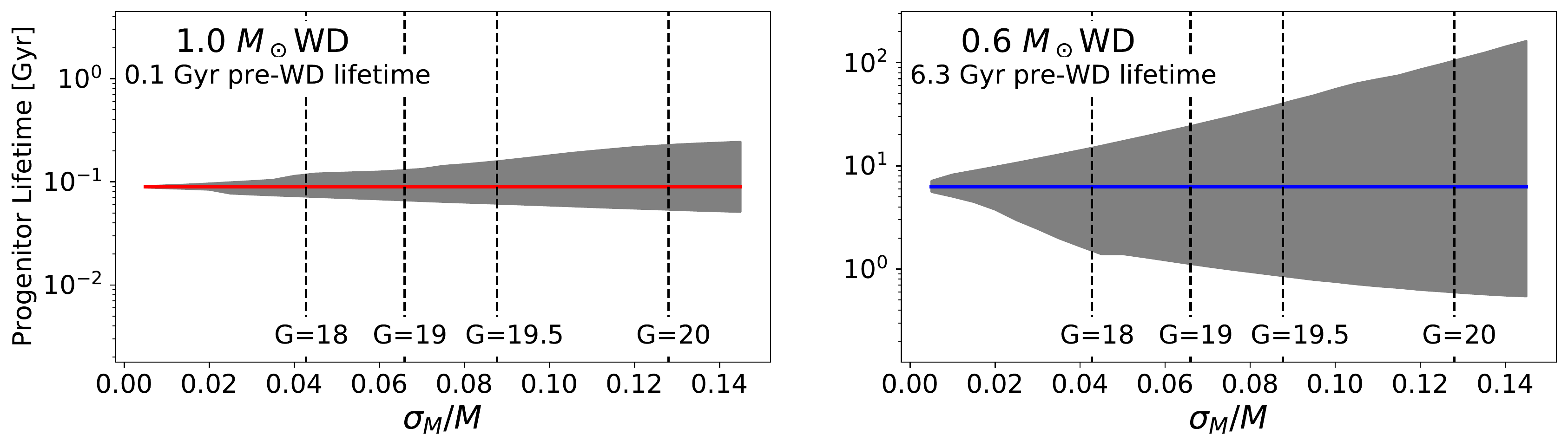}
     \caption{Visualization of how the precision on the ZAMS progenitor lifetime changes as a function of the WD mass uncertainty for a $1.0$\,$M_{\sun}$ WD (left) and $0.6$\,$M_{\sun}$ WD (right). The grey shaded regions represent the 1$\sigma$ uncertainties on the progenitor lifetimes. The axes are scaled to show the relative size of the uncertainties. The ZAMS progenitor lifetimes are given in the top left of each plot. The dashed vertical lines show the average relative uncertainty on the WD mass at a given $G$-band magnitude. The uncertainty on the WD mass is also affected by the precision of the parallax, and thus, for a given $G$-band magnitude there is some dispersion where more precise parallaxes give better mass constraints.}
     \label{fig:age_precision}
 \end{figure*}
  
This is especially true for the lower-mass WDs. For a typical $0.6$\,$M_{\sun}$ WD \citep{2007MNRAS.375.1315K,2016MNRAS.461.2100T}, the initial progenitor MS mass is around $1.2$\,$M_{\sun}$ which corresponds to a progenitor lifetime of $\approx 6$\,Gyr. A $5\%$ decrease in the WD mass results in a ZAMS progenitor mass of $0.88$\,$M_{\sun}$ and progenitor lifetime around 17\,Gyr. Thus, for WDs near this mass range, it is extremely difficult to constrain the total age without very precise WD mass estimates. This problem is compounded by the fact that the lower-mass region of the IFMR is the least constrained. Depending on the IFMR used, initial mass values for a 0.6\,\msun\ WD can range between $1.2-2$\,\msun\ which corresponds to progenitor lifetimes between $1.4-6$\,Gyr. This problem is also exaggerated when considering that for these low-mass WDs, the progenitor lifetimes are a major fraction of the total age.

We stress that low-mass WDs are still useful in providing a lower limit on total ages. A $0.6$\,$M_{\sun}$ WD with a 5\% error on its mass can provide a $1\sigma$ lower limit of 1.4 Gyr on the progenitor lifetime, modulo uncertainties on the IFMR used. This, in conjunction with a precise cooling age, can still provide useful information on these WDs and their companions.
  
Fortunately, the opposite is true for the higher-mass WDs. Higher-mass WDs ($>0.8$\,$M_{\sun}$) come from massive progenitors, which have short ZAMS progenitor lifetimes. Thus, even larger mass uncertainties for high-mass WDs do not result in large errors on the progenitor lifetime (see Figure~\ref{fig:age_precision}). Put another way, high-mass WDs can have large relative uncertainties in their progenitor lifetimes but still have relatively precise total ages. Additionally, the IFMR of these high-mass WDs are better empirically constrained from cluster WDs.  

\section{Coeval Analysis}\label{sec:results}

Before engaging in any analysis on the total ages, we remove systems where we know our methods described in the previous section return unreliable total ages. We apply cuts on the sample to remove all systems with on-sky separation $<$2\arcsec, as well as systems with  \texttt{R\_chance\_align}~$<$~0.1. We also remove systems with at least one component with $\mathrm{T_{eff}}<3200$\,K, as well as those with an inferred total age $>14$\,Gyr.

The cut to on-sky separation removes systems with significant blending of the photometry, which confuses the SED fits (154 of the WD pairs have on-sky separations $<2$\arcsec). We adopt the cut on the \texttt{R\_chance\_align} parameter provided in \citet{2021MNRAS.506.2269E}; cutting at 0.1 ensures the remaining systems have a high probability of being bound ($>90$\%). Although we cut on a $<10$\% chance alignment probability, we emphasize that the expected contamination fraction from chance alignments is much less than 10\%, because most of the pairs in the sample have chance alignment probabilities that are much lower than 10\%. From the actual chance alignment probabilities of the pairs in our sample, we estimate that there are $4\pm2$ chance alignments in the sample, corresponding to a contamination fraction of about 0.3\%. The wide WD+WD sample contains 1390 high fidelity pairs with \texttt{R\_chance\_align} $< 0.1$ (see Figure~\ref{fig:sample}).

The cut on effective temperature is employed because our model grid stops at 3000\,K. Since our fitting routine does not attempt to extrapolate, objects with reported temperatures near the cutoff will have underestimated uncertainties and their temperatures will be artificially hotter than their true temperatures. We also ignore all systems with a total age greater than that of the Universe. Many of these systems come from overluminous sources in the Gaia CMD (see Figure~\ref{fig:sample}). These sources will have low WD masses that, if they are actually so low mass, could not have formed from single-star evolution. At lower temperatures, the observed WD sample does not follow the cooling model trends which may result in these lower unreliable masses (see Figure~\ref{fig:sample}). There are a total of 670 WDs with inferred total ages $>14$ Gyr, which results in 548 pairs being removed.

After these cuts, the high-fidelity sample used for analysis contains 1272 WDs in 636 wide double WD pairs. The mean magnitude of this surviving sample is $G=19.3$\,mag.

\subsection{Age Comparison}\label{sec:age_compare}

\begin{figure}
    \centering
    \includegraphics[width=0.975\columnwidth]{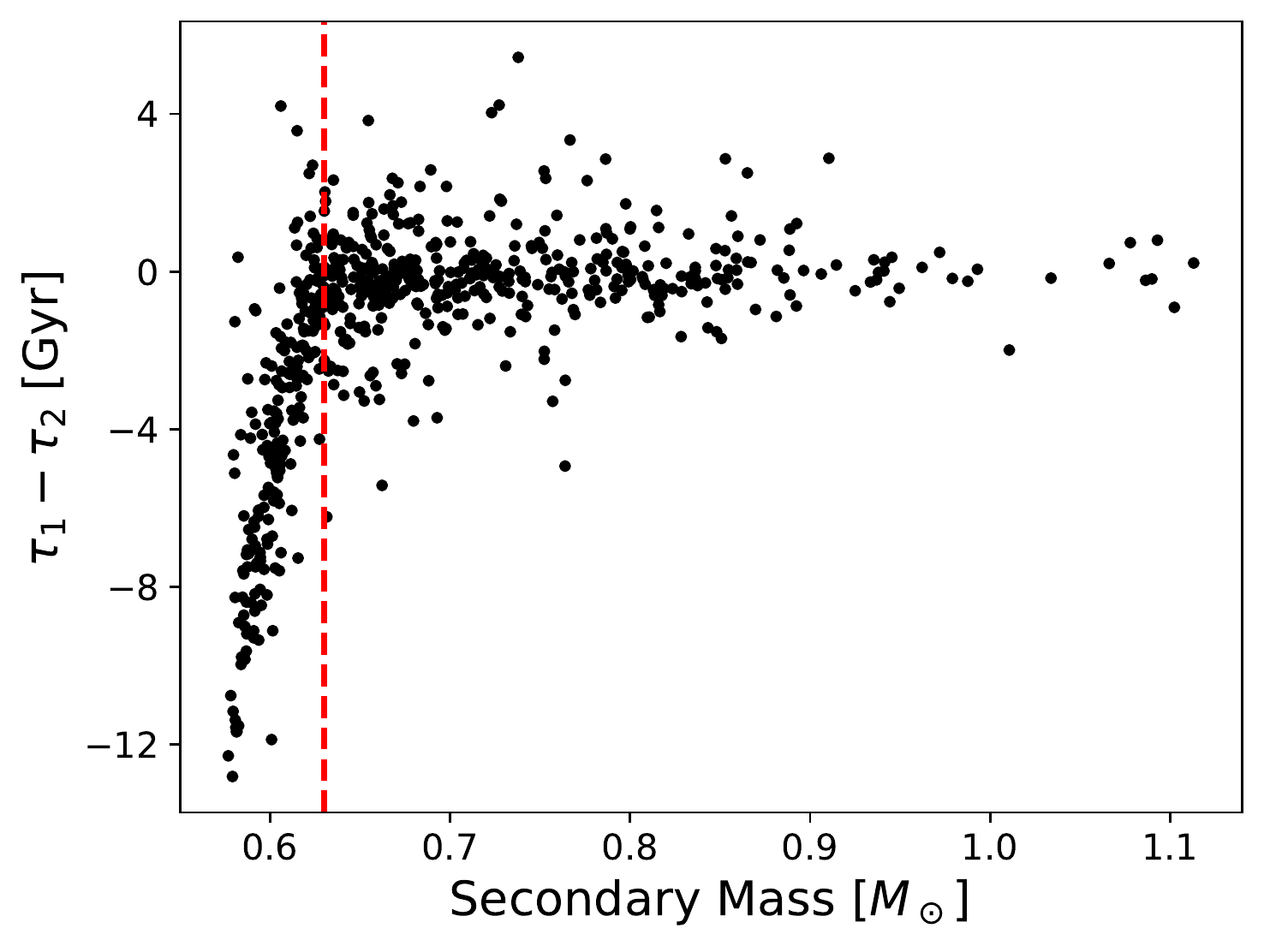}
    \caption{The total age difference in each binary as a function of the mass of the secondary in each binary (the least massive component) for our 1272 systems surviving the cuts outlined in Section~\ref{sec:results}. The red line shows the cut used to remove the poorly behaved low mass systems, whose behavior change drastically at $M < 0.63 M_\odot$. Since the secondary is always the least massive component, the cut shown by the red line removes all systems with WDs $< 0.63 M_\odot$.}
    \label{fig:masscut}
\end{figure}

The first test of the total ages made is to check the absolute age difference in each pair. Since we assume the stars in each wide pair in our sample are coeval, we expect them to return identical total ages. This absolute age difference is plotted in Figure~\ref{fig:masscut}. The total age difference between each component is typically close to zero until one of the components masses is below a value of $\sim0.63 M_\odot$. These systems do not provide very precise ages, and it is evident that the systematic uncertainty in these systems is too large to provide statistically significant comparisons. 

This leads us to remove any systems that have a WD with a mass less than $<0.63\ M_\odot$ from any further analysis.  These systems can still be useful to provide lower limits on the total age, as discussed in Section~\ref{sec:age_precision}, but we are not able to assess their reliability due to the large systematics in conjunction with their large uncertainties. After removing these systems, the number of wide WD+WD pairs is reduced to 423, with a total of 846 WDs.

The second test of the accuracy of the total ages is to compare the ages of the WDs in each binary in relation to the uncertainties on those total ages. To do this, we explore distributions of age agreement, which we define as
\begin{equation}
    \mathbf{\sigma_{1-2} = }\frac{\Delta \tau}{\sigma_{\Delta \tau}} = \frac{\tau_1 - \tau_2}{\sqrt{\sigma_{\tau_1}^2 + \sigma_{\tau_2}^2}}
\end{equation}
where $\tau_1$ and $\sigma_{\tau_1}$ are the total age and uncertainty on the total age of the primary (the most massive component of each binary) and $\tau_2$ and $\sigma_{\tau_2}$ are the total age and uncertainty on the total age of the less massive component. Systems with an age agreement ($\mathbf{\sigma_{1-2}}$) equal to 5 have individual component ages that disagree at a $5\sigma$ level. 
 
\begin{figure}
    \centering
    \includegraphics[width=0.975\columnwidth]{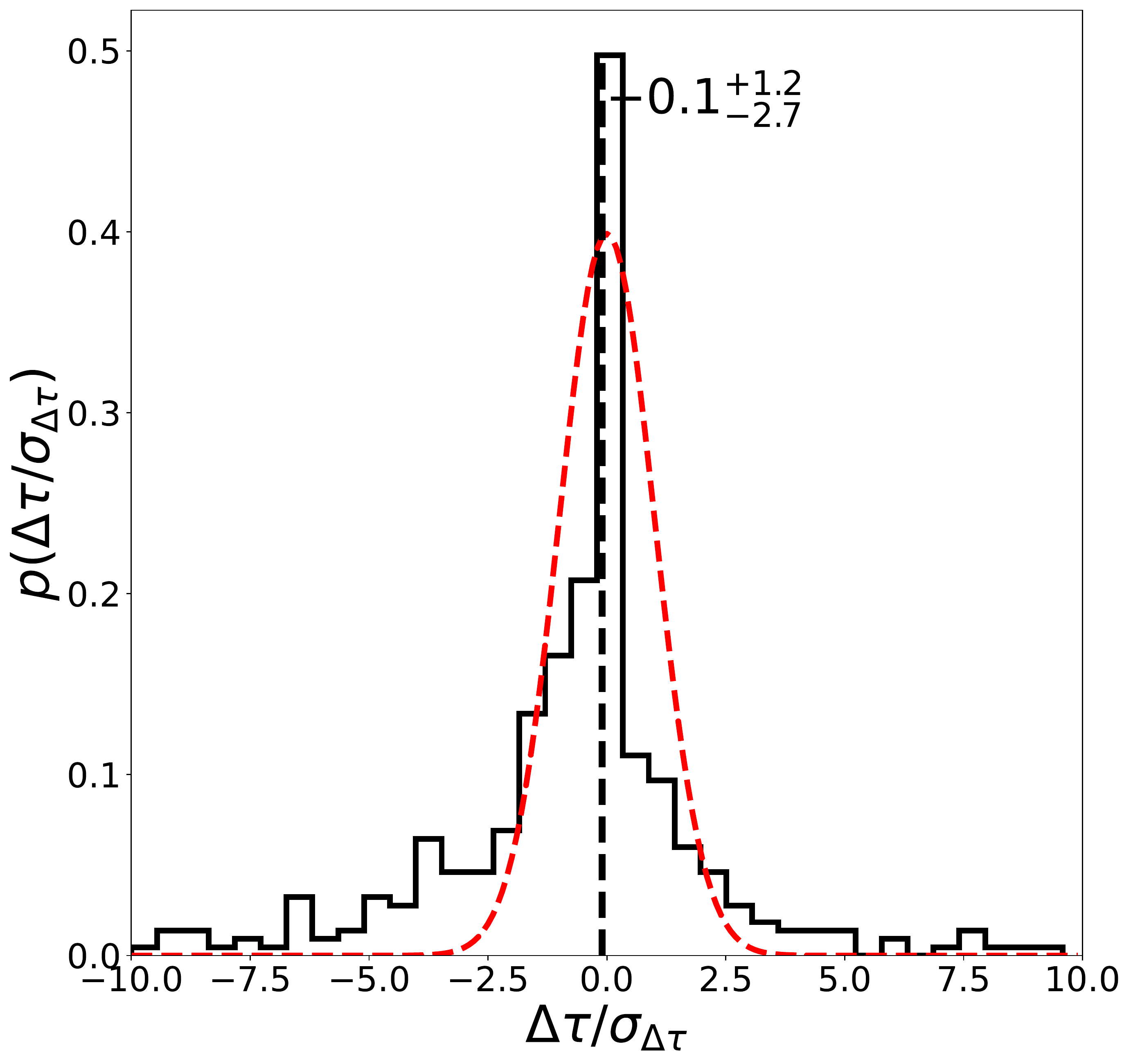}
    \caption{The distribution of $\sigma_{1-2}$ disagreement for the 423 wide WD+WD pairs in our sample that pass the cuts discussed in Section \ref{sec:results}. The red dashed line shows the expected distribution for a Gaussian with $\sigma=1$ for reference. The distribution is skewed towards negative values, where the primary is the more massive of the two components. The black dashed line represents the $50^{th}$ percentile and the lower and upper errors come from the $16^{th}$ and $84^{th}$ percentiles respectively.}
    \label{fig:diff_age}
\end{figure}

We show in Figure~\ref{fig:diff_age} the distribution of the $\sigma_{1-2}$ disagreement for the 423 high-fidelity pairs where both WDs have $M > 0.63 M_\odot$. The distribution exhibits a long tail out to large absolute values and has a notable bias towards negative values. Fewer than half (47\%) of the systems have total age agreement within $1\sigma$. The slight tendency towards negative $\sigma_{1-2}$ disagreement values shows that the more massive component of each pair has a younger total age on average compared to its companion. Possible causes of this asymmetry are further explored in Section~\ref{sec:ill-behaved}. The long tail of the distribution reveals that there is a population of at least 60 systems (14\%) with values of the $\sigma_{1-2}$ disagreement greater than five where the systematic errors ($\Delta\tau$) are non-negligible compared to the random errors reported ($\sigma_\tau$).

\subsection{Ill-Behaved Systems}
\label{sec:ill-behaved}

Substantial uncertainty in the total ages in these WDs comes from the process of converting the WD mass to a ZAMS progenitor lifetime. We remove the most model-dependent aspect in the determination of the WD's progenitor lifetime by simply considering that the IFMR is monotonic and increasing, such that a higher-mass WD comes from a higher-mass progenitor. This assumption may not hold for all WD masses, as some evidence of a kink is seen in the IFMR for a range of WD masses around $0.65$\,\msun\ (\citealt{2020NatAs...4.1102M}).
 
Still, for the vast majority of cases, the more massive WD in the binary comes from a more massive main-sequence star, and it's progenitor lifetime should be smaller than the progenitor lifetime of its companion. For a coeval binary pair, the cooling age of the more massive WD must be longer than the cooling age of its companion. Any systems where the more massive WD has a shorter cooling age than its lower mass companion are discrepant before invoking any theoretical IFMR or other process for determining a progenitor lifetime.

\begin{figure}
    \centering
    \includegraphics[width=0.97\columnwidth]{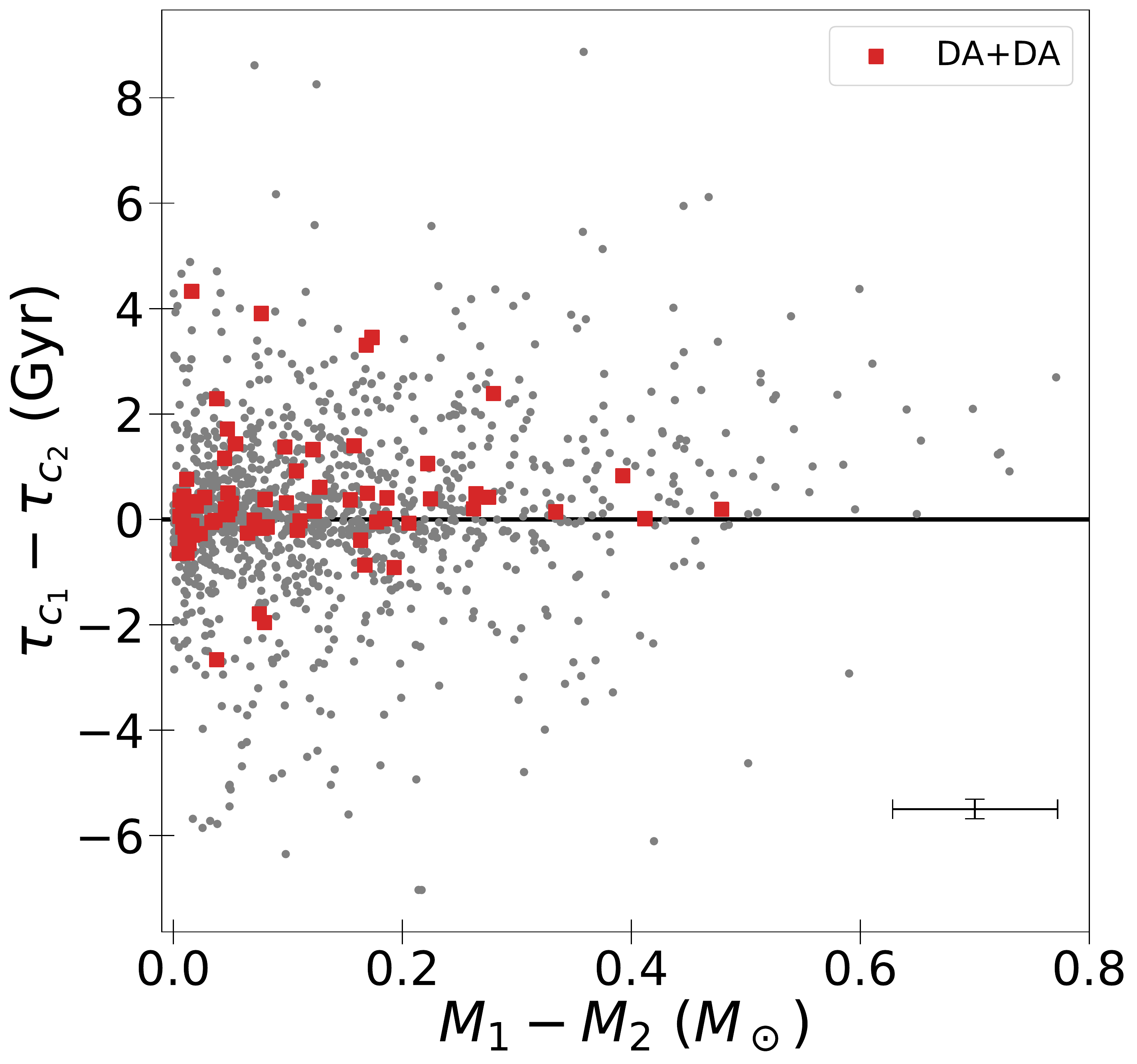}
    \caption{A comparison of the cooling age difference versus mass difference for the full wide binary sample (1259 grey points) and a subset of pairs where both WDs are known to be DA (79 red diamonds). A representative error bar is provided in the bottom right of the figure. A significant fraction of objects (43\% for the full sample, 35\% for the DA+DA systems) sit in the negative region, which implies the more massive component has a shorter cooling age, which is unphysical and likely implies a large fraction of wide WD+WD binaries were once triple systems.}
    \label{fig:dm_dtc}
\end{figure}

Our sample contains many systems where this is the case, as shown by Figure~\ref{fig:dm_dtc}. The figure includes systems that pass the temperature, separation, and chance alignment cuts discussed in Section~\ref{sec:results}, but we do not apply the total age and mass cut described in that section for this comparison. Roughly 43\% lie in the negative region of the plot, where the more massive WD has a shorter cooling age. This statistic requires we properly define the more massive WD (see representative uncertainty in Figure~\ref{fig:dm_dtc}); 21\% of our systems are more than 1$\sigma$ away from having the more massive component with a shorter cooling age. This is a conservative estimate for the number of these ill-behaved systems because the masses of the two WDs in each pair have a positive correlation. This is due to the fact that the two WDs are effectively at the same distance, and the uncertainty on the distance affects the determined uncertainties on the masses. If in fact the system is closer and therefore brighter, then both WDs would be brighter and the derived masses would both be lower, leaving the determination of the more-massive primary unchanged.

\begin{figure}
    \centering
    \includegraphics[width=0.97\columnwidth]{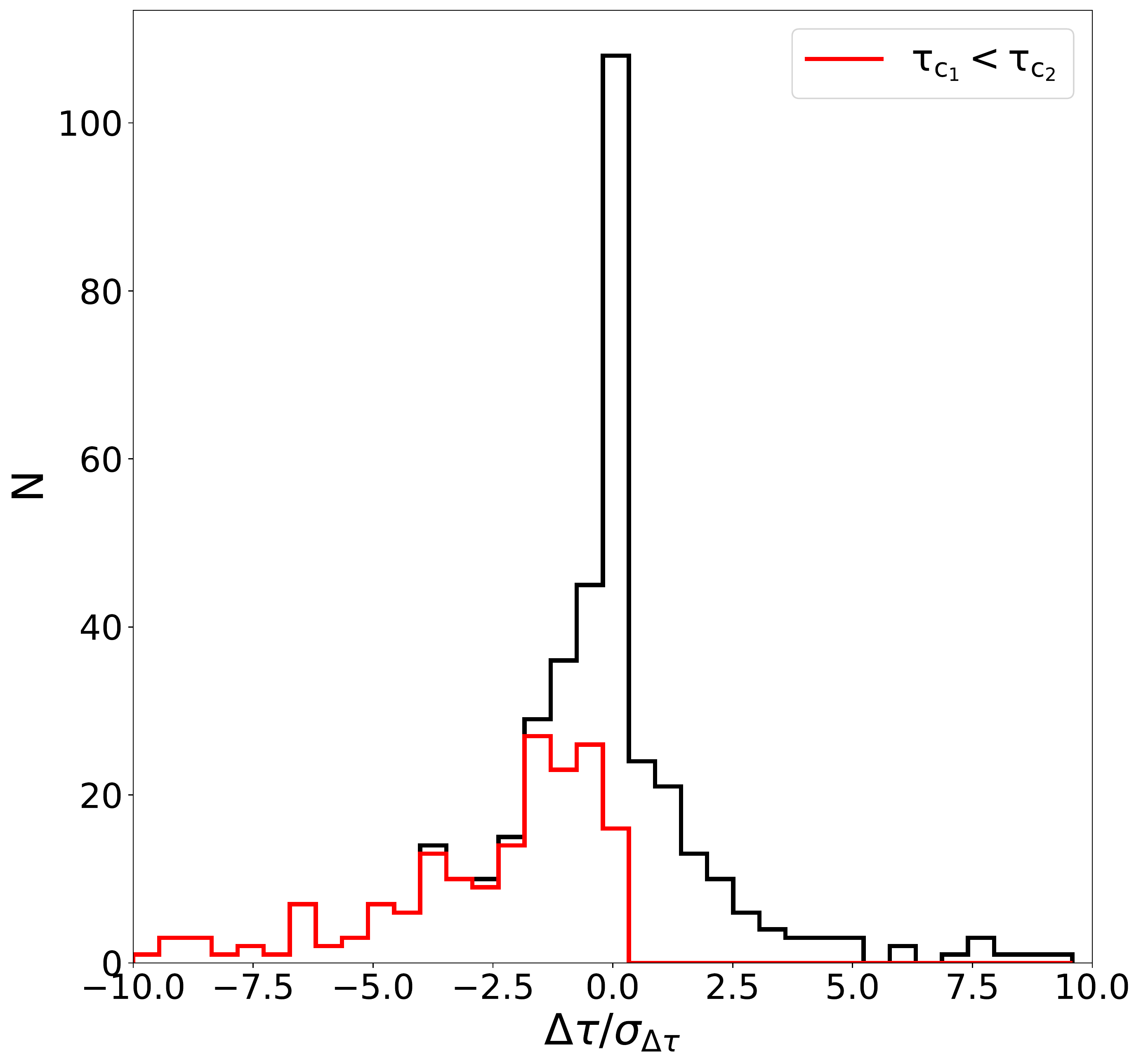}
    \caption{The distribution of $\sigma_{1-2}$ disagreement for the 423 wide binaries that pass the cuts outlined in Section \ref{sec:results} (black) compared to the distribution of $\sigma_{1-2}$ disagreement for the sample of systems where the more massive WD has a shorter cooling age (red). These systems can account for the bias towards negative values including the most discrepant systems.}
    \label{fig:diff_age_ill_behaved}
\end{figure}

Thus, somewhere between 21-43\% of the wide WD+WD systems have the more massive WD with a shorter cooling age. Given a monotonic and increasing IFMR, these systems will have large $\sigma_{1-2}$ disagreement values. The systems that have negative cooling age differences shown in Figure~\ref{fig:dm_dtc} show no bias towards any mass or temperature range. Removal of the 187 ill-behaved systems that pass the cuts described in Section~\ref{sec:results} results in a substantial fraction of systems in the long negative tail of the $\sigma_{1-2}$ disagreement distribution being removed (see Figure~\ref{fig:diff_age_ill_behaved}).

One possible explanation for this discrepancy is that we are assuming all of our WDs have a hydrogen dominated atmosphere. Applying a DA (hydrogen-dominated atmosphere) model to a non-DA WD can cause systematic errors on the order of 10-15\% on the WD mass (\citealt{2012ApJS..199...29G}) which can lead to the wrong determination of the primary. The objects in the sample that are known to have two pure DA WDs still show objects in the negative region where the higher mass WD has a shorter cooling age (see Figure~\ref{fig:dm_dtc}). The amount of wide DA+DAs that are ill-behaved is around 36\% of 74 systems (with 21\% being more than 1$\sigma$ discrepant). Systems where one of the WDs is known from spectroscopy to be a non-DA have 63\% out of 73 systems in the negative region (with 52\% being more than 1$\sigma$ discrepant).

Our spectroscopic follow-up using the LDT was designed to specifically target systems in the negative region of Figure~\ref{fig:dm_dtc} and thus biases our results to this region. %When excluding our own spectra taken with the LDT, the number of wide DA+DAs that are ill-behaved is 37\% of 59 systems (with 20\% being larger than 1$\sigma$), while systems where one of the WDs is not a pure DA have 54\% out of 57 systems in the negative region (with 42\% being more than 1$\sigma$ discrepant). 
An alternative way to determine if there is an excess of non-DA systems in the ill-behaved region of Figure \ref{fig:dm_dtc} without having to worry about the bias introduced by our follow-up observations is to compare relative rates of spectral types within the sample of the ill-behaved objects. There are 101 systems in the negative region and 156 of the 202 WDs have a spectral type determined. 66\% of these WDs are DAs. In magnitude limited samples, we expect $\approx65$\% of the WDs to be DAs \citep{2013ApJS..204....5K}. For the systems that are more than $1\sigma$ discrepant, 59\% of the WDs with known spectral types are DAs.

Although there is some evidence for an overabundance of non-DAs in the negative region of Figure \ref{fig:dm_dtc}, we conclude it is unlikely that a large population of non-DAs is the full explanation for the number of these ill-behaved systems. To avoid any contamination from non-DAs, we will refer to the values for the DA+DA subsample (21-36\%) when discussing the percentage of ill-behaved systems.

Instead, we propose that most of the ill-behaved systems are the descendants of triples that have an inner pair that either experienced a merger that effectively reset the age of the more massive WD or is currently unresolved. Through binary population modelling, \citet{2020A&A...636A..31T} find that assuming single star evolution for a merger remnant can result in underestimating the age by $3-5$ times on average, which could manifest as an anomalous cooling age like those in Figure~\ref{fig:dm_dtc}. They also find $10-30$\% of isolated WDs have experienced a merger in their past, usually before the WD phase is reached, which is similar to the percentage of ill-behaved systems in our sample ($21-36$\%). While that value is for single WDs once in binaries, it is a valuable number to anchor expectations.

If all the systems in the negative region of Figure~\ref{fig:dm_dtc} are or were previously triples, this implies that roughly 21-36\% of the wide WD+WD pairs in our sample started as a triple system and have either undergone a merger or have a close inner pair that is unresolved. This range is consistent with the expected binary and triple fraction ($30-35$\% and $10-20$\%, respectively) for WD progenitor stars that are $2-4$\,$M_{\sun}$ on the ZAMS \citep{2017ApJS..230...15M}. We also find that 50\% of the double WD pairs in the triple systems are in the negative region of Figure~\ref{fig:dm_dtc}. This would imply that these systems are descended from quadruple systems.

At least one well-studied merger remnant is included in our sample: the massive ($>1.1$\,$M_{\sun}$, \citealt{2010ANA...524A..36K}), strongly magnetic (up to 800\,MG, \citealt{1999ApJ...510L..37B}), rapidly rotating (725\,s, \citealt{2003ApJ...593.1040V}) object REJ0317-853  \citep{1997MNRAS.292..205F}. This WD is in a system with a $\sigma_{1-2}$ disagreement value of $-17$ and its effective temperature is at least twice that of its lower mass companion. We conclude that the majority of these ill-behaved systems are likely merger remnants or unresolved triples. Spectroscopic follow-up of these systems can help to disentangle the subset of these objects that are the result of a merger from the unresolved triples which can place empirical constraints on the triple fraction of WD progenitors.

\subsection{Dependence on Mass and Temperature} \label{sec:dependence}

The dependence of the age disagreement on the WD mass and temperature can provide insights into where our models are insufficient. A treatment and correction for these dependencies is crucial if WD ages are to be used on large scales to age various populations.

We explore the dependence of the age disagreement on the WD mass and temperature by splitting the systems into individual WDs, and inspecting the median of the absolute value of the age disagreement of each system in various bins of effective temperature and mass. In this analysis, we do not remove any of the ill-behaved systems discussed in Section~\ref{sec:ill-behaved}, since it is not usually feasible to separate these possible merger remnants when looking at an isolated WD or WD+MS system in the field.

\begin{figure}
    \centering
    \includegraphics[width=0.975\columnwidth]{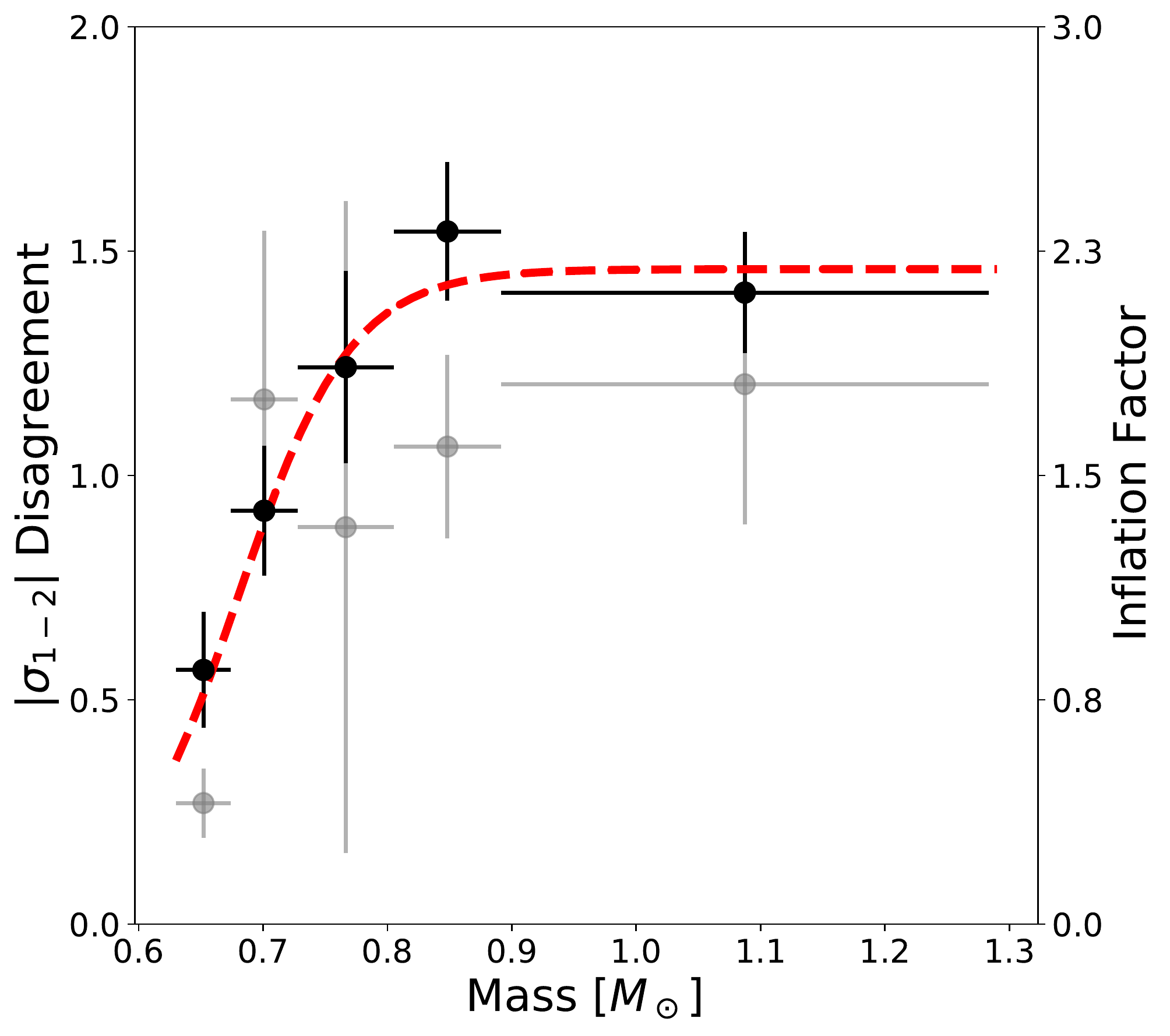}
    \caption{The absolute value of the total age disagreement and the resulting total age error inflation factors as a function of WD mass. Each point represents the median of the absolute value of the total age disagreement of a bin centered at the point. The total age error inflation factors are calculated by dividing this median by the expectation for a folded Gaussian distribution. The black points represent the full sample of 423 systems that survive the cuts described in Section~\ref{sec:results}. The grey background points represent the median age disagreement for a sub-sample of systems where both WDs have masses contained in each bin. Each bin contains 20\% of the sample, which is roughly 170 individual WDs per bin for the black points (the grey points are around $3-5$ times fewer). The red line is a fit to the black points that is described in Section~\ref{sec:dependence} to define an inflation factor; a coarse mapping for this fit can be found in Table~\ref{tab:inflation}.}
    \label{fig:agreement}
\end{figure}

We find that the age disagreement is effectively constant as a function of temperature, but shows a strong trend as a function of mass. This can be seen in Figure~\ref{fig:agreement}. The highest masses ($>0.8$\msun) have a median $|\sigma_{1-2}|$ disagreement value of $\approx{1.5}$ while the lowest mass bin ($<0.67$\msun) has a median value of $\approx{0.6}$. The uncertainties in the medians shown are found through a bootstrapping method with replacement. To ensure that this trend is the true underlying trend, we calculate the median age disagreement for a sub-sample of systems where the mass of both WDs in the system are in the same bin. This removes cross contamination from systems that have WDs with very different masses. The grey points in Figure~\ref{fig:agreement} show the trend for this sub-sample, and are consistent with the points for the whole sample.

This trend in mass can be caused by larger systematic errors in the total ages of massive WDs causing larger relative age differences in binaries containing them, or by an underestimation of the total age uncertainties of high mass WDs. We find that the relative age difference is essentially constant ($\approx{25\%}$) as a function of WD mass. This leads us to conclude that the uncertainties on the total ages of the high mass WDs in the sample are underestimated.

To determine the amount of inflation of the total age errors needed, we take the median age disagreement and divide by the expected median value of a folded Gaussian distribution with a standard deviation of one. These values can be seen as a second y-axis in Figure~\ref{fig:agreement}. Next, we fit these points with a sigmoid function, defined as,
\begin{equation}
    f_{\mathrm{inflate}}\left(M_\odot\right) = \frac{a}{1+\mathrm{e}^{-c (M - b)}}
\end{equation}
where $M$ is the mass of the WD and $a$, $b$, and $c$ are all constants. We fit this function to the inflation factor points shown in Figure~\ref{fig:agreement}. The best fit parameters are $a=2.2$, $b=0.68$, and $c=22$. The resultant fit can be seen in Figure~\ref{fig:agreement} as the red dashed line, and select points are given in Table~\ref{tab:inflation}. The total age error inflation factor for the lowest mass bin is 0.85, but we do not recommend deflating any total age errors. The uncertainty makes the point consistent with a total age error inflation factor of 1.0. Thus, we recommend using the fit function only for WDs with masses greater than $0.67 M_\odot$.

\begin{deluxetable}{cc|cc}
\label{tab:inflation}
\tablecaption{Total Age Error Inflation Factors}
\tablewidth{700pt}
\tabletypesize{\scriptsize}
\tablehead{\colhead{WD Mass $\left[M_\odot\right]$} & \colhead{$f_{\mathrm{inflate}}$} & \colhead{WD Mass $\left[M_\odot\right]$} & \colhead{$f_{\mathrm{inflate}}$}}
\startdata
0.67 & 1.00 & 0.74 & 1.74\\
 0.68 & 1.10 & 0.75 & 1.81\\
 0.69 & 1.20 & 0.76 & 1.88\\
 0.70 &  1.34 & 0.77 & 1.90\\
 0.71 & 1.45 & 0.78 & 2.00\\
 0.72 & 1.55 & 0.80 &  2.05\\
 0.73 & 1.65 & $>$0.90 &  2.20\\
\enddata
\end{deluxetable}

\begin{figure*}
    \centering
    \includegraphics[width=1.99\columnwidth]{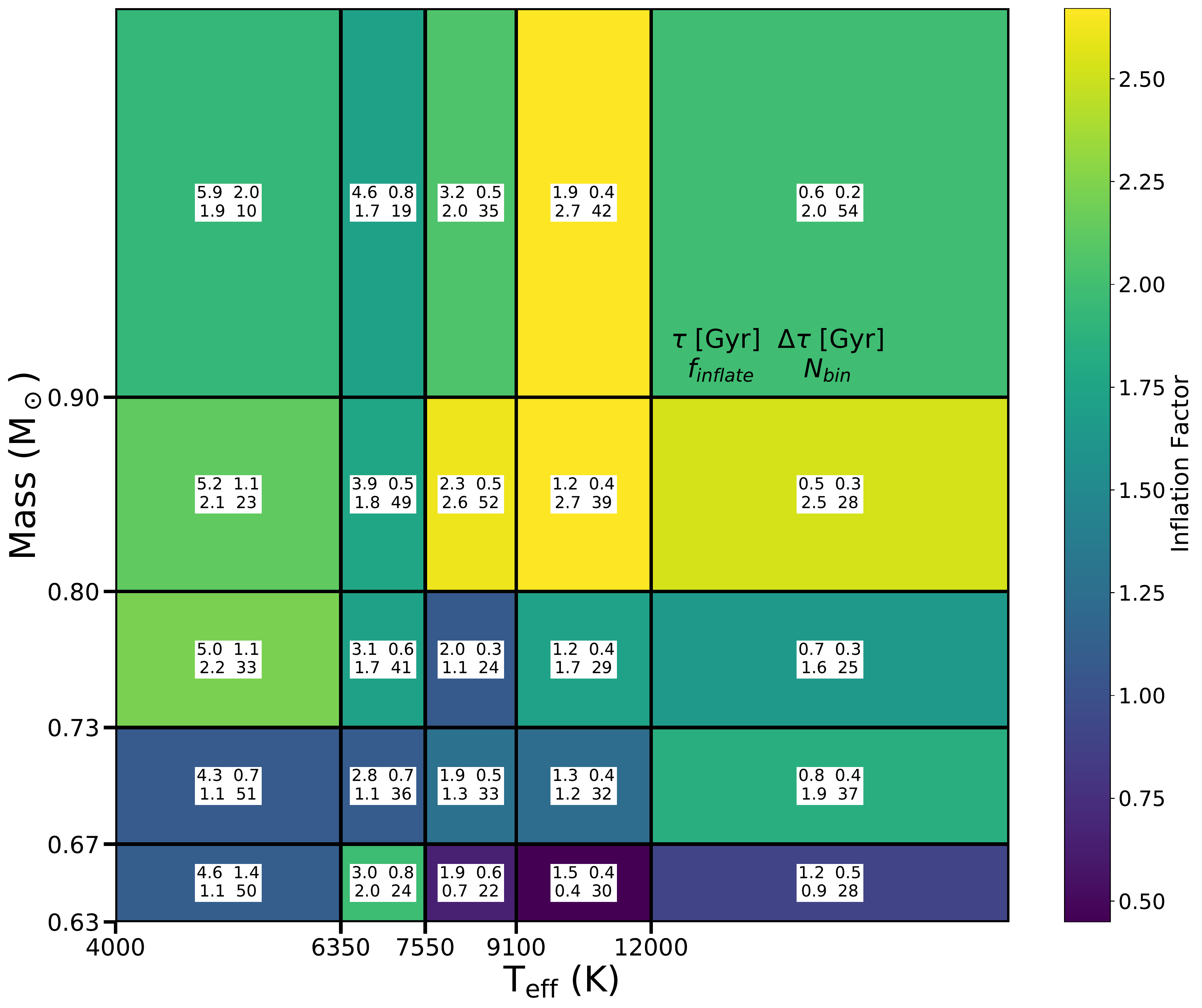}
    \caption{The inflation factor and median total age and age difference for 25 bins in mass and temperature of roughly similar size. Each box is colored by the inflation factor and the white box shows the median total age (Gyr, upper left), median total age difference (Gyr, upper right), the total age error inflation factor (lower left), and the number of WDs in the bin (lower right). The upper right bin also includes the parameters represented in the white boxes. For example, the 50 WDs in the bottom left bin ($4000-6350$\,K, $0.63-0.67$\,\msun) have a median total age of 4.6 Gyr, their ages differ from their companions by 1.4 Gyr on average, and their total age uncertainties need to be inflated by a factor of 1.2 for their ages to come into 1$\sigma$ agreement.}
    \label{fig:2D_agreement}
\end{figure*}

To further understand where WD ages are the most reliable, we split the parameter space into 25 bins in temperature and mass. For each bin, we check the median total age, absolute age difference, and total age error inflation factor needed. The results of this are shown in Figure~\ref{fig:2D_agreement}. The most problematic areas of parameter space are the higher masses, which is what is seen in Figure~\ref{fig:agreement}, and especially at the hotter temperatures where the cooling age contribution to the total age is minimized. These hot, high-mass regions of parameter space are where we would expect the merger remnants to be at their highest contamination rate. This is also where the Q-branch is located, which was first seen in the HR diagram in \emph{Gaia} DR2 \citep{2018A&A...616A..10G,2019ApJ...886..100C}, as well as the onset of crystallization \citep{2019Natur.565..202T}. The cooling models we use do include crystallization, but not other exotic cooling effects, such as sedimentation of neutron-rich ions (e.g., \citealt{2001ApJ...549L.219B,2020ApJ...902...93B,2021ApJ...911L...5B}).

The typical percent age difference across all of the bins is around $25\%$, ranging from 60\% in the worst case ($T_{\mathrm{eff}} > 12000$ K, $0.8$ \msun $ < M < 0.9$ \msun) and 15\% in the best case ($7550$ K $< T_{\mathrm{eff}} < 9100$ K, $0.73$ \msun $< M < 0.8$ \msun). The WDs with the hottest temperatures show larger percent age differences on the order of 35\%-60\% (0.2-0.5 Gyr) compared to the rest of parameter space. Thus, 25\% total age uncertainties are a representative approximation for most WD age estimates on an individual WD for WDs $>0.63$\,$M_\odot$ and temperatures cooler than 12,000~K. This median WD age precision compares well to similar results from an independent analysis of many of the same objects \citep{2021ApJS..253...58Q}. It should be noted that with additional information through spectroscopy we can achieve better age precision \citep{2022ApJ...929...26M}.

\section{Conclusions}

In this work, we constructed a sample of 1565 wide WD+WD binaries as well as 24 mostly new wide triple WD+WD+X systems to test the accuracy and precision of WD total age determinations when no spectral type information is known. We measured effective temperatures, surface gravities, and masses for all WDs in our sample by using hydrogen-dominated atmosphere models and fitting photometry from a variety of all-sky surveys. Using these atmospheric parameters, we determined the total ages of each WD through the use of WD cooling models, a theoretically motivated and observationally calibrated initial-to-final-mass relation, and stellar evolution model grids. 
  
Using a high-fidelity sample of wide WD+WD pairs with uncontaminated photometry described in Section~\ref{sec:results}, the main conclusions of this work can be summarized as follows:

\begin{itemize}
    
    \item We find 24 mostly new widely separated triples with at least two WDs, and possibly the second ever resolved triple-WD system. We find 21 of the 24 triples have a  hierarchical structure. The three triples not in a hierarchical structure will likely dissipate due to dynamical instability, but the ages of the WDs in these systems can provide useful lower limits on the dissipation timescale. We find that two of the non-heirarchical triple systems are $\approx{3}$ Gyr old, while the other is quite young at $\approx{90}$ Myr old.
    
    \item We find that the total ages of the lowest mass WDs in the sample ($<0.63 M_\odot$) are too uncertain to be used in the age comparisons conducted in this work. The absolute age difference of systems containing at least one lower-mass WD quickly diverges towards high values of a few Gyr. A lower limit on the total age can still be determined for these WDs, but we do not attempt to test the absolute reliability of their total ages. This is unfortunate, since the mean mass of field WDs is roughly 0.63\,\msun\ \citep{2016MNRAS.461.2100T, 2016MNRAS.455.3413K, 2019MNRAS.486.2169K}.
    
    \item When comparing the total ages of each component of the WD pairs, we find a significant fraction of systems ($21-43$\%) where the more massive WD has a shorter cooling age. We find that this trend holds even for a subsample of DA+DA systems, with 21-36\% having the more massive WD with a shorter cooling age. This is the opposite of what is expected for a monotonic and increasing initial-to-final-mass relation. We attribute this discrepancy to the presence of many merger remnants or unresolved triples in the sample, such that roughly $21-36$\% of our sample of currently wide WD+WD binaries started as bound triple systems. These WDs have large age disagreement values and can cause problems when determining ages for field WDs, but unresolved companions or a merger history is a problem for many techniques for determining stellar ages.
    
    \item We find that the level of age disagreement strongly depends on the mass of the WD. Wide binaries containing at least one higher-mass WD ($>$0.8\,\msun) generally have a lower level of agreement compared to those with lower-mass WDs, reflecting the fact that the formal age uncertainties are often underestimated for higher-mass WDs. We fit this dependence to determine the amount of error inflation on the total ages needed to improve the agreement. We find that errors on the total ages of the highest-mass WDs ($>$0.9\,\msun) need to be inflated, up to a factor of 2.2, when assuming that the WD is a DA. Fortunately, higher-mass WDs also have more precisely determined ages (median total age uncertainties of $\approx{0.2}$ Gyr), so they remain very useful chronometers.
    
    \item Looking at 25 bins in mass and effective temperature space in Figure~\ref{fig:agreement}, we find that total ages are roughly accurate at the 25\% level for WDs with masses $>$0.63\,\msun\ using only photometry. The WDs with the hottest temperatures have the largest inaccuracy, with uncertainties of 35\%-60\%, but these are relatively small absolute errors on the order of 0.3 Gyr. 
\end{itemize}

Due to the large sample size, we assumed that all the WDs in our sample are DA. When a spectral type is known, a more accurate total age can be determined and the recommendations discussed above are not immediately applicable. A large sample of wide double WDs with known spectral types needs to be investigated to determine the accuracy of WD total ages when the spectral type is known.

Future work is planned to quantify how  atmospheric parameters from spectroscopy can alleviate the systematic errors mentioned above. Spectroscopic observations do not suffer from the same problems as the photometric methods employed in this work, and should be more sensitive to the WD mass. A spectrum can also provide important clues as to whether a WD has a prior merger history or is part of an unresolved binary (for example, if it is strongly magnetic, or if it is radial-velocity variable). With our follow-up efforts using the LDT, we continue to increase the sample of wide WD+WD pairs with available spectra in hopes of gathering a large sample of wide double WDs with spectroscopically determined atmospheric parameters.

In general, we find that WD total ages agree very well in wide double WDs, but the uncertainties occasionally need to be inflated to compensate for systematics on the total ages, the presence of an unresolved companion, or merger history. Given the massive increase in the number of systems with a wide WD companion thanks to Gaia, we hope this result provides a first road map to providing more reliable stellar ages using these wide binaries.

%Mention something about next steps with WD+dM, spectroscopy, etc.

\section{Acknowledgements}
We would like to acknowledge Jeff Andrews for helpful discussions. T.M.H, J.J.H., and C.W. acknowledge support from the National Science Foundation under Grant No. AST-1908119. J.v.S. acknowledges support from the National Science Foundation under Grant No. AST-1908723. We thank Phil Muirhead and Adam Samuels for support with the quick-look spectroscopic tool.
 
This work has made use of data from the European Space Agency (ESA) mission
{\it Gaia} (\url{https://www.cosmos.esa.int/gaia}), processed by the {\it Gaia}
Data Processing and Analysis Consortium (DPAC, \url{https://www.cosmos.esa.int/web/gaia/dpac/consortium}). Funding for the DPAC
has been provided by national institutions, in particular the institutions
participating in the {\it Gaia} Multilateral Agreement. These results made use of the Lowell Discovery Telescope (LDT) at Lowell Observatory. Lowell is a private, non-profit institution dedicated to astrophysical research and public appreciation of astronomy and operates the LDT in partnership with Boston University, the University of Maryland, the University of Toledo, Northern Arizona University and Yale University. The upgrade of the DeVeny optical spectrograph has been funded by a generous grant from John and Ginger Giovale and by a grant from the Mt. Cuba Astronomical Foundation.


\bibliography{WDWD}{}

\begin{thebibliography}{}
\expandafter\ifx\csname natexlab\endcsname\relax\def\natexlab#1{#1}\fi
\providecommand{\url}[1]{\href{#1}{#1}}
\providecommand{\dodoi}[1]{doi:~\href{http://doi.org/#1}{\nolinkurl{#1}}}
\providecommand{\doeprint}[1]{\href{http://ascl.net/#1}{\nolinkurl{http://ascl.net/#1}}}
\providecommand{\doarXiv}[1]{\href{https://arxiv.org/abs/#1}{\nolinkurl{https://arxiv.org/abs/#1}}}

\bibitem[{{Andrews} {et~al.}(2015){Andrews}, {Ag{\"u}eros}, {Gianninas},
  {Kilic}, {Dhital}, \& {Anderson}}]{2015ApJ...815...63A}
{Andrews}, J.~J., {Ag{\"u}eros}, M.~A., {Gianninas}, A., {et~al.} 2015, \apj,
  815, 63, \dodoi{10.1088/0004-637X/815/1/63}

\bibitem[{{Astropy Collaboration} {et~al.}(2018){Astropy Collaboration},
  {Price-Whelan}, {Sip{\H{o}}cz}, {G{\"u}nther}, {Lim}, {Crawford}, {Conseil},
  {Shupe}, {Craig}, {Dencheva}, {Ginsburg}, {VanderPlas}, {Bradley},
  {P{\'e}rez-Su{\'a}rez}, {de Val-Borro}, {Aldcroft}, {Cruz}, {Robitaille},
  {Tollerud}, {Ardelean}, {Babej}, {Bach}, {Bachetti}, {Bakanov}, {Bamford},
  {Barentsen}, {Barmby}, {Baumbach}, {Berry}, {Biscani}, {Boquien}, {Bostroem},
  {Bouma}, {Brammer}, {Bray}, {Breytenbach}, {Buddelmeijer}, {Burke},
  {Calderone}, {Cano Rodr{\'\i}guez}, {Cara}, {Cardoso}, {Cheedella}, {Copin},
  {Corrales}, {Crichton}, {D'Avella}, {Deil}, {Depagne}, {Dietrich}, {Donath},
  {Droettboom}, {Earl}, {Erben}, {Fabbro}, {Ferreira}, {Finethy}, {Fox},
  {Garrison}, {Gibbons}, {Goldstein}, {Gommers}, {Greco}, {Greenfield},
  {Groener}, {Grollier}, {Hagen}, {Hirst}, {Homeier}, {Horton}, {Hosseinzadeh},
  {Hu}, {Hunkeler}, {Ivezi{\'c}}, {Jain}, {Jenness}, {Kanarek}, {Kendrew},
  {Kern}, {Kerzendorf}, {Khvalko}, {King}, {Kirkby}, {Kulkarni}, {Kumar},
  {Lee}, {Lenz}, {Littlefair}, {Ma}, {Macleod}, {Mastropietro}, {McCully},
  {Montagnac}, {Morris}, {Mueller}, {Mumford}, {Muna}, {Murphy}, {Nelson},
  {Nguyen}, {Ninan}, {N{\"o}the}, {Ogaz}, {Oh}, {Parejko}, {Parley}, {Pascual},
  {Patil}, {Patil}, {Plunkett}, {Prochaska}, {Rastogi}, {Reddy Janga},
  {Sabater}, {Sakurikar}, {Seifert}, {Sherbert}, {Sherwood-Taylor}, {Shih},
  {Sick}, {Silbiger}, {Singanamalla}, {Singer}, {Sladen}, {Sooley},
  {Sornarajah}, {Streicher}, {Teuben}, {Thomas}, {Tremblay}, {Turner},
  {Terr{\'o}n}, {van Kerkwijk}, {de la Vega}, {Watkins}, {Weaver}, {Whitmore},
  {Woillez}, {Zabalza}, \& {Astropy Contributors}}]{2018AJ....156..123A}
{Astropy Collaboration}, {Price-Whelan}, A.~M., {Sip{\H{o}}cz}, B.~M., {et~al.}
  2018, \aj, 156, 123, \dodoi{10.3847/1538-3881/aabc4f}

\bibitem[{{Barrientos} \& {Chanam{\'e}}(2021)}]{2021ApJ...923..181B}
{Barrientos}, M., \& {Chanam{\'e}}, J. 2021, \apj, 923, 181,
  \dodoi{10.3847/1538-4357/ac2f49}

\bibitem[{{Bauer} {et~al.}(2020){Bauer}, {Schwab}, {Bildsten}, \&
  {Cheng}}]{2020ApJ...902...93B}
{Bauer}, E.~B., {Schwab}, J., {Bildsten}, L., \& {Cheng}, S. 2020, \apj, 902,
  93, \dodoi{10.3847/1538-4357/abb5a5}

\bibitem[{{Baxter} {et~al.}(2014){Baxter}, {Dobbie}, {Parker}, {Casewell},
  {Lodieu}, {Burleigh}, {Lawrie}, {K{\"u}lebi}, {Koester}, \&
  {Holland}}]{2014MNRAS.440.3184B}
{Baxter}, R.~B., {Dobbie}, P.~D., {Parker}, Q.~A., {et~al.} 2014, \mnras, 440,
  3184, \dodoi{10.1093/mnras/stu464}

\bibitem[{{B{\'e}dard} {et~al.}(2020){B{\'e}dard}, {Bergeron}, {Brassard}, \&
  {Fontaine}}]{2020ApJ...901...93B}
{B{\'e}dard}, A., {Bergeron}, P., {Brassard}, P., \& {Fontaine}, G. 2020, \apj,
  901, 93, \dodoi{10.3847/1538-4357/abafbe}

\bibitem[{{Bergeron} {et~al.}(2019){Bergeron}, {Dufour}, {Fontaine}, {Coutu},
  {Blouin}, {Genest-Beaulieu}, {B{\'e}dard}, \&
  {Rolland}}]{2019ApJ...876...67B}
{Bergeron}, P., {Dufour}, P., {Fontaine}, G., {et~al.} 2019, \apj, 876, 67,
  \dodoi{10.3847/1538-4357/ab153a}

\bibitem[{{Bergeron} {et~al.}(1995){Bergeron}, {Wesemael}, \&
  {Beauchamp}}]{1995PASP..107.1047B}
{Bergeron}, P., {Wesemael}, F., \& {Beauchamp}, A. 1995, \pasp, 107, 1047,
  \dodoi{10.1086/133661}

\bibitem[{{Bergeron} {et~al.}(2011){Bergeron}, {Wesemael}, {Dufour},
  {Beauchamp}, {Hunter}, {Saffer}, {Gianninas}, {Ruiz}, {Limoges}, {Dufour},
  {Fontaine}, \& {Liebert}}]{2011ApJ...737...28B}
{Bergeron}, P., {Wesemael}, F., {Dufour}, P., {et~al.} 2011, \apj, 737, 28,
  \dodoi{10.1088/0004-637X/737/1/28}

\bibitem[{{Bida} {et~al.}(2014){Bida}, {Dunham}, {Massey}, \&
  {Roe}}]{2014SPIE.9147E..2NB}
{Bida}, T.~A., {Dunham}, E.~W., {Massey}, P., \& {Roe}, H.~G. 2014, in Society
  of Photo-Optical Instrumentation Engineers (SPIE) Conference Series, Vol.
  9147, Ground-based and Airborne Instrumentation for Astronomy V, ed. S.~K.
  {Ramsay}, I.~S. {McLean}, \& H.~{Takami}, 91472N, \dodoi{10.1117/12.2056872}

\bibitem[{{Bildsten} \& {Hall}(2001)}]{2001ApJ...549L.219B}
{Bildsten}, L., \& {Hall}, D.~M. 2001, \apjl, 549, L219, \dodoi{10.1086/319169}

\bibitem[{{Blouin} {et~al.}(2021){Blouin}, {Daligault}, \&
  {Saumon}}]{2021ApJ...911L...5B}
{Blouin}, S., {Daligault}, J., \& {Saumon}, D. 2021, \apjl, 911, L5,
  \dodoi{10.3847/2041-8213/abf14b}

\bibitem[{{Bressan} {et~al.}(2012){Bressan}, {Marigo}, {Girardi}, {Salasnich},
  {Dal Cero}, {Rubele}, \& {Nanni}}]{2012MNRAS.427..127B}
{Bressan}, A., {Marigo}, P., {Girardi}, L., {et~al.} 2012, \mnras, 427, 127,
  \dodoi{10.1111/j.1365-2966.2012.21948.x}

\bibitem[{{Burleigh} {et~al.}(1999){Burleigh}, {Jordan}, \&
  {Schweizer}}]{1999ApJ...510L..37B}
{Burleigh}, M.~R., {Jordan}, S., \& {Schweizer}, W. 1999, \apjl, 510, L37,
  \dodoi{10.1086/311794}

\bibitem[{{Catal{\'a}n} {et~al.}(2008){Catal{\'a}n}, {Isern},
  {Garc{\'\i}a-Berro}, \& {Ribas}}]{2008MNRAS.387.1693C}
{Catal{\'a}n}, S., {Isern}, J., {Garc{\'\i}a-Berro}, E., \& {Ribas}, I. 2008,
  \mnras, 387, 1693, \dodoi{10.1111/j.1365-2966.2008.13356.x}

\bibitem[{{Cheng} {et~al.}(2019){Cheng}, {Cummings}, \&
  {M{\'e}nard}}]{2019ApJ...886..100C}
{Cheng}, S., {Cummings}, J.~D., \& {M{\'e}nard}, B. 2019, \apj, 886, 100,
  \dodoi{10.3847/1538-4357/ab4989}

\bibitem[{{Choi} {et~al.}(2016){Choi}, {Dotter}, {Conroy}, {Cantiello},
  {Paxton}, \& {Johnson}}]{2016ApJ...823..102C}
{Choi}, J., {Dotter}, A., {Conroy}, C., {et~al.} 2016, \apj, 823, 102,
  \dodoi{10.3847/0004-637X/823/2/102}

\bibitem[{{Croom} {et~al.}(2004){Croom}, {Smith}, {Boyle}, {Shanks}, {Miller},
  {Outram}, \& {Loaring}}]{2004MNRAS.349.1397C}
{Croom}, S.~M., {Smith}, R.~J., {Boyle}, B.~J., {et~al.} 2004, \mnras, 349,
  1397, \dodoi{10.1111/j.1365-2966.2004.07619.x}

\bibitem[{{Cummings} {et~al.}(2019){Cummings}, {Kalirai}, {Choi}, {Georgy},
  {Tremblay}, \& {Ramirez-Ruiz}}]{2019ApJ...871L..18C}
{Cummings}, J.~D., {Kalirai}, J.~S., {Choi}, J., {et~al.} 2019, \apjl, 871,
  L18, \dodoi{10.3847/2041-8213/aafc2d}

\bibitem[{{Cummings} {et~al.}(2015){Cummings}, {Kalirai}, {Tremblay}, \&
  {Ramirez-Ruiz}}]{2015ApJ...807...90C}
{Cummings}, J.~D., {Kalirai}, J.~S., {Tremblay}, P.~E., \& {Ramirez-Ruiz}, E.
  2015, \apj, 807, 90, \dodoi{10.1088/0004-637X/807/1/90}

\bibitem[{{Cummings} {et~al.}(2016){Cummings}, {Kalirai}, {Tremblay}, \&
  {Ramirez-Ruiz}}]{2016ApJ...818...84C}
---. 2016, \apj, 818, 84, \dodoi{10.3847/0004-637X/818/1/84}

\bibitem[{{Cummings} {et~al.}(2018){Cummings}, {Kalirai}, {Tremblay},
  {Ramirez-Ruiz}, \& {Choi}}]{2018ApJ...866...21C}
{Cummings}, J.~D., {Kalirai}, J.~S., {Tremblay}, P.~E., {Ramirez-Ruiz}, E., \&
  {Choi}, J. 2018, \apj, 866, 21, \dodoi{10.3847/1538-4357/aadfd6}

\bibitem[{{deBoer} {et~al.}(2017){deBoer}, {G{\"o}rres}, {Wiescher}, {Azuma},
  {Best}, {Brune}, {Fields}, {Jones}, {Pignatari}, {Sayre}, {Smith}, {Timmes},
  \& {Uberseder}}]{2017RvMP...89c5007D}
{deBoer}, R.~J., {G{\"o}rres}, J., {Wiescher}, M., {et~al.} 2017, Reviews of
  Modern Physics, 89, 035007, \dodoi{10.1103/RevModPhys.89.035007}

\bibitem[{{Dobbie} {et~al.}(2012){Dobbie}, {Baxter}, {K{\"u}lebi}, {Parker},
  {Koester}, {Jordan}, {Lodieu}, \& {Euchner}}]{2012MNRAS.421..202D}
{Dobbie}, P.~D., {Baxter}, R., {K{\"u}lebi}, B., {et~al.} 2012, \mnras, 421,
  202, \dodoi{10.1111/j.1365-2966.2012.20291.x}

\bibitem[{{Doherty} {et~al.}(2015){Doherty}, {Gil-Pons}, {Siess}, {Lattanzio},
  \& {Lau}}]{2015MNRAS.446.2599D}
{Doherty}, C.~L., {Gil-Pons}, P., {Siess}, L., {Lattanzio}, J.~C., \& {Lau}, H.
  H.~B. 2015, \mnras, 446, 2599, \dodoi{10.1093/mnras/stu2180}

\bibitem[{{Dotter}(2016)}]{2016ApJS..222....8D}
{Dotter}, A. 2016, \apjs, 222, 8, \dodoi{10.3847/0067-0049/222/1/8}

\bibitem[{{Eisenstein} {et~al.}(2006){Eisenstein}, {Liebert}, {Koester},
  {Kleinmann}, {Nitta}, {Smith}, {Barentine}, {Brewington}, {Brinkmann},
  {Harvanek}, {Krzesi{\'n}ski}, {Neilsen}, {Long}, {Schneider}, \&
  {Snedden}}]{2006AJ....132..676E}
{Eisenstein}, D.~J., {Liebert}, J., {Koester}, D., {et~al.} 2006, \aj, 132,
  676, \dodoi{10.1086/504424}

\bibitem[{{El-Badry} \& {Rix}(2018)}]{2018MNRAS.480.4884E}
{El-Badry}, K., \& {Rix}, H.-W. 2018, \mnras, 480, 4884,
  \dodoi{10.1093/mnras/sty2186}

\bibitem[{{El-Badry} {et~al.}(2021){El-Badry}, {Rix}, \&
  {Heintz}}]{2021MNRAS.506.2269E}
{El-Badry}, K., {Rix}, H.-W., \& {Heintz}, T.~M. 2021, \mnras,
  \dodoi{10.1093/mnras/stab323}

\bibitem[{{El-Badry} {et~al.}(2018){El-Badry}, {Rix}, \&
  {Weisz}}]{2018ApJ...860L..17E}
{El-Badry}, K., {Rix}, H.-W., \& {Weisz}, D.~R. 2018, \apjl, 860, L17,
  \dodoi{10.3847/2041-8213/aaca9c}

\bibitem[{{Ferrario} {et~al.}(2015){Ferrario}, {de Martino}, \&
  {G{\"a}nsicke}}]{2015SSRv..191..111F}
{Ferrario}, L., {de Martino}, D., \& {G{\"a}nsicke}, B.~T. 2015, \ssr, 191,
  111, \dodoi{10.1007/s11214-015-0152-0}

\bibitem[{{Ferrario} {et~al.}(1997){Ferrario}, {Vennes}, {Wickramasinghe},
  {Bailey}, \& {Christian}}]{1997MNRAS.292..205F}
{Ferrario}, L., {Vennes}, S., {Wickramasinghe}, D.~T., {Bailey}, J.~A., \&
  {Christian}, D.~J. 1997, \mnras, 292, 205, \dodoi{10.1093/mnras/292.2.205}

\bibitem[{{Fields} {et~al.}(2016){Fields}, {Farmer}, {Petermann}, {Iliadis}, \&
  {Timmes}}]{2016ApJ...823...46F}
{Fields}, C.~E., {Farmer}, R., {Petermann}, I., {Iliadis}, C., \& {Timmes},
  F.~X. 2016, \apj, 823, 46, \dodoi{10.3847/0004-637X/823/1/46}

\bibitem[{{Finley} \& {Koester}(1997)}]{1997ApJ...489L..79F}
{Finley}, D.~S., \& {Koester}, D. 1997, \apjl, 489, L79, \dodoi{10.1086/310967}

\bibitem[{{Fontaine} {et~al.}(2001){Fontaine}, {Brassard}, \&
  {Bergeron}}]{2001PASP..113..409F}
{Fontaine}, G., {Brassard}, P., \& {Bergeron}, P. 2001, \pasp, 113, 409,
  \dodoi{10.1086/319535}

\bibitem[{{Foreman-Mackey} {et~al.}(2013){Foreman-Mackey}, {Hogg}, {Lang}, \&
  {Goodman}}]{2013PASP..125..306F}
{Foreman-Mackey}, D., {Hogg}, D.~W., {Lang}, D., \& {Goodman}, J. 2013, \pasp,
  125, 306, \dodoi{10.1086/670067}

\bibitem[{{Fouesneau} {et~al.}(2019){Fouesneau}, {Rix}, {von Hippel}, {Hogg},
  \& {Tian}}]{2019ApJ...870....9F}
{Fouesneau}, M., {Rix}, H.-W., {von Hippel}, T., {Hogg}, D.~W., \& {Tian}, H.
  2019, \apj, 870, 9, \dodoi{10.3847/1538-4357/aaee74}

\bibitem[{{Gaia Collaboration} {et~al.}(2016){Gaia Collaboration}, {Prusti},
  {de Bruijne}, {Brown}, {Vallenari}, {Babusiaux}, {Bailer-Jones}, {Bastian},
  {Biermann}, {Evans}, {Eyer}, {Jansen}, {Jordi}, {Klioner}, {Lammers},
  {Lindegren}, {Luri}, {Mignard}, {Milligan}, {Panem}, {Poinsignon},
  {Pourbaix}, {Randich}, {Sarri}, {Sartoretti}, {Siddiqui}, {Soubiran},
  {Valette}, {van Leeuwen}, {Walton}, {Aerts}, {Arenou}, {Cropper}, {Drimmel},
  {H{\o}g}, {Katz}, {Lattanzi}, {O'Mullane}, {Grebel}, {Holland}, {Huc},
  {Passot}, {Bramante}, {Cacciari}, {Casta{\~n}eda}, {Chaoul}, {Cheek}, {De
  Angeli}, {Fabricius}, {Guerra}, {Hern{\'a}ndez}, {Jean-Antoine-Piccolo},
  {Masana}, {Messineo}, {Mowlavi}, {Nienartowicz}, {Ord{\'o}{\~n}ez-Blanco},
  {Panuzzo}, {Portell}, {Richards}, {Riello}, {Seabroke}, {Tanga},
  {Th{\'e}venin}, {Torra}, {Els}, {Gracia-Abril}, {Comoretto},
  {Garcia-Reinaldos}, {Lock}, {Mercier}, {Altmann}, {Andrae}, {Astraatmadja},
  {Bellas-Velidis}, {Benson}, {Berthier}, {Blomme}, {Busso}, {Carry},
  {Cellino}, {Clementini}, {Cowell}, {Creevey}, {Cuypers}, {Davidson}, {De
  Ridder}, {de Torres}, {Delchambre}, {Dell'Oro}, {Ducourant}, {Fr{\'e}mat},
  {Garc{\'\i}a-Torres}, {Gosset}, {Halbwachs}, {Hambly}, {Harrison}, {Hauser},
  {Hestroffer}, {Hodgkin}, {Huckle}, {Hutton}, {Jasniewicz}, {Jordan},
  {Kontizas}, {Korn}, {Lanzafame}, {Manteiga}, {Moitinho}, {Muinonen},
  {Osinde}, {Pancino}, {Pauwels}, {Petit}, {Recio-Blanco}, {Robin}, {Sarro},
  {Siopis}, {Smith}, {Smith}, {Sozzetti}, {Thuillot}, {van Reeven}, {Viala},
  {Abbas}, {Abreu Aramburu}, {Accart}, {Aguado}, {Allan}, {Allasia},
  {Altavilla}, {{\'A}lvarez}, {Alves}, {Anderson}, {Andrei}, {Anglada Varela},
  {Antiche}, {Antoja}, {Ant{\'o}n}, {Arcay}, {Atzei}, {Ayache}, {Bach},
  {Baker}, {Balaguer-N{\'u}{\~n}ez}, {Barache}, {Barata}, {Barbier}, {Barblan},
  {Baroni}, {Barrado y Navascu{\'e}s}, {Barros}, {Barstow}, {Becciani},
  {Bellazzini}, {Bellei}, {Bello Garc{\'\i}a}, {Belokurov}, {Bendjoya},
  {Berihuete}, {Bianchi}, {Bienaym{\'e}}, {Billebaud}, {Blagorodnova},
  {Blanco-Cuaresma}, {Boch}, {Bombrun}, {Borrachero}, {Bouquillon}, {Bourda},
  {Bouy}, {Bragaglia}, {Breddels}, {Brouillet}, {Br{\"u}semeister},
  {Bucciarelli}, {Budnik}, {Burgess}, {Burgon}, {Burlacu}, {Busonero}, {Buzzi},
  {Caffau}, {Cambras}, {Campbell}, {Cancelliere}, {Cantat-Gaudin}, {Carlucci},
  {Carrasco}, {Castellani}, {Charlot}, {Charnas}, {Charvet}, {Chassat},
  {Chiavassa}, {Clotet}, {Cocozza}, {Collins}, {Collins}, {Costigan}, {Crifo},
  {Cross}, {Crosta}, {Crowley}, {Dafonte}, {Damerdji}, {Dapergolas}, {David},
  {David}, {De Cat}, {de Felice}, {de Laverny}, {De Luise}, {De March}, {de
  Martino}, {de Souza}, {Debosscher}, {del Pozo}, {Delbo}, {Delgado},
  {Delgado}, {di Marco}, {Di Matteo}, {Diakite}, {Distefano}, {Dolding}, {Dos
  Anjos}, {Drazinos}, {Dur{\'a}n}, {Dzigan}, {Ecale}, {Edvardsson}, {Enke},
  {Erdmann}, {Escolar}, {Espina}, {Evans}, {Eynard Bontemps}, {Fabre},
  {Fabrizio}, {Faigler}, {Falc{\~a}o}, {Farr{\`a}s Casas}, {Faye}, {Federici},
  {Fedorets}, {Fern{\'a}ndez-Hern{\'a}ndez}, {Fernique}, {Fienga}, {Figueras},
  {Filippi}, {Findeisen}, {Fonti}, {Fouesneau}, {Fraile}, {Fraser}, {Fuchs},
  {Furnell}, {Gai}, {Galleti}, {Galluccio}, {Garabato}, {Garc{\'\i}a-Sedano},
  {Gar{\'e}}, {Garofalo}, {Garralda}, {Gavras}, {Gerssen}, {Geyer}, {Gilmore},
  {Girona}, {Giuffrida}, {Gomes}, {Gonz{\'a}lez-Marcos},
  {Gonz{\'a}lez-N{\'u}{\~n}ez}, {Gonz{\'a}lez-Vidal}, {Granvik}, {Guerrier},
  {Guillout}, {Guiraud}, {G{\'u}rpide}, {Guti{\'e}rrez-S{\'a}nchez}, {Guy},
  {Haigron}, {Hatzidimitriou}, {Haywood}, {Heiter}, {Helmi}, {Hobbs},
  {Hofmann}, {Holl}, {Holland}, {Hunt}, {Hypki}, {Icardi}, {Irwin}, {Jevardat
  de Fombelle}, {Jofr{\'e}}, {Jonker}, {Jorissen}, {Julbe}, {Karampelas},
  {Kochoska}, {Kohley}, {Kolenberg}, {Kontizas}, {Koposov}, {Kordopatis},
  {Koubsky}, {Kowalczyk}, {Krone-Martins}, {Kudryashova}, {Kull}, {Bachchan},
  {Lacoste-Seris}, {Lanza}, {Lavigne}, {Le Poncin-Lafitte}, {Lebreton},
  {Lebzelter}, {Leccia}, {Leclerc}, {Lecoeur-Taibi}, {Lemaitre}, {Lenhardt},
  {Leroux}, {Liao}, {Licata}, {Lindstr{\o}m}, {Lister}, {Livanou}, {Lobel},
  {L{\"o}ffler}, {L{\'o}pez}, {Lopez-Lozano}, {Lorenz}, {Loureiro},
  {MacDonald}, {Magalh{\~a}es Fernandes}, {Managau}, {Mann}, {Mantelet},
  {Marchal}, {Marchant}, {Marconi}, {Marie}, {Marinoni}, {Marrese},
  {Marschalk{\'o}}, {Marshall}, {Mart{\'\i}n-Fleitas}, {Martino}, {Mary},
  {Matijevi{\v{c}}}, {Mazeh}, {McMillan}, {Messina}, {Mestre}, {Michalik},
  {Millar}, {Miranda}, {Molina}, {Molinaro}, {Molinaro}, {Moln{\'a}r},
  {Moniez}, {Montegriffo}, {Monteiro}, {Mor}, {Mora}, {Morbidelli}, {Morel},
  {Morgenthaler}, {Morley}, {Morris}, {Mulone}, {Muraveva}, {Musella},
  {Narbonne}, {Nelemans}, {Nicastro}, {Noval}, {Ord{\'e}novic},
  {Ordieres-Mer{\'e}}, {Osborne}, {Pagani}, {Pagano}, {Pailler}, {Palacin},
  {Palaversa}, {Parsons}, {Paulsen}, {Pecoraro}, {Pedrosa}, {Pentik{\"a}inen},
  {Pereira}, {Pichon}, {Piersimoni}, {Pineau}, {Plachy}, {Plum}, {Poujoulet},
  {Pr{\v{s}}a}, {Pulone}, {Ragaini}, {Rago}, {Rambaux}, {Ramos-Lerate},
  {Ranalli}, {Rauw}, {Read}, {Regibo}, {Renk}, {Reyl{\'e}}, {Ribeiro},
  {Rimoldini}, {Ripepi}, {Riva}, {Rixon}, {Roelens}, {Romero-G{\'o}mez},
  {Rowell}, {Royer}, {Rudolph}, {Ruiz-Dern}, {Sadowski}, {Sagrist{\`a}
  Sell{\'e}s}, {Sahlmann}, {Salgado}, {Salguero}, {Sarasso}, {Savietto},
  {Schnorhk}, {Schultheis}, {Sciacca}, {Segol}, {Segovia}, {Segransan},
  {Serpell}, {Shih}, {Smareglia}, {Smart}, {Smith}, {Solano}, {Solitro},
  {Sordo}, {Soria Nieto}, {Souchay}, {Spagna}, {Spoto}, {Stampa}, {Steele},
  {Steidelm{\"u}ller}, {Stephenson}, {Stoev}, {Suess}, {S{\"u}veges}, {Surdej},
  {Szabados}, {Szegedi-Elek}, {Tapiador}, {Taris}, {Tauran}, {Taylor},
  {Teixeira}, {Terrett}, {Tingley}, {Trager}, {Turon}, {Ulla}, {Utrilla},
  {Valentini}, {van Elteren}, {Van Hemelryck}, {van Leeuwen}, {Varadi},
  {Vecchiato}, {Veljanoski}, {Via}, {Vicente}, {Vogt}, {Voss}, {Votruba},
  {Voutsinas}, {Walmsley}, {Weiler}, {Weingrill}, {Werner}, {Wevers},
  {Whitehead}, {Wyrzykowski}, {Yoldas}, {{\v{Z}}erjal}, {Zucker}, {Zurbach},
  {Zwitter}, {Alecu}, {Allen}, {Allende Prieto}, {Amorim},
  {Anglada-Escud{\'e}}, {Arsenijevic}, {Azaz}, {Balm}, {Beck}, {Bernstein},
  {Bigot}, {Bijaoui}, {Blasco}, {Bonfigli}, {Bono}, {Boudreault}, {Bressan},
  {Brown}, {Brunet}, {Bunclark}, {Buonanno}, {Butkevich}, {Carret}, {Carrion},
  {Chemin}, {Ch{\'e}reau}, {Corcione}, {Darmigny}, {de Boer}, {de Teodoro}, {de
  Zeeuw}, {Delle Luche}, {Domingues}, {Dubath}, {Fodor}, {Fr{\'e}zouls},
  {Fries}, {Fustes}, {Fyfe}, {Gallardo}, {Gallegos}, {Gardiol}, {Gebran},
  {Gomboc}, {G{\'o}mez}, {Grux}, {Gueguen}, {Heyrovsky}, {Hoar}, {Iannicola},
  {Isasi Parache}, {Janotto}, {Joliet}, {Jonckheere}, {Keil}, {Kim},
  {Klagyivik}, {Klar}, {Knude}, {Kochukhov}, {Kolka}, {Kos}, {Kutka}, {Lainey},
  {LeBouquin}, {Liu}, {Loreggia}, {Makarov}, {Marseille}, {Martayan},
  {Martinez-Rubi}, {Massart}, {Meynadier}, {Mignot}, {Munari}, {Nguyen},
  {Nordlander}, {Ocvirk}, {O'Flaherty}, {Olias Sanz}, {Ortiz}, {Osorio},
  {Oszkiewicz}, {Ouzounis}, {Palmer}, {Park}, {Pasquato}, {Peltzer}, {Peralta},
  {P{\'e}turaud}, {Pieniluoma}, {Pigozzi}, {Poels}, {Prat}, {Prod'homme},
  {Raison}, {Rebordao}, {Risquez}, {Rocca-Volmerange}, {Rosen}, {Ruiz-Fuertes},
  {Russo}, {Sembay}, {Serraller Vizcaino}, {Short}, {Siebert}, {Silva},
  {Sinachopoulos}, {Slezak}, {Soffel}, {Sosnowska}, {Strai{\v{z}}ys}, {ter
  Linden}, {Terrell}, {Theil}, {Tiede}, {Troisi}, {Tsalmantza}, {Tur},
  {Vaccari}, {Vachier}, {Valles}, {Van Hamme}, {Veltz}, {Virtanen}, {Wallut},
  {Wichmann}, {Wilkinson}, {Ziaeepour}, \& {Zschocke}}]{2016A&A...595A...1G}
{Gaia Collaboration}, {Prusti}, T., {de Bruijne}, J.~H.~J., {et~al.} 2016,
  \aap, 595, A1, \dodoi{10.1051/0004-6361/201629272}

\bibitem[{{Gaia Collaboration} {et~al.}(2018{\natexlab{a}}){Gaia
  Collaboration}, {Brown}, {Vallenari}, {Prusti}, {de Bruijne}, {Babusiaux},
  {Bailer-Jones}, {Biermann}, {Evans}, {Eyer}, {Jansen}, {Jordi}, {Klioner},
  {Lammers}, {Lindegren}, {Luri}, {Mignard}, {Panem}, {Pourbaix}, {Randich},
  {Sartoretti}, {Siddiqui}, {Soubiran}, {van Leeuwen}, {Walton}, {Arenou},
  {Bastian}, {Cropper}, {Drimmel}, {Katz}, {Lattanzi}, {Bakker}, {Cacciari},
  {Casta{\~n}eda}, {Chaoul}, {Cheek}, {De Angeli}, {Fabricius}, {Guerra},
  {Holl}, {Masana}, {Messineo}, {Mowlavi}, {Nienartowicz}, {Panuzzo},
  {Portell}, {Riello}, {Seabroke}, {Tanga}, {Th{\'e}venin}, {Gracia-Abril},
  {Comoretto}, {Garcia-Reinaldos}, {Teyssier}, {Altmann}, {Andrae}, {Audard},
  {Bellas-Velidis}, {Benson}, {Berthier}, {Blomme}, {Burgess}, {Busso},
  {Carry}, {Cellino}, {Clementini}, {Clotet}, {Creevey}, {Davidson}, {De
  Ridder}, {Delchambre}, {Dell'Oro}, {Ducourant},
  {Fern{\'a}ndez-Hern{\'a}ndez}, {Fouesneau}, {Fr{\'e}mat}, {Galluccio},
  {Garc{\'\i}a-Torres}, {Gonz{\'a}lez-N{\'u}{\~n}ez}, {Gonz{\'a}lez-Vidal},
  {Gosset}, {Guy}, {Halbwachs}, {Hambly}, {Harrison}, {Hern{\'a}ndez},
  {Hestroffer}, {Hodgkin}, {Hutton}, {Jasniewicz}, {Jean-Antoine-Piccolo},
  {Jordan}, {Korn}, {Krone-Martins}, {Lanzafame}, {Lebzelter}, {L{\"o}ffler},
  {Manteiga}, {Marrese}, {Mart{\'\i}n-Fleitas}, {Moitinho}, {Mora}, {Muinonen},
  {Osinde}, {Pancino}, {Pauwels}, {Petit}, {Recio-Blanco}, {Richards},
  {Rimoldini}, {Robin}, {Sarro}, {Siopis}, {Smith}, {Sozzetti}, {S{\"u}veges},
  {Torra}, {van Reeven}, {Abbas}, {Abreu Aramburu}, {Accart}, {Aerts},
  {Altavilla}, {{\'A}lvarez}, {Alvarez}, {Alves}, {Anderson}, {Andrei},
  {Anglada Varela}, {Antiche}, {Antoja}, {Arcay}, {Astraatmadja}, {Bach},
  {Baker}, {Balaguer-N{\'u}{\~n}ez}, {Balm}, {Barache}, {Barata}, {Barbato},
  {Barblan}, {Barklem}, {Barrado}, {Barros}, {Barstow}, {Bartholom{\'e}
  Mu{\~n}oz}, {Bassilana}, {Becciani}, {Bellazzini}, {Berihuete}, {Bertone},
  {Bianchi}, {Bienaym{\'e}}, {Blanco-Cuaresma}, {Boch}, {Boeche}, {Bombrun},
  {Borrachero}, {Bossini}, {Bouquillon}, {Bourda}, {Bragaglia}, {Bramante},
  {Breddels}, {Bressan}, {Brouillet}, {Br{\"u}semeister}, {Brugaletta},
  {Bucciarelli}, {Burlacu}, {Busonero}, {Butkevich}, {Buzzi}, {Caffau},
  {Cancelliere}, {Cannizzaro}, {Cantat-Gaudin}, {Carballo}, {Carlucci},
  {Carrasco}, {Casamiquela}, {Castellani}, {Castro-Ginard}, {Charlot},
  {Chemin}, {Chiavassa}, {Cocozza}, {Costigan}, {Cowell}, {Crifo}, {Crosta},
  {Crowley}, {Cuypers}, {Dafonte}, {Damerdji}, {Dapergolas}, {David}, {David},
  {de Laverny}, {De Luise}, {De March}, {de Martino}, {de Souza}, {de Torres},
  {Debosscher}, {del Pozo}, {Delbo}, {Delgado}, {Delgado}, {Di Matteo},
  {Diakite}, {Diener}, {Distefano}, {Dolding}, {Drazinos}, {Dur{\'a}n},
  {Edvardsson}, {Enke}, {Eriksson}, {Esquej}, {Eynard Bontemps}, {Fabre},
  {Fabrizio}, {Faigler}, {Falc{\~a}o}, {Farr{\`a}s Casas}, {Federici},
  {Fedorets}, {Fernique}, {Figueras}, {Filippi}, {Findeisen}, {Fonti},
  {Fraile}, {Fraser}, {Fr{\'e}zouls}, {Gai}, {Galleti}, {Garabato},
  {Garc{\'\i}a-Sedano}, {Garofalo}, {Garralda}, {Gavel}, {Gavras}, {Gerssen},
  {Geyer}, {Giacobbe}, {Gilmore}, {Girona}, {Giuffrida}, {Glass}, {Gomes},
  {Granvik}, {Gueguen}, {Guerrier}, {Guiraud}, {Guti{\'e}rrez-S{\'a}nchez},
  {Haigron}, {Hatzidimitriou}, {Hauser}, {Haywood}, {Heiter}, {Helmi}, {Heu},
  {Hilger}, {Hobbs}, {Hofmann}, {Holland}, {Huckle}, {Hypki}, {Icardi},
  {Jan{\ss}en}, {Jevardat de Fombelle}, {Jonker}, {Juh{\'a}sz}, {Julbe},
  {Karampelas}, {Kewley}, {Klar}, {Kochoska}, {Kohley}, {Kolenberg},
  {Kontizas}, {Kontizas}, {Koposov}, {Kordopatis}, {Kostrzewa-Rutkowska},
  {Koubsky}, {Lambert}, {Lanza}, {Lasne}, {Lavigne}, {Le Fustec}, {Le
  Poncin-Lafitte}, {Lebreton}, {Leccia}, {Leclerc}, {Lecoeur-Taibi},
  {Lenhardt}, {Leroux}, {Liao}, {Licata}, {Lindstr{\o}m}, {Lister}, {Livanou},
  {Lobel}, {L{\'o}pez}, {Managau}, {Mann}, {Mantelet}, {Marchal}, {Marchant},
  {Marconi}, {Marinoni}, {Marschalk{\'o}}, {Marshall}, {Martino}, {Marton},
  {Mary}, {Massari}, {Matijevi{\v{c}}}, {Mazeh}, {McMillan}, {Messina},
  {Michalik}, {Millar}, {Molina}, {Molinaro}, {Moln{\'a}r}, {Montegriffo},
  {Mor}, {Morbidelli}, {Morel}, {Morris}, {Mulone}, {Muraveva}, {Musella},
  {Nelemans}, {Nicastro}, {Noval}, {O'Mullane}, {Ord{\'e}novic},
  {Ord{\'o}{\~n}ez-Blanco}, {Osborne}, {Pagani}, {Pagano}, {Pailler},
  {Palacin}, {Palaversa}, {Panahi}, {Pawlak}, {Piersimoni}, {Pineau}, {Plachy},
  {Plum}, {Poggio}, {Poujoulet}, {Pr{\v{s}}a}, {Pulone}, {Racero}, {Ragaini},
  {Rambaux}, {Ramos-Lerate}, {Regibo}, {Reyl{\'e}}, {Riclet}, {Ripepi}, {Riva},
  {Rivard}, {Rixon}, {Roegiers}, {Roelens}, {Romero-G{\'o}mez}, {Rowell},
  {Royer}, {Ruiz-Dern}, {Sadowski}, {Sagrist{\`a} Sell{\'e}s}, {Sahlmann},
  {Salgado}, {Salguero}, {Sanna}, {Santana-Ros}, {Sarasso}, {Savietto},
  {Schultheis}, {Sciacca}, {Segol}, {Segovia}, {S{\'e}gransan}, {Shih},
  {Siltala}, {Silva}, {Smart}, {Smith}, {Solano}, {Solitro}, {Sordo}, {Soria
  Nieto}, {Souchay}, {Spagna}, {Spoto}, {Stampa}, {Steele},
  {Steidelm{\"u}ller}, {Stephenson}, {Stoev}, {Suess}, {Surdej}, {Szabados},
  {Szegedi-Elek}, {Tapiador}, {Taris}, {Tauran}, {Taylor}, {Teixeira},
  {Terrett}, {Teyssandier}, {Thuillot}, {Titarenko}, {Torra Clotet}, {Turon},
  {Ulla}, {Utrilla}, {Uzzi}, {Vaillant}, {Valentini}, {Valette}, {van Elteren},
  {Van Hemelryck}, {van Leeuwen}, {Vaschetto}, {Vecchiato}, {Veljanoski},
  {Viala}, {Vicente}, {Vogt}, {von Essen}, {Voss}, {Votruba}, {Voutsinas},
  {Walmsley}, {Weiler}, {Wertz}, {Wevers}, {Wyrzykowski}, {Yoldas},
  {{\v{Z}}erjal}, {Ziaeepour}, {Zorec}, {Zschocke}, {Zucker}, {Zurbach}, \&
  {Zwitter}}]{2018A&A...616A...1G}
{Gaia Collaboration}, {Brown}, A.~G.~A., {Vallenari}, A., {et~al.}
  2018{\natexlab{a}}, \aap, 616, A1, \dodoi{10.1051/0004-6361/201833051}

\bibitem[{{Gaia Collaboration} {et~al.}(2018{\natexlab{b}}){Gaia
  Collaboration}, {Babusiaux}, {van Leeuwen}, {Barstow}, {Jordi}, {Vallenari},
  {Bossini}, {Bressan}, {Cantat-Gaudin}, {van Leeuwen}, {Brown}, {Prusti}, {de
  Bruijne}, {Bailer-Jones}, {Biermann}, {Evans}, {Eyer}, {Jansen}, {Klioner},
  {Lammers}, {Lindegren}, {Luri}, {Mignard}, {Panem}, {Pourbaix}, {Randich},
  {Sartoretti}, {Siddiqui}, {Soubiran}, {Walton}, {Arenou}, {Bastian},
  {Cropper}, {Drimmel}, {Katz}, {Lattanzi}, {Bakker}, {Cacciari},
  {Casta{\~n}eda}, {Chaoul}, {Cheek}, {De Angeli}, {Fabricius}, {Guerra},
  {Holl}, {Masana}, {Messineo}, {Mowlavi}, {Nienartowicz}, {Panuzzo},
  {Portell}, {Riello}, {Seabroke}, {Tanga}, {Th{\'e}venin}, {Gracia-Abril},
  {Comoretto}, {Garcia-Reinaldos}, {Teyssier}, {Altmann}, {Andrae}, {Audard},
  {Bellas-Velidis}, {Benson}, {Berthier}, {Blomme}, {Burgess}, {Busso},
  {Carry}, {Cellino}, {Clementini}, {Clotet}, {Creevey}, {Davidson}, {De
  Ridder}, {Delchambre}, {Dell'Oro}, {Ducourant},
  {Fern{\'a}ndez-Hern{\'a}ndez}, {Fouesneau}, {Fr{\'e}mat}, {Galluccio},
  {Garc{\'\i}a-Torres}, {Gonz{\'a}lez-N{\'u}{\~n}ez}, {Gonz{\'a}lez-Vidal},
  {Gosset}, {Guy}, {Halbwachs}, {Hambly}, {Harrison}, {Hern{\'a}ndez},
  {Hestroffer}, {Hodgkin}, {Hutton}, {Jasniewicz}, {Jean-Antoine-Piccolo},
  {Jordan}, {Korn}, {Krone-Martins}, {Lanzafame}, {Lebzelter}, {L{\"o}ffler},
  {Manteiga}, {Marrese}, {Mart{\'\i}n-Fleitas}, {Moitinho}, {Mora}, {Muinonen},
  {Osinde}, {Pancino}, {Pauwels}, {Petit}, {Recio-Blanco}, {Richards},
  {Rimoldini}, {Robin}, {Sarro}, {Siopis}, {Smith}, {Sozzetti}, {S{\"u}veges},
  {Torra}, {van Reeven}, {Abbas}, {Abreu Aramburu}, {Accart}, {Aerts},
  {Altavilla}, {{\'A}lvarez}, {Alvarez}, {Alves}, {Anderson}, {Andrei},
  {Anglada Varela}, {Antiche}, {Antoja}, {Arcay}, {Astraatmadja}, {Bach},
  {Baker}, {Balaguer-N{\'u}{\~n}ez}, {Balm}, {Barache}, {Barata}, {Barbato},
  {Barblan}, {Barklem}, {Barrado}, {Barros}, {Bartholom{\'e} Mu{\~n}oz},
  {Bassilana}, {Becciani}, {Bellazzini}, {Berihuete}, {Bertone}, {Bianchi},
  {Bienaym{\'e}}, {Blanco-Cuaresma}, {Boch}, {Boeche}, {Bombrun}, {Borrachero},
  {Bouquillon}, {Bourda}, {Bragaglia}, {Bramante}, {Breddels}, {Brouillet},
  {Br{\"u}semeister}, {Brugaletta}, {Bucciarelli}, {Burlacu}, {Busonero},
  {Butkevich}, {Buzzi}, {Caffau}, {Cancelliere}, {Cannizzaro}, {Carballo},
  {Carlucci}, {Carrasco}, {Casamiquela}, {Castellani}, {Castro-Ginard},
  {Charlot}, {Chemin}, {Chiavassa}, {Cocozza}, {Costigan}, {Cowell}, {Crifo},
  {Crosta}, {Crowley}, {Cuypers}, {Dafonte}, {Damerdji}, {Dapergolas}, {David},
  {David}, {de Laverny}, {De Luise}, {De March}, {de Martino}, {de Souza}, {de
  Torres}, {Debosscher}, {del Pozo}, {Delbo}, {Delgado}, {Delgado}, {Diakite},
  {Diener}, {Distefano}, {Dolding}, {Drazinos}, {Dur{\'a}n}, {Edvardsson},
  {Enke}, {Eriksson}, {Esquej}, {Eynard Bontemps}, {Fabre}, {Fabrizio},
  {Faigler}, {Falc{\~a}o}, {Farr{\`a}s Casas}, {Federici}, {Fedorets},
  {Fernique}, {Figueras}, {Filippi}, {Findeisen}, {Fonti}, {Fraile}, {Fraser},
  {Fr{\'e}zouls}, {Gai}, {Galleti}, {Garabato}, {Garc{\'\i}a-Sedano},
  {Garofalo}, {Garralda}, {Gavel}, {Gavras}, {Gerssen}, {Geyer}, {Giacobbe},
  {Gilmore}, {Girona}, {Giuffrida}, {Glass}, {Gomes}, {Granvik}, {Gueguen},
  {Guerrier}, {Guiraud}, {Guti{\'e}}, {Haigron}, {Hatzidimitriou}, {Hauser},
  {Haywood}, {Heiter}, {Helmi}, {Heu}, {Hilger}, {Hobbs}, {Hofmann}, {Holland},
  {Huckle}, {Hypki}, {Icardi}, {Jan{\ss}en}, {Jevardat de Fombelle}, {Jonker},
  {Juh{\'a}sz}, {Julbe}, {Karampelas}, {Kewley}, {Klar}, {Kochoska}, {Kohley},
  {Kolenberg}, {Kontizas}, {Kontizas}, {Koposov}, {Kordopatis},
  {Kostrzewa-Rutkowska}, {Koubsky}, {Lambert}, {Lanza}, {Lasne}, {Lavigne}, {Le
  Fustec}, {Le Poncin-Lafitte}, {Lebreton}, {Leccia}, {Leclerc},
  {Lecoeur-Taibi}, {Lenhardt}, {Leroux}, {Liao}, {Licata}, {Lindstr{\o}m},
  {Lister}, {Livanou}, {Lobel}, {L{\'o}pez}, {Managau}, {Mann}, {Mantelet},
  {Marchal}, {Marchant}, {Marconi}, {Marinoni}, {Marschalk{\'o}}, {Marshall},
  {Martino}, {Marton}, {Mary}, {Massari}, {Matijevi{\v{c}}}, {Mazeh},
  {McMillan}, {Messina}, {Michalik}, {Millar}, {Molina}, {Molinaro},
  {Moln{\'a}r}, {Montegriffo}, {Mor}, {Morbidelli}, {Morel}, {Morris},
  {Mulone}, {Muraveva}, {Musella}, {Nelemans}, {Nicastro}, {Noval},
  {O'Mullane}, {Ord{\'e}novic}, {Ord{\'o}{\~n}ez-Blanco}, {Osborne}, {Pagani},
  {Pagano}, {Pailler}, {Palacin}, {Palaversa}, {Panahi}, {Pawlak},
  {Piersimoni}, {Pineau}, {Plachy}, {Plum}, {Poggio}, {Poujoulet},
  {Pr{\v{s}}a}, {Pulone}, {Racero}, {Ragaini}, {Rambaux}, {Ramos-Lerate},
  {Regibo}, {Reyl{\'e}}, {Riclet}, {Ripepi}, {Riva}, {Rivard}, {Rixon},
  {Roegiers}, {Roelens}, {Romero-G{\'o}mez}, {Rowell}, {Royer}, {Ruiz-Dern},
  {Sadowski}, {Sagrist{\`a} Sell{\'e}s}, {Sahlmann}, {Salgado}, {Salguero},
  {Sanna}, {Santana-Ros}, {Sarasso}, {Savietto}, {Schultheis}, {Sciacca},
  {Segol}, {Segovia}, {S{\'e}gransan}, {Shih}, {Siltala}, {Silva}, {Smart},
  {Smith}, {Solano}, {Solitro}, {Sordo}, {Soria Nieto}, {Souchay}, {Spagna},
  {Spoto}, {Stampa}, {Steele}, {Steidelm{\"u}ller}, {Stephenson}, {Stoev},
  {Suess}, {Surdej}, {Szabados}, {Szegedi-Elek}, {Tapiador}, {Taris}, {Tauran},
  {Taylor}, {Teixeira}, {Terrett}, {Teyssandier}, {Thuillot}, {Titarenko},
  {Torra Clotet}, {Turon}, {Ulla}, {Utrilla}, {Uzzi}, {Vaillant}, {Valentini},
  {Valette}, {van Elteren}, {Van Hemelryck}, {Vaschetto}, {Vecchiato},
  {Veljanoski}, {Viala}, {Vicente}, {Vogt}, {von Essen}, {Voss}, {Votruba},
  {Voutsinas}, {Walmsley}, {Weiler}, {Wertz}, {Wevers}, {Wyrzykowski},
  {Yoldas}, {{\v{Z}}erjal}, {Ziaeepour}, {Zorec}, {Zschocke}, {Zucker},
  {Zurbach}, \& {Zwitter}}]{2018A&A...616A..10G}
{Gaia Collaboration}, {Babusiaux}, C., {van Leeuwen}, F., {et~al.}
  2018{\natexlab{b}}, \aap, 616, A10, \dodoi{10.1051/0004-6361/201832843}

\bibitem[{{Gaia Collaboration} {et~al.}(2021){Gaia Collaboration}, {Brown},
  {Vallenari}, {Prusti}, {de Bruijne}, {Babusiaux}, {Biermann}, {Creevey},
  {Evans}, {Eyer}, {Hutton}, {Jansen}, {Jordi}, {Klioner}, {Lammers},
  {Lindegren}, {Luri}, {Mignard}, {Panem}, {Pourbaix}, {Randich}, {Sartoretti},
  {Soubiran}, {Walton}, {Arenou}, {Bailer-Jones}, {Bastian}, {Cropper},
  {Drimmel}, {Katz}, {Lattanzi}, {van Leeuwen}, {Bakker}, {Cacciari},
  {Casta{\~n}eda}, {De Angeli}, {Ducourant}, {Fabricius}, {Fouesneau},
  {Fr{\'e}mat}, {Guerra}, {Guerrier}, {Guiraud}, {Jean-Antoine Piccolo},
  {Masana}, {Messineo}, {Mowlavi}, {Nicolas}, {Nienartowicz}, {Pailler},
  {Panuzzo}, {Riclet}, {Roux}, {Seabroke}, {Sordo}, {Tanga}, {Th{\'e}venin},
  {Gracia-Abril}, {Portell}, {Teyssier}, {Altmann}, {Andrae}, {Bellas-Velidis},
  {Benson}, {Berthier}, {Blomme}, {Brugaletta}, {Burgess}, {Busso}, {Carry},
  {Cellino}, {Cheek}, {Clementini}, {Damerdji}, {Davidson}, {Delchambre},
  {Dell'Oro}, {Fern{\'a}ndez-Hern{\'a}ndez}, {Galluccio}, {Garc{\'\i}a-Lario},
  {Garcia-Reinaldos}, {Gonz{\'a}lez-N{\'u}{\~n}ez}, {Gosset}, {Haigron},
  {Halbwachs}, {Hambly}, {Harrison}, {Hatzidimitriou}, {Heiter},
  {Hern{\'a}ndez}, {Hestroffer}, {Hodgkin}, {Holl}, {Jan{\ss}en}, {Jevardat de
  Fombelle}, {Jordan}, {Krone-Martins}, {Lanzafame}, {L{\"o}ffler}, {Lorca},
  {Manteiga}, {Marchal}, {Marrese}, {Moitinho}, {Mora}, {Muinonen}, {Osborne},
  {Pancino}, {Pauwels}, {Petit}, {Recio-Blanco}, {Richards}, {Riello},
  {Rimoldini}, {Robin}, {Roegiers}, {Rybizki}, {Sarro}, {Siopis}, {Smith},
  {Sozzetti}, {Ulla}, {Utrilla}, {van Leeuwen}, {van Reeven}, {Abbas}, {Abreu
  Aramburu}, {Accart}, {Aerts}, {Aguado}, {Ajaj}, {Altavilla}, {{\'A}lvarez},
  {{\'A}lvarez Cid-Fuentes}, {Alves}, {Anderson}, {Anglada Varela}, {Antoja},
  {Audard}, {Baines}, {Baker}, {Balaguer-N{\'u}{\~n}ez}, {Balbinot}, {Balog},
  {Barache}, {Barbato}, {Barros}, {Barstow}, {Bartolom{\'e}}, {Bassilana},
  {Bauchet}, {Baudesson-Stella}, {Becciani}, {Bellazzini}, {Bernet}, {Bertone},
  {Bianchi}, {Blanco-Cuaresma}, {Boch}, {Bombrun}, {Bossini}, {Bouquillon},
  {Bragaglia}, {Bramante}, {Breedt}, {Bressan}, {Brouillet}, {Bucciarelli},
  {Burlacu}, {Busonero}, {Butkevich}, {Buzzi}, {Caffau}, {Cancelliere},
  {C{\'a}novas}, {Cantat-Gaudin}, {Carballo}, {Carlucci}, {Carnerero},
  {Carrasco}, {Casamiquela}, {Castellani}, {Castro-Ginard}, {Castro Sampol},
  {Chaoul}, {Charlot}, {Chemin}, {Chiavassa}, {Cioni}, {Comoretto}, {Cooper},
  {Cornez}, {Cowell}, {Crifo}, {Crosta}, {Crowley}, {Dafonte}, {Dapergolas},
  {David}, {David}, {de Laverny}, {De Luise}, {De March}, {De Ridder}, {de
  Souza}, {de Teodoro}, {de Torres}, {del Peloso}, {del Pozo}, {Delbo},
  {Delgado}, {Delgado}, {Delisle}, {Di Matteo}, {Diakite}, {Diener},
  {Distefano}, {Dolding}, {Eappachen}, {Edvardsson}, {Enke}, {Esquej}, {Fabre},
  {Fabrizio}, {Faigler}, {Fedorets}, {Fernique}, {Fienga}, {Figueras},
  {Fouron}, {Fragkoudi}, {Fraile}, {Franke}, {Gai}, {Garabato},
  {Garcia-Gutierrez}, {Garc{\'\i}a-Torres}, {Garofalo}, {Gavras}, {Gerlach},
  {Geyer}, {Giacobbe}, {Gilmore}, {Girona}, {Giuffrida}, {Gomel}, {Gomez},
  {Gonzalez-Santamaria}, {Gonz{\'a}lez-Vidal}, {Granvik},
  {Guti{\'e}rrez-S{\'a}nchez}, {Guy}, {Hauser}, {Haywood}, {Helmi}, {Hidalgo},
  {Hilger}, {H{\l}adczuk}, {Hobbs}, {Holland}, {Huckle}, {Jasniewicz},
  {Jonker}, {Juaristi Campillo}, {Julbe}, {Karbevska}, {Kervella}, {Khanna},
  {Kochoska}, {Kontizas}, {Kordopatis}, {Korn}, {Kostrzewa-Rutkowska},
  {Kruszy{\'n}ska}, {Lambert}, {Lanza}, {Lasne}, {Le Campion}, {Le Fustec},
  {Lebreton}, {Lebzelter}, {Leccia}, {Leclerc}, {Lecoeur-Taibi}, {Liao},
  {Licata}, {Lindstr{\o}m}, {Lister}, {Livanou}, {Lobel}, {Madrero Pardo},
  {Managau}, {Mann}, {Marchant}, {Marconi}, {Marcos Santos}, {Marinoni},
  {Marocco}, {Marshall}, {Martin Polo}, {Mart{\'\i}n-Fleitas}, {Masip},
  {Massari}, {Mastrobuono-Battisti}, {Mazeh}, {McMillan}, {Messina},
  {Michalik}, {Millar}, {Mints}, {Molina}, {Molinaro}, {Moln{\'a}r},
  {Montegriffo}, {Mor}, {Morbidelli}, {Morel}, {Morris}, {Mulone}, {Munoz},
  {Muraveva}, {Murphy}, {Musella}, {Noval}, {Ord{\'e}novic}, {Orr{\`u}},
  {Osinde}, {Pagani}, {Pagano}, {Palaversa}, {Palicio}, {Panahi}, {Pawlak},
  {Pe{\~n}alosa Esteller}, {Penttil{\"a}}, {Piersimoni}, {Pineau}, {Plachy},
  {Plum}, {Poggio}, {Poretti}, {Poujoulet}, {Pr{\v{s}}a}, {Pulone}, {Racero},
  {Ragaini}, {Rainer}, {Raiteri}, {Rambaux}, {Ramos}, {Ramos-Lerate}, {Re
  Fiorentin}, {Regibo}, {Reyl{\'e}}, {Ripepi}, {Riva}, {Rixon}, {Robichon},
  {Robin}, {Roelens}, {Rohrbasser}, {Romero-G{\'o}mez}, {Rowell}, {Royer},
  {Rybicki}, {Sadowski}, {Sagrist{\`a} Sell{\'e}s}, {Sahlmann}, {Salgado},
  {Salguero}, {Samaras}, {Sanchez Gimenez}, {Sanna}, {Santove{\~n}a},
  {Sarasso}, {Schultheis}, {Sciacca}, {Segol}, {Segovia}, {S{\'e}gransan},
  {Semeux}, {Shahaf}, {Siddiqui}, {Siebert}, {Siltala}, {Slezak}, {Smart},
  {Solano}, {Solitro}, {Souami}, {Souchay}, {Spagna}, {Spoto}, {Steele},
  {Steidelm{\"u}ller}, {Stephenson}, {S{\"u}veges}, {Szabados}, {Szegedi-Elek},
  {Taris}, {Tauran}, {Taylor}, {Teixeira}, {Thuillot}, {Tonello}, {Torra},
  {Torra}, {Turon}, {Unger}, {Vaillant}, {van Dillen}, {Vanel}, {Vecchiato},
  {Viala}, {Vicente}, {Voutsinas}, {Weiler}, {Wevers}, {Wyrzykowski}, {Yoldas},
  {Yvard}, {Zhao}, {Zorec}, {Zucker}, {Zurbach}, \&
  {Zwitter}}]{2021A&A...649A...1G}
{Gaia Collaboration}, {Brown}, A.~G.~A., {Vallenari}, A., {et~al.} 2021, \aap,
  649, A1, \dodoi{10.1051/0004-6361/202039657}

\bibitem[{{Gentile Fusillo} {et~al.}(2019){Gentile Fusillo}, {Tremblay},
  {G{\"a}nsicke}, {Manser}, {Cunningham}, {Cukanovaite}, {Hollands}, {Marsh},
  {Raddi}, {Jordan}, {Toonen}, {Geier}, {Barstow}, \&
  {Cummings}}]{2019MNRAS.482.4570G}
{Gentile Fusillo}, N.~P., {Tremblay}, P.-E., {G{\"a}nsicke}, B.~T., {et~al.}
  2019, \mnras, 482, 4570, \dodoi{10.1093/mnras/sty3016}

\bibitem[{{Gentile Fusillo} {et~al.}(2021){Gentile Fusillo}, {Tremblay},
  {Cukanovaite}, {Vorontseva}, {Lallement}, {Hollands}, {G{\"a}nsicke},
  {Burdge}, {McCleery}, \& {Jordan}}]{2021MNRAS.508.3877G}
{Gentile Fusillo}, N.~P., {Tremblay}, P.~E., {Cukanovaite}, E., {et~al.} 2021,
  \mnras, 508, 3877, \dodoi{10.1093/mnras/stab2672}

\bibitem[{{Giammichele} {et~al.}(2012){Giammichele}, {Bergeron}, \&
  {Dufour}}]{2012ApJS..199...29G}
{Giammichele}, N., {Bergeron}, P., \& {Dufour}, P. 2012, \apjs, 199, 29,
  \dodoi{10.1088/0067-0049/199/2/29}

\bibitem[{{Gianninas} {et~al.}(2011){Gianninas}, {Bergeron}, \&
  {Ruiz}}]{2011ApJ...743..138G}
{Gianninas}, A., {Bergeron}, P., \& {Ruiz}, M.~T. 2011, \apj, 743, 138,
  \dodoi{10.1088/0004-637X/743/2/138}

\bibitem[{{Goodman} \& {Weare}(2010)}]{2010CAMCS...5...65G}
{Goodman}, J., \& {Weare}, J. 2010, Communications in Applied Mathematics and
  Computational Science, 5, 65, \dodoi{10.2140/camcos.2010.5.65}

\bibitem[{{Guo} {et~al.}(2015){Guo}, {Zhao}, {Tziamtzis}, {Liu}, {Li}, {Zhang},
  {Hou}, \& {Wang}}]{2015MNRAS.454.2787G}
{Guo}, J., {Zhao}, J., {Tziamtzis}, A., {et~al.} 2015, \mnras, 454, 2787,
  \dodoi{10.1093/mnras/stv2104}

\bibitem[{{Hut} \& {Bahcall}(1983)}]{1983ApJ...268..319H}
{Hut}, P., \& {Bahcall}, J.~N. 1983, \apj, 268, 319, \dodoi{10.1086/160956}

\bibitem[{{Kalirai} {et~al.}(2008){Kalirai}, {Hansen}, {Kelson}, {Reitzel},
  {Rich}, \& {Richer}}]{2008ApJ...676..594K}
{Kalirai}, J.~S., {Hansen}, B. M.~S., {Kelson}, D.~D., {et~al.} 2008, \apj,
  676, 594, \dodoi{10.1086/527028}

\bibitem[{{Kepler} {et~al.}(2007){Kepler}, {Kleinman}, {Nitta}, {Koester},
  {Castanheira}, {Giovannini}, {Costa}, \& {Althaus}}]{2007MNRAS.375.1315K}
{Kepler}, S.~O., {Kleinman}, S.~J., {Nitta}, A., {et~al.} 2007, \mnras, 375,
  1315, \dodoi{10.1111/j.1365-2966.2006.11388.x}

\bibitem[{{Kepler} {et~al.}(2015){Kepler}, {Pelisoli}, {Koester}, {Ourique},
  {Kleinman}, {Romero}, {Nitta}, {Eisenstein}, {Costa}, {K{\"u}lebi}, {Jordan},
  {Dufour}, {Giommi}, \& {Rebassa-Mansergas}}]{2015MNRAS.446.4078K}
{Kepler}, S.~O., {Pelisoli}, I., {Koester}, D., {et~al.} 2015, \mnras, 446,
  4078, \dodoi{10.1093/mnras/stu2388}

\bibitem[{{Kepler} {et~al.}(2016){Kepler}, {Pelisoli}, {Koester}, {Ourique},
  {Romero}, {Reindl}, {Kleinman}, {Eisenstein}, {Valois}, \&
  {Amaral}}]{2016MNRAS.455.3413K}
---. 2016, \mnras, 455, 3413, \dodoi{10.1093/mnras/stv2526}

\bibitem[{{Kepler} {et~al.}(2019){Kepler}, {Pelisoli}, {Koester}, {Reindl},
  {Geier}, {Romero}, {Ourique}, {Oliveira}, \& {Amaral}}]{2019MNRAS.486.2169K}
---. 2019, \mnras, 486, 2169, \dodoi{10.1093/mnras/stz960}

\bibitem[{{Kilic} {et~al.}(2019){Kilic}, {Bergeron}, {Dame}, {Hambly},
  {Rowell}, \& {Crawford}}]{2019MNRAS.482..965K}
{Kilic}, M., {Bergeron}, P., {Dame}, K., {et~al.} 2019, \mnras, 482, 965,
  \dodoi{10.1093/mnras/sty2755}

\bibitem[{{Kilic} {et~al.}(2020){Kilic}, {Bergeron}, {Kosakowski}, {Brown},
  {Ag{\"u}eros}, \& {Blouin}}]{2020ApJ...898...84K}
{Kilic}, M., {Bergeron}, P., {Kosakowski}, A., {et~al.} 2020, \apj, 898, 84,
  \dodoi{10.3847/1538-4357/ab9b8d}

\bibitem[{{Kiman} {et~al.}(2021){Kiman}, {Faherty}, {Cruz}, {Gagn{\'e}},
  {Angus}, {Schmidt}, {Mann}, {Bardalez Gagliuffi}, \&
  {Rice}}]{2021AJ....161..277K}
{Kiman}, R., {Faherty}, J.~K., {Cruz}, K.~L., {et~al.} 2021, \aj, 161, 277,
  \dodoi{10.3847/1538-3881/abf561}

\bibitem[{{Kleinman} {et~al.}(2013){Kleinman}, {Kepler}, {Koester}, {Pelisoli},
  {Pe{\c{c}}anha}, {Nitta}, {Costa}, {Krzesinski}, {Dufour}, {Lachapelle},
  {Bergeron}, {Yip}, {Harris}, {Eisenstein}, {Althaus}, \&
  {C{\'o}rsico}}]{2013ApJS..204....5K}
{Kleinman}, S.~J., {Kepler}, S.~O., {Koester}, D., {et~al.} 2013, \apjs, 204,
  5, \dodoi{10.1088/0067-0049/204/1/5}

\bibitem[{{Koester}(2010)}]{2010MmSAI..81..921K}
{Koester}, D. 2010, \memsai, 81, 921

\bibitem[{{Koester} {et~al.}(2009){Koester}, {Voss}, {Napiwotzki},
  {Christlieb}, {Homeier}, {Lisker}, {Reimers}, \&
  {Heber}}]{2009ANA...505..441K}
{Koester}, D., {Voss}, B., {Napiwotzki}, R., {et~al.} 2009, \aap, 505, 441,
  \dodoi{10.1051/0004-6361/200912531}

\bibitem[{{Kraus} \& {Hillenbrand}(2009)}]{2009ApJ...704..531K}
{Kraus}, A.~L., \& {Hillenbrand}, L.~A. 2009, \apj, 704, 531,
  \dodoi{10.1088/0004-637X/704/1/531}

\bibitem[{{K{\"u}lebi} {et~al.}(2010){K{\"u}lebi}, {Jordan}, {Nelan},
  {Bastian}, \& {Altmann}}]{2010ANA...524A..36K}
{K{\"u}lebi}, B., {Jordan}, S., {Nelan}, E., {Bastian}, U., \& {Altmann}, M.
  2010, \aap, 524, A36, \dodoi{10.1051/0004-6361/201015237}

\bibitem[{{Liebert} {et~al.}(2005){Liebert}, {Bergeron}, \&
  {Holberg}}]{2005ApJS..156...47L}
{Liebert}, J., {Bergeron}, P., \& {Holberg}, J.~B. 2005, \apjs, 156, 47,
  \dodoi{10.1086/425738}

\bibitem[{{Limoges} {et~al.}(2013){Limoges}, {L{\'e}pine}, \&
  {Bergeron}}]{2013AJ....145..136L}
{Limoges}, M.~M., {L{\'e}pine}, S., \& {Bergeron}, P. 2013, \aj, 145, 136,
  \dodoi{10.1088/0004-6256/145/5/136}

\bibitem[{{Marigo} {et~al.}(2020){Marigo}, {Cummings}, {Curtis}, {Kalirai},
  {Chen}, {Tremblay}, {Ramirez-Ruiz}, {Bergeron}, {Bladh}, {Bressan},
  {Girardi}, {Pastorelli}, {Trabucchi}, {Cheng}, {Aringer}, \&
  {Tio}}]{2020NatAs...4.1102M}
{Marigo}, P., {Cummings}, J.~D., {Curtis}, J.~L., {et~al.} 2020, Nature
  Astronomy, 4, 1102, \dodoi{10.1038/s41550-020-1132-1}

\bibitem[{{Marrese} {et~al.}(2021){Marrese}, {Marinoni}, {Fabrizio}, \&
  {Altavilla}}]{2021gdr3.reptE...9M}
{Marrese}, P.~M., {Marinoni}, S., {Fabrizio}, M., \& {Altavilla}, G. 2021,
  {Gaia EDR3 documentation Chapter 9: Cross-match with external catalogues},
  Gaia EDR3 documentation

\bibitem[{{Martin} {et~al.}(2021){Martin}, {El-Badry}, {Hod{\v{z}}i{\'c}},
  {Triaud}, {Angus}, {Birky}, {Foreman-Mackey}, {Hedges}, {Montet}, {Murphy},
  {Santerne}, {Stassun}, {Stephan}, {Wang}, {Benni}, {Krushinsky}, {Chazov},
  {Mishevskiy}, {Ziegler}, {Soubkiou}, {Benkhaldoun}, {Boisse}, {Battley},
  {Miller}, {Caldwell}, {Collins}, {Henze}, {Guerrero}, {Jenkins}, {Latham},
  {Levine}, {McDermott}, {Mullally}, {Ricker}, {Seager}, {Shporer},
  {Vanderburg}, {Vanderspek}, \& {Winn}}]{2021MNRAS.507.4132M}
{Martin}, D.~V., {El-Badry}, K., {Hod{\v{z}}i{\'c}}, V.~K., {et~al.} 2021,
  \mnras, 507, 4132, \dodoi{10.1093/mnras/stab2129}

\bibitem[{{McCook} \& {Sion}(1999)}]{1999ApJS..121....1M}
{McCook}, G.~P., \& {Sion}, E.~M. 1999, \apjs, 121, 1, \dodoi{10.1086/313186}

\bibitem[{{Moe} \& {Di Stefano}(2017)}]{2017ApJS..230...15M}
{Moe}, M., \& {Di Stefano}, R. 2017, \apjs, 230, 15,
  \dodoi{10.3847/1538-4365/aa6fb6}

\bibitem[{{Morgan} {et~al.}(2012){Morgan}, {West}, {Garc{\'e}s}, {Catal{\'a}n},
  {Dhital}, {Fuchs}, \& {Silvestri}}]{2012AJ....144...93M}
{Morgan}, D.~P., {West}, A.~A., {Garc{\'e}s}, A., {et~al.} 2012, \aj, 144, 93,
  \dodoi{10.1088/0004-6256/144/4/93}

\bibitem[{{Moss} {et~al.}(2022){Moss}, {von Hippel}, {Robinson}, {El-Badry},
  {Stenning}, {van Dyk}, {Fouesneau}, {Bailer-Jones}, {Jeffery}, {Sargent},
  {Kloc}, \& {Moticska}}]{2022ApJ...929...26M}
{Moss}, A., {von Hippel}, T., {Robinson}, E., {et~al.} 2022, \apj, 929, 26,
  \dodoi{10.3847/1538-4357/ac5ac0}

\bibitem[{{Paxton} {et~al.}(2011){Paxton}, {Bildsten}, {Dotter}, {Herwig},
  {Lesaffre}, \& {Timmes}}]{2011ApJS..192....3P}
{Paxton}, B., {Bildsten}, L., {Dotter}, A., {et~al.} 2011, \apjs, 192, 3,
  \dodoi{10.1088/0067-0049/192/1/3}

\bibitem[{{Paxton} {et~al.}(2013){Paxton}, {Cantiello}, {Arras}, {Bildsten},
  {Brown}, {Dotter}, {Mankovich}, {Montgomery}, {Stello}, {Timmes}, \&
  {Townsend}}]{2013ApJS..208....4P}
{Paxton}, B., {Cantiello}, M., {Arras}, P., {et~al.} 2013, \apjs, 208, 4,
  \dodoi{10.1088/0067-0049/208/1/4}

\bibitem[{{Paxton} {et~al.}(2015){Paxton}, {Marchant}, {Schwab}, {Bauer},
  {Bildsten}, {Cantiello}, {Dessart}, {Farmer}, {Hu}, {Langer}, {Townsend},
  {Townsley}, \& {Timmes}}]{2015ApJS..220...15P}
{Paxton}, B., {Marchant}, P., {Schwab}, J., {et~al.} 2015, \apjs, 220, 15,
  \dodoi{10.1088/0067-0049/220/1/15}

\bibitem[{{Perpiny{\`a}-Vall{\`e}s} {et~al.}(2019){Perpiny{\`a}-Vall{\`e}s},
  {Rebassa-Mansergas}, {G{\"a}nsicke}, {Toonen}, {Hermes}, {Gentile Fusillo},
  \& {Tremblay}}]{2019MNRAS.483..901P}
{Perpiny{\`a}-Vall{\`e}s}, M., {Rebassa-Mansergas}, A., {G{\"a}nsicke}, B.~T.,
  {et~al.} 2019, \mnras, 483, 901, \dodoi{10.1093/mnras/sty3149}

\bibitem[{{Qiu} {et~al.}(2021){Qiu}, {Tian}, {Wang}, {Nie}, {von Hippel},
  {Liu}, {Fouesneau}, \& {Rix}}]{2021ApJS..253...58Q}
{Qiu}, D., {Tian}, H.-J., {Wang}, X.-D., {et~al.} 2021, \apjs, 253, 58,
  \dodoi{10.3847/1538-4365/abe468}

\bibitem[{{Rybizki} {et~al.}(2021){Rybizki}, {Green}, {Rix}, {Demleitner},
  {Zari}, {Udalski}, {Smart}, \& {Gould}}]{2021arXiv210111641R}
{Rybizki}, J., {Green}, G., {Rix}, H.-W., {et~al.} 2021, arXiv e-prints,
  arXiv:2101.11641.
\newblock \doarXiv{2101.11641}

\bibitem[{{Salaris} {et~al.}(2009){Salaris}, {Serenelli}, {Weiss}, \& {Miller
  Bertolami}}]{2009ApJ...692.1013S}
{Salaris}, M., {Serenelli}, A., {Weiss}, A., \& {Miller Bertolami}, M. 2009,
  \apj, 692, 1013, \dodoi{10.1088/0004-637X/692/2/1013}

\bibitem[{{Schlafly} \& {Finkbeiner}(2011)}]{2011ApJ...737..103S}
{Schlafly}, E.~F., \& {Finkbeiner}, D.~P. 2011, \apj, 737, 103,
  \dodoi{10.1088/0004-637X/737/2/103}

\bibitem[{{Sion} {et~al.}(1991){Sion}, {Oswalt}, {Liebert}, \&
  {Hintzen}}]{1991AJ....101.1476S}
{Sion}, E.~M., {Oswalt}, T.~D., {Liebert}, J., \& {Hintzen}, P. 1991, \aj, 101,
  1476, \dodoi{10.1086/115779}

\bibitem[{{Temmink} {et~al.}(2020){Temmink}, {Toonen}, {Zapartas}, {Justham},
  \& {G{\"a}nsicke}}]{2020A&A...636A..31T}
{Temmink}, K.~D., {Toonen}, S., {Zapartas}, E., {Justham}, S., \&
  {G{\"a}nsicke}, B.~T. 2020, \aap, 636, A31,
  \dodoi{10.1051/0004-6361/201936889}

\bibitem[{{Tian} {et~al.}(2020){Tian}, {El-Badry}, {Rix}, \& {Gould}}]{Tian}
{Tian}, H.-J., {El-Badry}, K., {Rix}, H.-W., \& {Gould}, A. 2020, \apjs, 246,
  4, \dodoi{10.3847/1538-4365/ab54c4}

\bibitem[{{Toonen} {et~al.}(2021){Toonen}, {Boekholt}, \& {Portegies
  Zwart}}]{2021arXiv210804272T}
{Toonen}, S., {Boekholt}, T.~C.~N., \& {Portegies Zwart}, S. 2021, arXiv
  e-prints, arXiv:2108.04272.
\newblock \doarXiv{2108.04272}

\bibitem[{{Torres} {et~al.}(2021){Torres}, {Rebassa-Mansergas}, {Camisassa}, \&
  {Raddi}}]{2021MNRAS.502.1753T}
{Torres}, S., {Rebassa-Mansergas}, A., {Camisassa}, M.~E., \& {Raddi}, R. 2021,
  \mnras, 502, 1753, \dodoi{10.1093/mnras/stab079}

\bibitem[{{Tremblay} {et~al.}(2011){Tremblay}, {Bergeron}, \&
  {Gianninas}}]{2011ApJ...730..128T}
{Tremblay}, P.~E., {Bergeron}, P., \& {Gianninas}, A. 2011, \apj, 730, 128,
  \dodoi{10.1088/0004-637X/730/2/128}

\bibitem[{{Tremblay} {et~al.}(2016){Tremblay}, {Cummings}, {Kalirai},
  {G{\"a}nsicke}, {Gentile-Fusillo}, \& {Raddi}}]{2016MNRAS.461.2100T}
{Tremblay}, P.~E., {Cummings}, J., {Kalirai}, J.~S., {et~al.} 2016, \mnras,
  461, 2100, \dodoi{10.1093/mnras/stw1447}

\bibitem[{{Tremblay} {et~al.}(2014){Tremblay}, {Kalirai}, {Soderblom},
  {Cignoni}, \& {Cummings}}]{2014ApJ...791...92T}
{Tremblay}, P.~E., {Kalirai}, J.~S., {Soderblom}, D.~R., {Cignoni}, M., \&
  {Cummings}, J. 2014, \apj, 791, 92, \dodoi{10.1088/0004-637X/791/2/92}

\bibitem[{{Tremblay} {et~al.}(2019){Tremblay}, {Fontaine}, {Gentile Fusillo},
  {Dunlap}, {G{\"a}nsicke}, {Hollands}, {Hermes}, {Marsh}, {Cukanovaite}, \&
  {Cunningham}}]{2019Natur.565..202T}
{Tremblay}, P.-E., {Fontaine}, G., {Gentile Fusillo}, N.~P., {et~al.} 2019,
  \nat, 565, 202, \dodoi{10.1038/s41586-018-0791-x}

\bibitem[{{Tremblay} {et~al.}(2020){Tremblay}, {Hollands}, {Gentile Fusillo},
  {McCleery}, {Izquierdo}, {G{\"a}nsicke}, {Cukanovaite}, {Koester}, {Brown},
  {Charpinet}, {Cunningham}, {Farihi}, {Giammichele}, {van Grootel}, {Hermes},
  {Hoskin}, {Jordan}, {Kepler}, {Kleinman}, {Manser}, {Marsh}, {de Martino},
  {Nitta}, {Parsons}, {Pelisoli}, {Raddi}, {Rebassa-Mansergas}, {Ren},
  {Schreiber}, {Silvotti}, {Toloza}, {Toonen}, \&
  {Torres}}]{2020MNRAS.497..130T}
{Tremblay}, P.~E., {Hollands}, M.~A., {Gentile Fusillo}, N.~P., {et~al.} 2020,
  \mnras, 497, 130, \dodoi{10.1093/mnras/staa1892}

\bibitem[{{Vennes} {et~al.}(2003){Vennes}, {Schmidt}, {Ferrario}, {Christian},
  {Wickramasinghe}, \& {Kawka}}]{2003ApJ...593.1040V}
{Vennes}, S., {Schmidt}, G.~D., {Ferrario}, L., {et~al.} 2003, \apj, 593, 1040,
  \dodoi{10.1086/376728}

\bibitem[{{White} \& {Ghez}(2001)}]{2001ApJ...556..265W}
{White}, R.~J., \& {Ghez}, A.~M. 2001, \apj, 556, 265, \dodoi{10.1086/321542}

\bibitem[{{Winget} {et~al.}(1987){Winget}, {Hansen}, {Liebert}, {van Horn},
  {Fontaine}, {Nather}, {Kepler}, \& {Lamb}}]{1987ApJ...315L..77W}
{Winget}, D.~E., {Hansen}, C.~J., {Liebert}, J., {et~al.} 1987, \apjl, 315,
  L77, \dodoi{10.1086/184864}

\bibitem[{{Zhang} {et~al.}(2020){Zhang}, {Liu}, {Hermes}, {Magnier}, {Marley},
  {Tremblay}, {Tucker}, {Do}, {Payne}, \& {Shappee}}]{2020ApJ...891..171Z}
{Zhang}, Z., {Liu}, M.~C., {Hermes}, J.~J., {et~al.} 2020, \apj, 891, 171,
  \dodoi{10.3847/1538-4357/ab765c}

\bibitem[{{Zhao} {et~al.}(2012){Zhao}, {Oswalt}, {Willson}, {Wang}, \&
  {Zhao}}]{2012ApJ...746..144Z}
{Zhao}, J.~K., {Oswalt}, T.~D., {Willson}, L.~A., {Wang}, Q., \& {Zhao}, G.
  2012, \apj, 746, 144, \dodoi{10.1088/0004-637X/746/2/144}

\end{thebibliography}
\bibliographystyle{aasjournal}


% \pagebreak
\appendix
\restartappendixnumbering

\section{Double WD Pair Catalog and Triples Catalog}
The full catalog of wide double WD pairs with total ages and atmospheric parameters can be accessed at \href{https://sites.bu.edu/buwd/files/2022/05/Table_A1.csv}{https://sites.bu.edu/buwd/files/2022/05/Table\_A1.csv}. Table~\ref{tab:description} gives a description of the columns of the catalog. Table~\ref{tab:triples} shows the 23 WD+WD+MS and WD+WD+WD triple found in this work.

\startlongtable
\begin{deluxetable*}{ccc}
\tabletypesize{\footnotesize}
\tablecaption{ Double WD Pair Catalog Description \label{tab:description}}
\tablecolumns{50}
\tablehead{
\colhead{Column} & \colhead{Units} & \colhead{Description}}
\startdata
System\_Name &  & name of the system; determined by the source\_id of the brightest component\\
source\_id &  & \emph{Gaia} EDR3 source identifier\\
ra & deg & right ascension (J2016)\\
dec & deg & declination (J2016)\\
pmra & mas $yr^{-1}$ & proper motion in the direction of right ascension\\
pmra\_error & mas $yr^{-1}$ & uncertainty on the proper motion in the direction of right ascension\\
pmdec &  mas $yr^{-1}$ & proper motion in the direction of declination\\
pmdec\_error & mas $yr^{-1}$ & uncertainty on the proper motion in the direction of declination\\
parallax & mas & parallax\\
parallax\_error & mas & uncertainty on the parallax\\
phot\_g\_mean\_mag & mag & \emph{Gaia} EDR3 G-band magnitude\\
phot\_g\_mean\_flux &  e\textsuperscript{-} s\textsuperscript{-1} & \emph{Gaia} EDR3 G-band flux\\
phot\_g\_mean\_flux\_error & e\textsuperscript{-} s\textsuperscript{-1} & uncertainty on the \emph{Gaia} EDR3 G-band flux\\
phot\_bp\_mean\_mag & mag & \emph{Gaia} EDR3 G\textsubscript{BP}-band magnitude\\
phot\_bp\_mean\_flux &  e\textsuperscript{-} s\textsuperscript{-1} & \emph{Gaia} EDR3 G\textsubscript{BP}-band flux\\
phot\_bp\_mean\_flux\_error &  e\textsuperscript{-} s\textsuperscript{-1} & uncertainty on the \emph{Gaia} EDR3 G\textsubscript{BP}-band flux\\
phot\_rp\_mean\_mag & mag & \emph{Gaia} EDR3 G\textsubscript{RP}-band magnitude\\
phot\_rp\_mean\_flux &  e\textsuperscript{-} s\textsuperscript{-1} & \emph{Gaia} EDR3 G-band flux\\
phot\_rp\_mean\_flux\_error &  e\textsuperscript{-} s\textsuperscript{-1} & uncertainty on the \emph{Gaia} EDR3 G\textsubscript{RP}-band flux\\
u\_sdss & mag & SDSS u-band magnitude\\
g\_sdss & mag & SDSS g-band magnitude\\
r\_sdss & mag & SDSS r-band magnitude\\
i\_sdss & mag & SDSS i-band magnitude\\
z\_sdss & mag & SDSS z-band magnitude\\
e\_u\_sdss & mag & uncertainty on the SDSS u-band magnitude\\
e\_g\_sdss & mag & uncertainty on the SDSS g-band magnitude\\
e\_r\_sdss & mag & uncertainty on the SDSS r-band magnitude\\
e\_i\_sdss & mag & uncertainty on the SDSS i-band magnitude\\
e\_z\_sdss & mag & uncertainty on the SDSS z-band magnitude\\
u\_smap & mag & SkyMapper u-band magnitude\\
v\_smap & mag & SkyMapper v-band magnitude\\
g\_smap & mag & SkyMapper g-band magnitude\\
r\_smap & mag & SkyMapper r-band magnitude\\
i\_smap & mag & SkyMapper i-band magnitude\\
z\_smap & mag & SkyMapper z-band magnitude\\
e\_u\_smap & mag & uncertainty on the SkyMapper u-band magnitude\\
e\_v\_smap & mag & uncertainty on the SkyMapper v-band magnitude\\
e\_g\_smap & mag & uncertainty on the SkyMapper g-band magnitude\\
e\_r\_smap & mag & uncertainty on the SkyMapper r-band magnitude\\
e\_i\_smap & mag & uncertainty on the SkyMapper i-band magnitude\\
e\_z\_smap & mag & uncertainty on the SkyMapper z-band magnitude\\
g\_pstarr & mag & PanSTARRS g-band magnitude\\
r\_pstarr & mag & PanSTARRS r-band magnitude\\
i\_pstarr & mag & PanSTARRS i-band magnitude\\
z\_pstarr & mag & PanSTARRS z-band magnitude\\
y\_pstarr & mag & PanSTARRS y-band magnitude\\
e\_g\_pstarr & mag & uncertainty on the PanSTARRS g-band magnitude\\
e\_r\_pstarr & mag & uncertainty on the PanSTARRS r-band magnitude\\
e\_i\_pstarr & mag & uncertainty on the PanSTARRS i-band magnitude\\
e\_z\_pstarr & mag & uncertainty on the PanSTARRS z-band magnitude\\
e\_y\_pstarr & mag & uncertainty on the PanSTARRS y-band magnitude\\
j\_tmass & mag & 2MASS J-band magnitude\\ 
h\_tmass & mag & 2MASS H-band magnitude\\ 
ks\_tmass & mag & 2MASS Ks-band magnitude\\
e\_j\_tmass & mag & uncertainty on the 2MASS J-band magnitude\\ 
e\_h\_tmass & mag & uncertainty on the 2MASS H-band magnitude\\ 
e\_ks\_tmass & mag & uncertainty on the 2MASS Ks-band magnitude\\
LoggH & [cm s\textsuperscript{-2}] & log of the surface gravity from a DA model with a thick H layer\\
e\_LoggH\_upper & [cm s\textsuperscript{-2}] & upper uncertainty on the log of the surface gravity from a DA model\\
e\_LoggH\_lower & [cm s\textsuperscript{-2}] & lower uncertainty on the log of the surface gravity from a DA model\\
TeffH & K & effective temperature from a DA model\\
e\_TeffH\_upper & K & upper uncertainty on the effective temperature from a DA model\\
e\_TeffH\_lower & K & lower uncertainty on the effective temperature from a DA model\\
MassH & $M_\odot$ & mass from a DA model\\
e\_MassH\_upper & $M_\odot$ & upper uncertainty on the mass from a DA model\\
e\_MassH\_lower & $M_\odot$ & lower uncertainty on the mass from a DA model\\
cool\_ageH & Gyr & cooling age for a DA model\\
e\_cool\_ageH\_upper & Gyr & upper uncertainty on the cooling age for a DA model\\
e\_cool\_ageH\_lower & Gyr & lower uncertainty on the cooling age for a DA model\\
init\_mass & $M_\odot$ & initial zero-age main-sequence mass\\
e\_init\_mass\_upper & $M_\odot$ & upper uncertainty on the initial mass\\
e\_init\_mass\_lower & $M_\odot$ & lower uncertainty on the initial mass\\
tot\_age & Gyr & total age of the WD\\
e\_tot\_age\_upper & Gyr & upper uncertainty on the total age\\
e\_tot\_age\_lower & Gyr & lower uncertainty on the total age\\
sp\_type & & spectral type of the WD, if available\\
sp\_type\_source & & source of spectral type\tablenotemark{*}\\
wtd\_par & mas & weighted mean of the parallax of the system\\
e\_wtd\_par & mas & uncertainty on the weighted mean of the parallax of the system\\
sep\_AU & AU & projected on-sky separation for the pair\\
R\_chance\_align &  & chance alignment variable from \cite{2021MNRAS.506.2269E}\\
\enddata
\tablecomments{Columns dealing with properties of the individual WDs will have ``\_1" or ``\_2" at the end to denote the primary and secondary, respectively, where the primary has been defined as the more massive WD. The final 26 rows contain the wide double WD pairs from the triples found in this paper.}

\tablenotetext{*}{Spectral types in this work are collected from \citet{1991AJ....101.1476S, 1997ApJ...489L..79F, 1999ApJS..121....1M, 2004MNRAS.349.1397C, 2009ANA...505..441K, 2010ANA...524A..36K, 2011ApJ...730..128T, 2011ApJ...737...28B, 2011ApJ...743..138G, 2012MNRAS.421..202D, 2013ApJS..204....5K, 2013AJ....145..136L, 2014MNRAS.440.3184B, 2015ApJ...815...63A, 2015MNRAS.446.4078K, 2015SSRv..191..111F, 2015MNRAS.454.2787G, 2016MNRAS.455.3413K, 2019MNRAS.482.4570G, 2020ApJ...898...84K, 2020MNRAS.497..130T}}
\end{deluxetable*}

\pagebreak

\begin{longrotatetable}
\begin{deluxetable*}{cccccccccccc}
\label{tab:triples}
\tablecaption{WD+WD+X Triples}
\tablewidth{700pt}
\tabletypesize{\scriptsize}
\tablehead{
\colhead{System Name} & \colhead{source\_id} & 
\colhead{RA} & \colhead{Dec} & 
\colhead{$\mu_{RA}$} & \colhead{$\mu_{Dec}$} & 
\colhead{$\omega$} & \colhead{G} & \colhead{Stellar} &
\colhead{$\theta_{12}$} & \colhead{$\theta_{13}$} & \colhead{$\theta_{23}$} \\ 
\colhead{} & \colhead{(EDR3)} & \colhead{(J2016 $^\circ$)} & \colhead{(J2016 $^\circ$)} & 
\colhead{(mas yr$^{-1}$)} & \colhead{(mas yr$^{-1}$)} & \colhead{(mas)} & \colhead{(mag)} &
\colhead{Type} & \colhead{(AU)} & \colhead{(AU)} & \colhead{(AU)}
} 
\startdata 
         \tstrut WDWDMS & 105617338913173888 & 30.6841382 & 26.0846896 & $-21.52 \pm 0.08$ & $9.97 \pm 0.08$ & $3.75 \pm 0.08$ & 15.4 & MS & & &\\
         EDR3 & 105617304552984576 & 30.6768988 & 26.0867710 & $-21.55 \pm 0.39$ & $9.83 \pm 0.38$ & $3.93 \pm 0.34$ & 19.4 & WD & $6550$ & $6330$ & $386$\\
         105617338913173888 & 105617338913173760 & 30.6772592 & 26.0870098 & $-21.64 \pm 0.48$ & $9.80 \pm 0.44$ & $3.56 \pm 0.38$ & 19.5 & WD & & & \bstrut \\
         \hline
         \tstrut WDWDMS & 1463032750563461248 & 198.5937866 & 30.8450953 & $-47.76 \pm 0.01$ &  $26.49 \pm 0.01$ & $7.17 \pm 0.02$ & 11.9 & MS & & &\\
         EDR3 & 1463032819282938112 & 198.5894261 & 30.8474817 & $-48.30 \pm 0.09$ & $27.43 \pm 0.08$ & $7.16 \pm 0.11$ & 17.8 & WD & $2230$ & $2010$ & $385$\\
         1463032750563461248 & 1463032819282938240 & 198.5902718 & 30.8477281 & $-48.02 \pm 0.11$ & $25.96 \pm 0.10$ & $7.26 \pm 0.14$ & 18.1 & WD & & & \bstrut \\
         \hline
         \tstrut WDWDMS & 1932475596596345088 & 340.4531174 & 40.4721954 & $47.54 \pm 0.01$ & $140.41 \pm 0.01$ & $12.50 \pm 0.01$ & 12.4 & MS & & &\\
         EDR3 & 1932475596593097344 & 340.4527195 & 40.4734028 & $43.98 \pm 0.08$ & $141.16 \pm 0.09$ & $12.41 \pm 0.10$ & 17.6 & WD & $359$ & $5380$ & $5340$\\
         1932475596596345088 & 1932475562233361536 & 340.4758520 & 40.4792377 & $45.46 \pm 0.15$ & $140.12 \pm 0.19$ & $12.55 \pm 0.21$ & 18.9 & WD & & & \bstrut \\
         \hline
         \tstrut WDWDMS & 2063683029262340608 & 311.2472539 & 39.4521779 & $-18.82 \pm 0.17$ & $-34.76 \pm 0.20$ & $5.74 \pm 0.17$ & 18.9 & MS & & &\\
         EDR3 & 2063684201794288512 & 311.1961772 & 39.5036734 & $-18.42 \pm 0.21$ & $-34.48 \pm 0.23$ & $5.25 \pm 0.20$ & 19.2 & WD & 43200 & 43400 & 400\\
         2063683029262340608 & 2063684201788243712 & 311.1954031 & 39.5037395 & $-18.65 \pm 0.23$ & $-35.06 \pm 0.26$ & $5.06 \pm 0.22$ & 19.4 & WD & & & \bstrut \\
         \hline
         \tstrut WDWDMS & 2477316902742801152 & 22.1010462 & -8.3944721 & $-60.79 \pm 0.02$ & $-170.47 \pm 0.01$ & $9.41 \pm 0.02$ & 9.2 & MS & & &\\
         EDR3 & 2477317211979980544 & 22.1036359 & -8.3822745 & $-62.28 0.31$ & $-171.74 \pm 0.16$ & $10.75 \pm 0.28$ & 18.6 & WD & 4765 & 4879 & 128\\
         2477316902742801152 & 2477317211980446208 & 22.1038481 & -8.3820147 & $-61.01 \pm 0.86$ & $-170.16 \pm 0.17$ & $9.74 \pm 0.39$ & 18.8 & WD & & & \bstrut \\
         \hline
         \tstrut WDWDMS & 3197507047683918976 & 61.4674880 & -5.3325341 & $15.68 \pm 0.54$ & $-21.82 \pm 0.57$ & $3.95 \pm 0.59$ & 20.1 & WD & & &\\
         EDR3 & 3197506983261219328 & 61.4666405 & -5.3345423 & $17.00 \pm 0.74$ & $-22.07 \pm 0.65$ & $5.79 \pm 0.77$ & 20.3 & WD & 1700 & 18000 & 17000\\
         3197507047683918976 & 3197506944605295488 & 61.4588283 & -5.3547354 & $15.25 \pm 0.78$ & $-22.32 \pm 0.73$ & $4.70 \pm 0.85$ & 20.5 & MS & & & \bstrut\\
         \hline
         \tstrut WDWDMS & 3218914298562400640 & 88.5970332 & 0.6795522 & $-3.25 \pm 0.22$ & $-12.71 \pm 0.19$ & $3.64 \pm 0.22$ & 18.8 & WD & & & \\
         EDR3 & 3218914298560464768 & 88.5960159  & 0.6820507 & $-3.39 \pm 0.45$ & $-12.21 \pm 0.39$ & $4.50 \pm 0.45$ & 19.7 & WD & 2600 & 36000 & 38000\\
         3218914298562400640 & 3218912889811507072 & 88.6192839 & 0.6487874 & $-2.52 \pm 0.48$ & $-12.21 \pm 0.41$ & $3.78 \pm 0.49$ & 19.7 & MS & & & \bstrut\\
         \hline
         \tstrut WDWDMS & 3389527411923077376 & 82.8666421 & 13.4102532 & $56.74 \pm 0.02$ & $-19.96 \pm 0.01$ & $6.29 \pm 0.02$ & 9.3 & MS & & &\\
         EDR3 & 3389527338906751744 & 82.8642176 & 13.3994875 & $56.72 \pm 0.52$ & $-19.40 \pm 0.33$ & $6.70 \pm 0.49$ & 19.9 & WD & 6300 & 6690 & 390\\
         3389527411923077376 & 3389527343202017280 & 82.8640568 & 13.3988239 & $55.58 \pm 0.65$ & $-19.04 \pm 0.40$ & $6.33 \pm 0.59$ & 20.1 & WD & & & \bstrut \\
         \hline
         \tstrut WDWDMS & 338979267341825408 & 35.0462274 & 41.0692488 & $16.16 \pm 0.02$ & $-43.19 \pm 0.02$ & $6.11 \pm 0.02$ & 13.1 & MS & & &\\
         EDR3 & 338979301700437120 & 35.0674487 & 41.0816494 & $16.58 \pm 0.71$ & $-44.35 \pm 0.68$ & $6.91 \pm 0.57$ & 20.1 & WD & 11920 & 12370 & 460\\
         338979267341825408 & 338979297405358848 & 35.0681580 & 41.0822196 & $15.83 \pm 0.94$ & $-44.05 \pm 0.88$ & $5.67 \pm 0.72$ & 20.4 & WD & & & \bstrut \\
         \hline
         \tstrut WDWDMS & 3681220906103896192 & 188.0529329 & -4.1563234 & $-150.63 \pm 0.02$ & $0.21 \pm 0.02$ & $11.30 \pm 0.02$ & 8.8 & MS & & &\\
         EDR3 & 3681220906102862720 & 188.0547401 & -4.1571420 & $-153.55 \pm 0.07$ & $1.45 \pm 0.06$ & $11.35 \pm 0.07$ & 15.7 & WD & 630 & 750 & 208\\
         3681220906103896192 & 3681220906103896064 & 188.0547785 & -4.1577956 & $-151.04 \pm 0.06$ & $3.74 \pm 0.05$ & $11.33 \pm 0.06$ & 15.7 & WD & & & \bstrut \\
         \hline
         \tstrut WDWDMS & 4438147044497125248 & 243.7861861 & 6.3673258 & $-55.58 \pm 0.05$ & $21.32 \pm 0.04$ & $3.82 \pm 0.06$ & 16.2 & MS & & &\\ 
         EDR3 & 4438147418158270848 & 243.7956106 & 6.3658639 & $-55.44 \pm 0.65$ & $20.13 \pm 0.52$ & $3.41 \pm 0.61$ & 20.2 & WD & 8900 & 9500 & 554\\ 
         4438147044497125248 & 4438147422454430464 & 243.7961101 & 6.3655519 & $-54.09 \pm 0.81$ & $20.39 \pm 0.60$ & $3.74 \pm 0.66$ & 20.4 & WD & & & \bstrut \\
         \hline
         \tstrut WDWDMS & 4987518647789031040 & 13.6994498 & -40.0797643 & $10.22 \pm 0.01$ & $32.19 \pm 0.01$ & $5.27 \pm 0.01$ & 12.8 & MS & & &\\ 
         EDR3 & 4987518682147617024 & 13.6869636 & -40.0802902 & $10.16 \pm 0.42$ & $31.98 \pm 0.49$ & $5.93 \pm 0.51$ & 20.2 & WD & 6540 & 6390 & 285\\ 
         4987518647789031040 & 4987518682148769280 & 13.6872776 & -40.0806311 & $10.64 \pm 0.53$ & $32.40 \pm 0.68$ & $5.08 \pm 0.69$ & 20.5 & WD & & & \bstrut \\
         \hline
         \tstrut WDWDMS & 5041268170632888704 & 20.2704431 & -23.8818893 & $81.43 \pm 0.02$ & $-84.14 \pm 0.01$ & $8.48 \pm 0.02$ & 14.1 & MS & & &\\ 
         EDR3 & 5041268200697001600 & 20.2798805 & -23.8794056 & $82.22 \pm 0.45$ & $-83.36 \pm 0.33$ & $9.04 \pm 0.41$ & 19.8 & WD & 3810 & 4080 & 278\\ 
         5041268170632888704 & 5041268204992049664 & 20.2805981 & -23.8793989 & $82.86 \pm 0.64$ & $-85.21 \pm 0.46$ & $8.88 \pm 0.62$ & 20.3 & WD & & & \bstrut \\
         \hline
         \tstrut WDWDMS & 5250998123862710272 & 143.2102271 & -61.9153951 & $91.80 \pm 0.07$ & $-118.74 \pm 0.05$ & $8.43 \pm 0.04$ & 16.4 & MS & & &\\ 
         EDR3 & 5250998119556563200 & 143.2106552 & -61.9129685 & $92.89 \pm 0.23$ & $-117.90 \pm 0.15$ & $8.54 \pm 0.14$ & 18.9 & WD & 1038 & 18650 & 17670\\ 
         5250998123862710272 & 5251022416193428096 & 143.1869479 & -61.8730341 & $92.32 \pm 0.69$ & $-117.36 \pm 0.47$ & $8.89 \pm 0.41$ & 20.2 & WD & & & \bstrut \\ 
         \hline
         \tstrut WDWDMS & 5455352324194328192 & 158.0412243 & -29.9751592 & $72.86 \pm 0.01$ & $-5.94 \pm 0.02$ & $7.89 \pm 0.02$ & 10.5 & MS & & &\\ 
         EDR3 & 5455352702151342592 & 158.0283938 & -29.9535991 & $72.95 \pm 0.06$ & $-6.34 \pm 0.10$ & $7.78 \pm 0.10$ & 17.4 & WD & 11080 & 10020 & 1077\\ 
         5455352324194328192 & 5455352697853121280 & 158.0291298 & -29.9558703 & $73.17 \pm 0.23$ & $-5.65 \pm 0.41$ & $7.42 \pm 0.41$ & 19.7 & WD & & & \bstrut \\
         \hline
         \tstrut WDWDMS & 5560517931332969600 & 111.8889790 & -41.3676372 & $-154.05 \pm 0.01$ & $-4.86 \pm 0.01$ & $9.65 \pm 0.01$ & 11.5 & MS & & &\\ 
         EDR3 & 5560518000052459520 & 111.9127938 & -41.3679211 & $-153.72 \pm 0.06$ & $-5.43 \pm 0.06$ & $9.61 \pm 0.06$ & 17.1 & WD & 6671 & 6676 & 559\\ 
         5560517931332969600 & 5560518000048274304 & 111.9127590 & -41.3664248 & $-153.02 \pm 0.79$ & $-6.17 \pm 0.91$ & $9.91 \pm 0.76$ & 20.5 & WD & & & \bstrut \\
         \hline
         \tstrut WDWDMS & 5874024705460836992 & 224.4952433 & -63.2877651 & $-33.67 \pm 0.02$ & $-54.39 \pm 0.02$ & $8.60 \pm 0.02$ & 8.1 & MS & & &\\ 
         EDR3 & 5874024769842933760 & 224.5481622 & -63.2928712 & $-34.04 \pm 0.06$ & $-54.37 \pm 0.06$ & $8.49 \pm 0.06$ & 16.8 & WD & 10190 & 10440 & 299\\ 
         5874024705460836992 & 5874024774146311808 & 224.5496572 & -63.2926298 & $-33.33 \pm 0.09$ & $-53.16 \pm 0.09$ & $8.75 \pm 0.09$ & 17.5 & WD & & & \bstrut \\
         \hline
         \tstrut WDWDMS & 6087696442178470400 & 202.4274557 & -44.0483557 & $-14.23 \pm 0.08$ & $-5.96 \pm 0.06$ & $3.89 \pm 0.07$ & 16.8 & MS & & &\\ 
         EDR3 & 6087695613244135936 & 202.4327137 & -44.0750059 & $-14.20 \pm 0.20$ & $-6.22 \pm 0.17$ & $3.48 \pm 0.21$ & 18.9 & WD & 25200 & 25800 & 840\\ 
         6087696442178470400 & 6087695617537911040 & 202.4318915 & -44.0756854 & $-15.18 \pm 0.33$ & $-5.69 \pm 0.26$ & $3.53 \pm 0.31$ & 19.6 & WD & & & \bstrut \\
         \hline
         \tstrut WDWDMS & 6145125411962111104 & 184.8137056 & -42.3393993 & $-25.88 \pm 0.02$ & $9.91 \pm 0.02$ & $2.88 \pm 0.03$ & 15.4 & MS & & &\\ 
         EDR3 & 6145125411959049856 & 184.8084702 & -42.3439378 & $-25.86 \pm 0.26$ & $10.15 \pm 0.20$ & $3.35 \pm 0.32$ & 19.5 & WD & 7450 & 7930 & 482\\ 
         6145125411962111104 & 6145125411962108416 & 184.8080984 & -42.3442087 & $-25.93 \pm 0.34$ & $9.64 \pm 0.31$ & $3.03 \pm 0.48$ & 19.7 & WD & & & \bstrut \\
         \hline
         \tstrut WDWDMS & 6756233901661753472 & 288.8382875 & -32.3347021 & $15.70 \pm 0.02$ & $4.10 \pm 0.01$ & $6.05 \pm 0.02$ & 11.8 & MS & & &\\ 
         EDR3 & 6756233905958975360 & 288.8369795 & -32.3330247 & $15.28 \pm 0.36$ & $4.63 \pm 0.29$ & $6.67 \pm 0.34$ & 19.4 & WD & 1195 & 2191 & 1757\\ 
         6756233901661753472 & 6756233905958976512 & 288.8339392 & -32.3344810 & $16.28 \pm 0.52$ & $2.76 \pm 0.44$ & $5.71 \pm 0.47$ & 19.9 & WD & & & \bstrut \\
         \hline
         \tstrut WDWDMS & 763543514763336832 & 169.8371661 & 37.0721947 & $90.75 \pm 0.01$ & $-53.81 \pm 0.02$ & $8.82 \pm 0.02$ & 12.0 & MS & & &\\ 
         EDR3 & 763543312899793664 & 169.8497518 & 37.0715253 & $90.80 \pm 0.48$ & $-54.38 \pm 0.58$ & $9.27 \pm 0.51$ & 19.8 & WD & 4108 & 3937 & 423\\ 
         763543514763336832 & 763543308609191936 & 169.8490934 & 37.0706329 & $89.68 \pm 0.48$ & $-54.14 \pm 0.59$ & $8.93 \pm 0.50$ & 19.9 & WD & & & \bstrut \\
         \hline
         \tstrut WDWDMS & 779305808516516864 & 161.4150757 & 40.0952785 & $-72.91 \pm 0.03$ & $21.17 \pm 0.03$ & $6.99 \pm 0.03$ & 14.9 & MS & & &\\ 
         EDR3 & 779305808516517376 & 161.4055376 & 40.0933527 & $-74.03 \pm 0.22$ & $21.61 \pm 0.25$ & $7.23 \pm 0.27$ & 18.8 & WD & 3880 & 4050 & 183\\ 
         779305808516516864 & 779305808516231552 & 161.4050805 & 40.0934117 & $-71.23 \pm 0.25$ & $21.35 \pm 0.24$ & $7.36 \pm 0.29$ & 18.9 & WD & & & \bstrut \\
         \hline
         \tstrut WDWDMS & 929001808377561216 & 121.6861312 & 44.7481737 & $3.01 \pm 0.03$ & $-3.41 \pm 0.03$ & $4.99 \pm 0.03$ & 15.5 & MS & & &\\ 
         EDR3 & 929001804082135168 & 121.6837451 & 44.7508697 & $3.08 \pm 0.15$ & $-3.38 \pm 0.11$ & $4.93 \pm 0.14$ & 18.2 & WD & 2230 & 2710 & 1023\\
         929001808377561216 & 929001804082135296 & 121.6818686 & 44.7503870 & $3.46 \pm 0.21$ & $-2.99 \pm 0.16$ & $5.00 \pm 0.20$ & 18.8 & WD & & & \bstrut \\
         \hline
         \tstrut WDWDWD & 4190499986125543168 & 298.3999651 & -10.3254921 & $-11.48 \pm 0.07$ & $-16.52 \pm 0.04$ & $7.75 \pm 0.06$ & 16.3 & WD & & &\\ 
         EDR3 & 4190499986125543296 & 298.4001112 & -10.3248602 & $-10.92 \pm 0.07$ & $-15.78 \pm 0.04$ & $7.66 \pm 0.06$ & 16.4 & WD & 302 & 6280 & 6490\\ 
         4190499986125543168 & 4190500054845023488 & 298.3879711 & -10.3319738 & $-10.96 \pm 0.11$ & $-16.14 \pm 0.06$ & $7.76 \pm 0.10$ & 17.3 & WD & & & \bstrut \\
\enddata
\end{deluxetable*}
\end{longrotatetable}

\section{Shifted IFMR}

The table used for the initial-to-final-mass relation in Figure~\ref{fig:IFMR_fit} is detailed here in Table~\ref{tab:IFMR}, adapted from \citet{2016ApJ...823...46F}. The WD mass is $M\textsubscript{f}$ and the ZAMS mass is $M\textsubscript{i}$.

\begin{deluxetable}{cc|cc}[h!]
\label{tab:IFMR}
\tablecaption{Shifted MESA IFMR}
%\tablewidth{700pt}
\tabletypesize{\scriptsize}
\tablehead{\colhead{$M\textsubscript{i}$ $\left[M_\odot\right]$} & \colhead{$M\textsubscript{f}$ $\left[M_\odot\right]$} & \colhead{$M\textsubscript{i}$ $\left[M_\odot\right]$} & \colhead{$M\textsubscript{f}$ $\left[M_\odot\right]$}
}
\startdata
1.0 & 0.583 & 3.5 & 0.802\\
1.1 & 0.594 & 3.6 & 0.828\\
1.2 & 0.601 & 3.7 & 0.848\\
1.3 & 0.608 & 3.8 & 0.868\\
1.4 & 0.613 & 3.9 & 0.880\\
1.5 & 0.615 & 4.0 & 0.888\\
1.6 & 0.619 & 4.1 & 0.895\\
1.7 & 0.621 & 4.2 & 0.901\\
1.8 & 0.624 & 4.3 & 0.906\\
1.9 & 0.625 & 4.4 & 0.913\\
2.0 & 0.629 & 4.5 & 0.919\\
2.1 & 0.634 & 4.6 & 0.927\\
2.2 & 0.640 & 4.7 & 0.930\\
2.3 & 0.645 & 4.8 & 0.939\\
2.4 & 0.651 & 4.9 & 0.955\\
2.5 & 0.657 & 5.0 & 0.960\\
2.6 & 0.664 & 5.1 & 0.957\\
2.7 & 0.672 & 5.2 & 0.965\\
2.8 & 0.682 & 5.3 & 0.974\\
2.9 & 0.697 & 5.4 & 0.983\\
3.0 & 0.709 & 5.5 & 0.994\\
3.1 & 0.724 & 5.6 & 1.006\\
3.2 & 0.743 & 5.7 & 1.020\\
3.3 & 0.763 & 5.9 & 1.023\\
3.4 & 0.780 & 6.0 & 1.033\\
\enddata
\end{deluxetable}

\end{document}